\documentclass[twocolumn]{aastex63}

\usepackage{amssymb}	
\usepackage{multirow}   
\usepackage{courier}
\usepackage{lipsum}
\usepackage{longtable}

\usepackage{rotating, graphicx}

\usepackage{color}
\usepackage{url}
\usepackage{ulem}
\usepackage{multirow}
\usepackage{amsmath}
\usepackage{wasysym}
\usepackage{newtxtext,newtxmath}

\usepackage[T1]{fontenc}

\DeclareRobustCommand{\VAN}[3]{#2}
\let\VANthebibliography\thebibliography
\def\thebibliography{\DeclareRobustCommand{\VAN}[3]{##3}\VANthebibliography}

\def\apjl{ApJ}%
\def\apjs{ApJS}%
\def\ao{Appl.~Opt.}%
\def\aap{A\&A}%
\newcommand\C   {\textsc{Clumpy}}

\shorttitle{BASS XLI: MIR emission lines}
\shortauthors{M. Bierschenk et al.}

\begin{document}

\title{BASS XLI: the correlation between Mid-infrared emission lines and Active Galactic Nuclei emission}
\author{M. Bierschenk}
\affiliation{Department of Physics and Astronomy, George Mason University, 4400 University Drive, MSN 3F3, Fairfax, VA 22030, USA}
\author[0000-0001-5231-2645]{C. Ricci}
\affiliation{Instituto de Estudios Astrof\'{\i}sicos, Facultad de Ingenier\'{\i}a y Ciencias, Universidad Diego Portales, Avenida Ejercito Libertador 441, Santiago, Chile}
\affiliation{Kavli Institute for Astronomy and Astrophysics, Peking University, Beijing 100871, China}
\author[0000-0001-8433-550X]{M. J. Temple}
\affiliation{Instituto de Estudios Astrof\'{\i}sicos, Facultad de Ingenier\'{\i}a y Ciencias, Universidad Diego Portales, Avenida Ejercito Libertador 441, Santiago, Chile}
\author[0000-0003-2277-2354]{S. Satyapal}
\affiliation{Department of Physics and Astronomy, George Mason University, 4400 University Drive, MSN 3F3, Fairfax, VA 22030, USA}
\author[0000-0003-1051-6564]{J. Cann}
\affiliation{X-ray Astrophysics Laboratory, NASA Goddard Space Flight Center, Code 662, Greenbelt, MD 20771, USA}
\affiliation{Oak Ridge Associated Universities, NASA NPP Program, Oak Ridge, TN 37831, USA}
\author[0000-0002-9707-1037]{Y. Xie}
\affiliation{Kavli Institute for Astronomy and Astrophysics, Peking University, Beijing 100871, China}
\author[0000-0002-8604-1158]{Y. Diaz}
\affiliation{Instituto de Estudios Astrof\'{\i}sicos, Facultad de Ingenier\'{\i}a y Ciencias, Universidad Diego Portales, Avenida Ejercito Libertador 441, Santiago, Chile}
\author[0000-0002-4377-903X]{K. Ichikawa}
\affiliation{Astronomical Institute, Tohoku University, Aramaki, Aoba-ku, Sendai, Miyagi 980-8578, Japan}
\affiliation{Frontier Research Institute for Interdisciplinary Sciences, Tohoku University, Sendai 980-8578, Japan}
\author[0000-0002-7998-9581]{M. J. Koss}
\affiliation{Eureka Scientific, 2452 Delmer Street Suite 100, Oakland, CA 94602-3017, USA}
\affiliation{Space Science Institute, 4750 Walnut Street, Suite 205, Boulder, CO 80301, USA}
\author[0000-0002-8686-8737]{F. E. Bauer}
\affiliation{Instituto de Astrof{\'{\i}}sica, Facultad de F{\'{i}}sica, Pontificia Universidad Cat{\'{o}}lica de Chile, Av. Vicuña Mackenna 4860, Macul Santiago,  7820436, Chile}
\affiliation{Centro de Astroingenier{\'{\i}}a, Facultad de F{\'{i}}sica, Pontificia Universidad Cat{\'{o}}lica de Chile, Av. Vicuña Mackenna 4860, Macul Santiago, 7820436, Chile}
\affiliation{Millennium Institute of Astrophysics, Nuncio Monse{\~{n}}or S{\'{o}}tero Sanz 100, Of 104, Providencia, Santiago, Chile}
\affiliation{Space Science Institute, 4750 Walnut Street, Suite 205, Boulder, CO 80301, USA}
\author[0000-0003-0006-8681]{A. Rojas}
\affiliation{Centro de Astronom\'ia (CITEVA), Universidad de Antofagasta, Avenida Angamos 601, Antofagasta, Chile}
\affiliation{Instituto de Estudios Astrof\'{\i}sicos, Facultad de Ingenier\'{\i}a y Ciencias, Universidad Diego Portales, Avenida Ejercito Libertador 441, Santiago, Chile}
\author[0000-0002-2603-2639]{D. Kakkad}
\affiliation{Space Telescope Science Institute, 3700 San Martin Drive, Baltimore, MD 21218, USA}
\author[0000-0003-3450-6483]{A. Tortosa}
\affiliation{INAF -- Osservatorio Astronomico di Roma, Via Frascati 33, I–00078 Monte Porzio Catone, Italy}
\author[0000-0001-5742-5980]{F. Ricci}
\affiliation{Dipartimento di Matematica e Fisica, Universita Roma Tre, via della Vasca Navale 84, I-00146 Roma, Italy}
\author[0000-0002-7962-5446]{R. Mushotzky}
\affiliation{Department of Astronomy, University of Maryland, College Park, MD 20742, USA}
\affiliation{Joint Space-Science Institute, University of Maryland, College Park, MD 20742, USA}
\author[0000-0002-6808-2052]{T. Kawamuro}
\affiliation{RIKEN Cluster for Pioneering Research, 2-1 Hirosawa, Wako, Saitama 351-0198, Japan}
\author[0009-0007-9018-1077]{K. K. Gupta}
\affiliation{STAR Institute, Li\`ege Universit\'e, Quartier Agora - All\'ee du six Ao\^ut, 19c B-4000 Li\`ege, Belgium}
\affiliation{Sterrenkundig Observatorium, Universiteit Gent, Krijgslaan 281 S9, B-9000 Gent, Belgium}
\author[0000-0002-3683-7297]{B. Trakhtenbrot}
\affiliation{School of Physics and Astronomy, Tel Aviv University, Tel Aviv 69978, Israel}
\author[0000-0001-9910-3234]{C.S. Chang}
\affiliation{Joint ALMA Observatory, Avenida Alonso de Cordova 3107, Vitacura 7630355, Santiago, Chile}
\author[0000-0002-1321-1320]{R. Riffel}
\affiliation{Departamento de Astronomia, Instituto de F\'\i sica, Universidade Federal do Rio Grande do Sul, CP 15051, 91501-970, Porto Alegre, RS, Brazil}
\affiliation{Instituto de Astrof\'\i sica de Canarias, Calle V\'\i a L\'actea s/n, E-38205 La Laguna, Tenerife, Spain}
\author[0000-0002-5037-951X]{K. Oh}
\affiliation{Korea Astronomy \& Space Science institute, 776, Daedeokdae-ro, Yuseong-gu, Daejeon 34055, Republic of Korea\\}
\author[0000-0002-4226-8959]{F. Harrison}
\affiliation{Cahill Center for Astronomy and Astrophysics, California Institute of Technology, Pasadena, CA 91125, USA}
\author[0000-0003-2284-8603]{M. Powell}
\affiliation{Kavli Institute for Particle Astrophysics and Cosmology, Stanford University, 452 Lomita Mall, Stanford, CA 94305, USA}
\affiliation{Department of Physics, Stanford University, 382 Via Pueblo Mall, Stanford, CA 94305, USA}
\author[0000-0003-2686-9241]{D. Stern}
\affiliation{Jet Propulsion Laboratory, California Institute of Technology, 4800 Oak Grove Drive, MS 169-224, Pasadena, CA 91109, USA}
\author[0000-0002-0745-9792]{C. M. Urry}
\affiliation{Yale Center for Astronomy \& Astrophysics, Physics Department, PO Box 208120, New Haven, CT 06520-8120, USA}

\correspondingauthor{Claudio Ricci}
\email{claudio.ricci@mail.udp.cl}

\begin{abstract}
We analyze the \textit{Spitzer} spectra of 140 active galactic nuclei (AGN) detected in the hard X-rays (14-195\,keV) by the Burst Alert Telescope (BAT) on board \textit{Swift}. This sample allows us to probe several orders of magnitude in black hole masses ($10^6-10^9 M_{\odot}$), Eddington ratios ($10^{-3}-1$), X-ray luminosities ($10^{42}-10^{45}\rm\,erg\,s^{-1}$), and X-ray column densities ($10^{20}-10^{24}\rm\,cm^{-2}$). The AGN emission is expected to be the dominant source of ionizing photons with energies $\gtrsim50$\,eV, and therefore high-ionization mid-infrared (MIR) emission lines such as [{\textsc Ne\,V}] 14.32, 24.32\,$\mu$m and [{\textsc O\,IV}] 25.89\,$\mu$m are predicted to be good proxies of AGN activity, and robust against obscuration effects. 
We find high detection rates ($\gtrsim85-90$ per cent) for the mid-infrared coronal emission lines in our AGN sample. The luminosities of these lines are correlated with the 14--150\,keV luminosity (with a typical scatter of $\sigma \sim 0.4-0.5$\,dex), strongly indicating that the mid-infrared coronal line emission is driven by AGN activity. Interestingly, we find that the coronal lines are more tightly correlated to the bolometric luminosity ($\sigma \sim 0.2-0.3$\,dex), calculated from careful analysis of the spectral energy distribution, than to the X-ray luminosity. We find that the relationship between the coronal line strengths and $L_{14-150\rm\,keV}$ is independent of black hole mass, Eddington ratio and X-ray column density. This confirms that the mid-infrared coronal lines can be used as unbiased tracers of the AGN power for X-ray luminosities in the $10^{42}-10^{45}\rm\,erg\,s^{-1}$ range.
\end{abstract}	
               

          \setcounter{footnote}{0}

\section{Introduction}

It is well established that supermassive black holes (SMBHs) lie at the center of most massive galaxies, and that they could play an important role in the evolution of their host galaxy (e.g., \citealp{1995ARA&A..33..581K}; \citealp{1998AJ....115.2285M}; \citealp{2000ApJ...539L..13G}; \citealp{2013ARA&A..51..511K}). SMBHs can grow by accreting gas and dust from their host galaxies.  These accreting SMBHs, or active galactic nuclei (AGN), are some of the brightest objects in the Universe, emitting large amounts of radiation across the entire electromagnetic spectrum (e.g., \citealp{1994ApJS...95....1E}, \citealp{doi:10.1146/annurev.astro.45.051806.110546}, \citealp{10.1111/j.1365-2966.2011.18448.x}). Some of the most luminous AGN have been found to have bolometric luminosities up to $10^{48}\, \mathrm{erg}~\mathrm{s}^{-1}$ (e.g., \citealp{Assef:2015zr,Diaz-Santos:2016no,Bischetti:2017uw,Martocchia:2017rj,Zappacosta:2020lw}).
Surrounding the SMBH and its accretion flow is a pc-scale anisotropic structure of gas and dust, often referred to as the torus, which absorbs and reprocesses the radiation from the accretion disk (e.g., \citealp{1993ARA&A..31..473A}, \citealp{1995PASP..107..803U},  \citealp{2015ARA&A..53..365N}, \citealp{2017NatAs...1..679R}).
According to the simplest AGN unification model (e.g., \citealp{1993ARA&A..31..473A}), depending on the viewing angle with respect to the obscuring material, light from the AGN can suffer from strong extinction. About 70\% of the AGN in the local Universe are obscured by gas with column densities $N_{\rm H}\geq 10^{22}\rm\,cm^{-2}$ (e.g., \citealp{Ricci:2015fk,Ricci:2017pm}), and a significant fraction ($\sim 20-30\%$) of them are obscured by Compton-thick material ($N_{\rm H}\geq 10^{24}\rm\,cm^{-2}$; \citealp{2011ApJ...728...58B,Ricci:2015fk,Torres-Alba:2021kl,Tanimoto:2022id}). 

One of the best ways to identify AGN is to select them in the hard X-ray band (${>}10$\,keV), where contamination from the host galaxy is negligible. Hard X-rays are also not strongly affected by obscuration, at least up to $N_\textrm{H} \sim 10^{24}$\,cm$^{-2}$ (e.g., \citealp{Ricci:2015fk}), and therefore they can provide an almost unbiased sample of AGN in the local universe. However, hard X-rays can be attenuated for $ N_\textrm{H} \gtrsim 10^{24}\rm\,cm^{-2}$, so the most heavily obscured AGN can be easily missed by hard X-ray surveys \citep{Ricci:2015fk}. 
Obscured AGN can also be identified using optical emission line ratios (e.g., \citealp{1981PASP...93....5B}, \citealp{1987ApJS...63..295V}, \citealp{2001ApJ...556..121K}, \citealp{2003MNRAS.346.1055K}), although it is important to stress that a significant fraction of hard X-ray selected AGN may not be classified as AGN by optical emission line spectroscopy due to contamination from star formation or merger-associated dust obscuration (e.g., \citealp{Ricci:2017aa,Koss:2018cw,Yamada:2021vz}). 

Another method to identify AGN is via the detection of forbidden e mission lines with ionization potentials (IPs) $\gtrsim$54\,eV (i.e. the double ionization edge of Helium), commonly referred to as `coronal lines' (CLs). In AGN, the coronal lines are emitted in the so-called coronal-line region, likely located in the narrow line region and beyond the broad line region (e.g., \citealp{Negus:2023rs}). The presence of coronal lines can serve as an important indicator of AGN activity, as they are extremely difficult to produce through stellar processes and are more likely created by AGN photoionization (e.g., \citealp{Sturm:2002tw,Abel:2008kx,Goulding:2009dq,Satyapal:2007nx,Satyapal:2008uq,Satyapal:2009me,Pereira-Santaella:2010kx,Gruppioni:2016sy,2018MNRAS.480.5203M,Cann:2018am,Spinoglio:2022of,Feltre:2023lz,McKaig24,Zhang:2024wj,Hermosa-Munoz:2024wd}). However, some authors have suggested that coronal line emission could also be enhanced by shocks, driven by the interaction of a radio jet and the interstellar medium, by outflowing material or associated to mergers (e.g., \citealp{Izotov:2012kv,Mazzalay:2013co, 2017MNRAS.465..906R, Rodriguez-Ardila:2020ii,Fonseca-Faria:2021rw,Negus:2021sv,Campos21,Campos23}). 
With the recent launch of the {\it James Webb Space Telescope} (\textit{JWST}; \citealp{Gardner:2006gd}), CLs have gained significant relevance. Now is an ideal time to revisit the use of infrared CLs as proxies of AGN activity, and to refine our understanding of how CL emission varies across the broader AGN population.

\begin{figure*}
    \centering
 \includegraphics[width=0.48\textwidth]{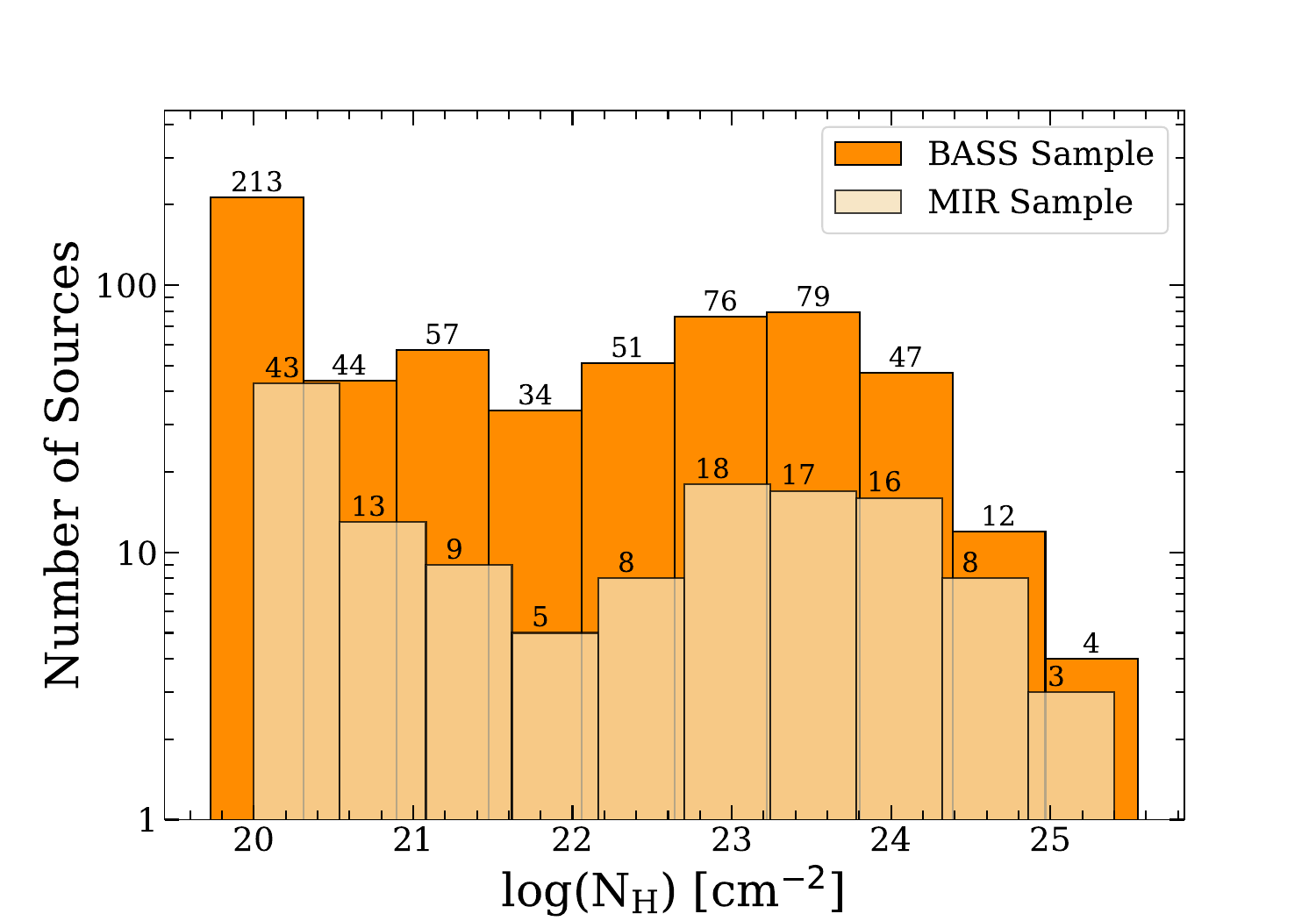}
   \includegraphics[width=0.48\textwidth]{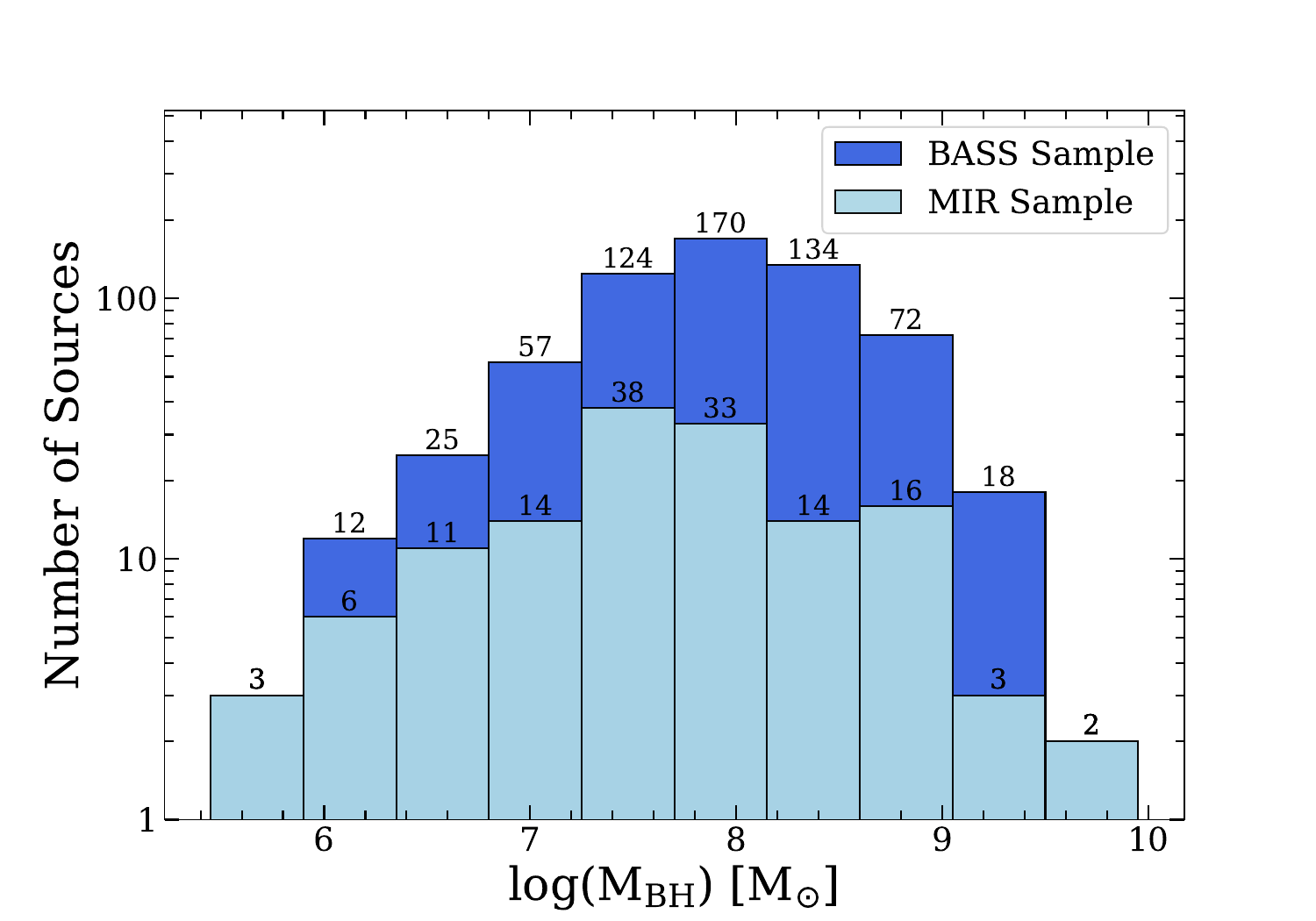} 
 \includegraphics[width=0.48\textwidth]{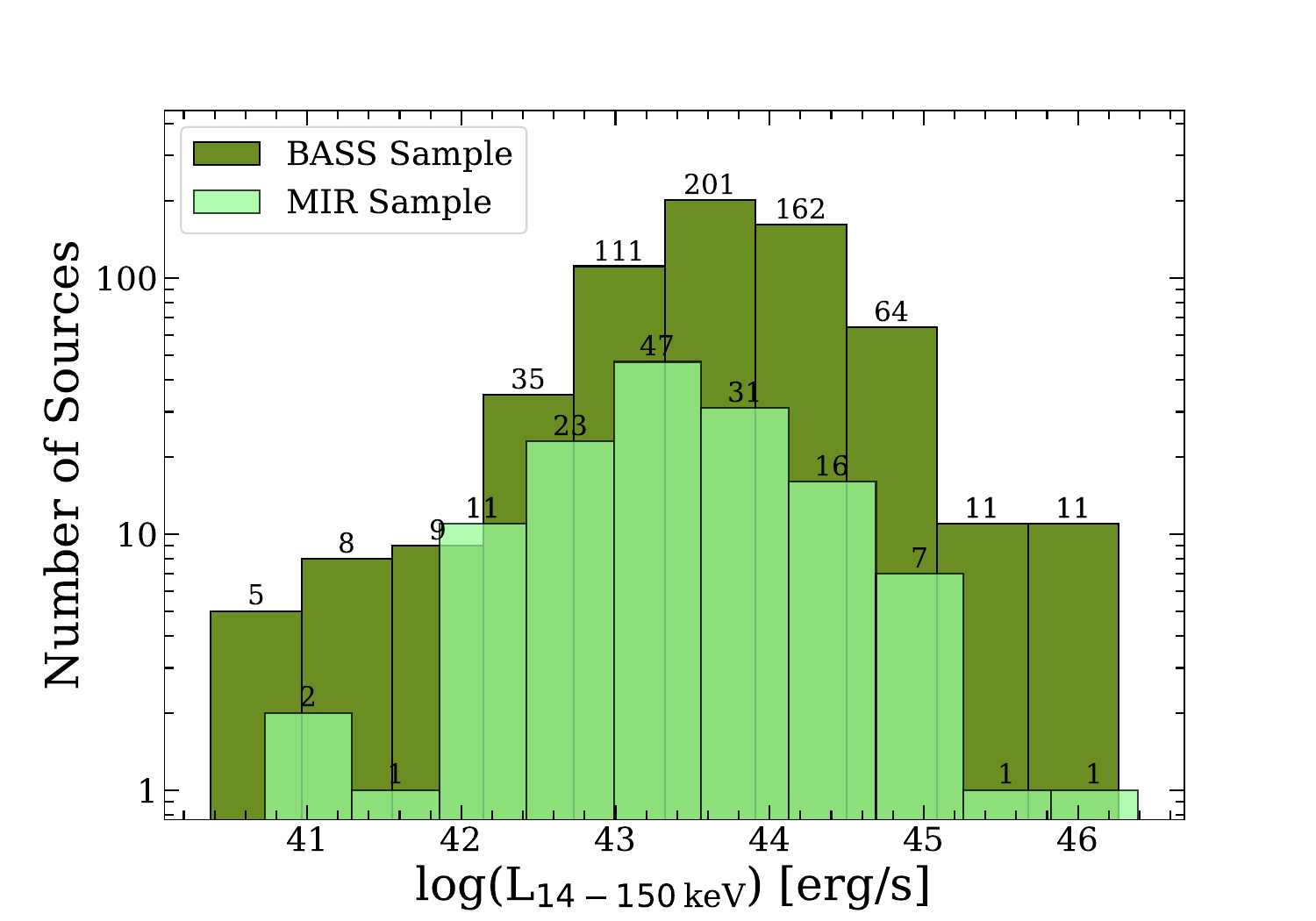} 
    \includegraphics[width=0.48\textwidth]{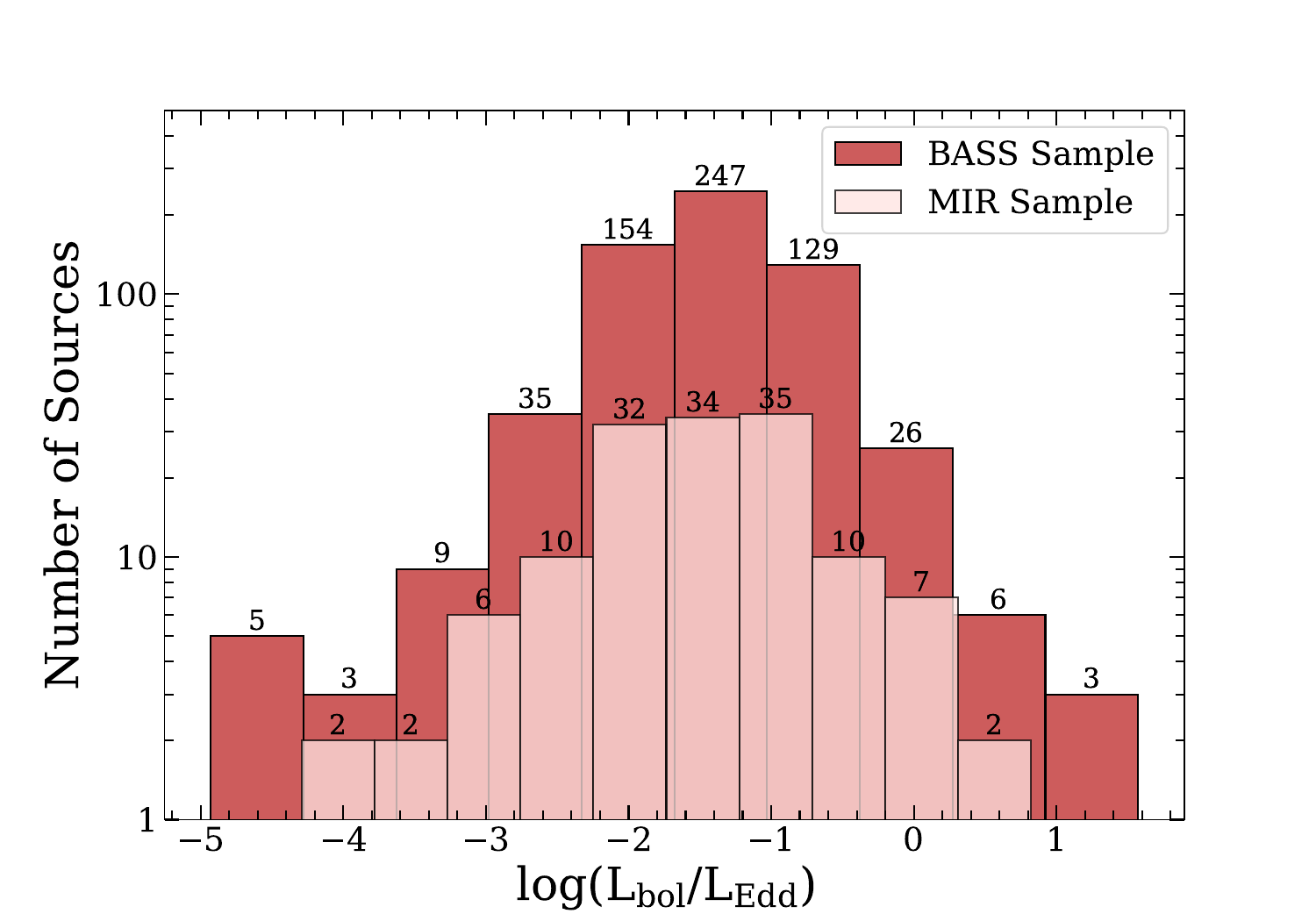} 
    \caption{The number of sources as a function of obscuring column density [log(N$_\textrm{H}$)], black hole mass [log(M$_\textrm{BH}$)], intrinsic 14--150\,keV X-ray luminosity [log($L_\textrm{14-150\,keV}$)], and Eddington ratio (log($\lambda_{\rm Edd}$)] for the parent BASS sample (darker shades) and the sample with MIR spectra used in this work (lighter shades).}
    \label{fig:Sample_AGN_properties}
\end{figure*}

Constraining the bolometric luminosity of an AGN is crucial to understand the energetics of such systems and to infer their accretion rates. Since most AGN are obscured (with no direct constraints on the dominant UV emission), estimating their intrinsic luminosity can be complicated. Several empirical relations linking X-ray luminosities to optical lines (e.g., \citealp{1999ApJS..121..473B}), mid-infrared continuum (MIR; \citealp{2009A&A...502..457G,2015MNRAS.454..766A}) and more recently the nuclear 100\,GHz \citep{Ricci:2023vy} and 200\,GHz \citep{Kawamuro:2022mo,Kawamuro:2023ux} mm continuum luminosities, have been used for this purpose. \citet{1999ApJS..121..473B} showed that the luminosity of [{\textsc O\,III}]\,5007\AA\ is correlated with X-ray luminosity (see also 
\citealp{2015MNRAS.454.3622B}, \citealp{Ueda2015},\citealp{Oh:2017di,Oh:2022fh}), albeit with a large ($\sim0.6-0.7$\,dex) scatter. Some of these correlations with photoionized optical lines can, in fact, be strongly affected by obscuration, AGN variability, contamination from stellar sources, and biased samples, which would greatly enhance their scatter. Near-infrared (NIR) CLs of hard X-ray selected AGN (e.g., [Si VI] 1.96$\mu$m), have been shown to be significantly more tightly correlated to the X-ray luminosity, (with a scatter of 0.37\,dex) than the optical [O III] $\lambda$ 5007 line (0.71 dex), possibly due to obscuration or the higher ionization required (e.g., \citealp{Lamperti:2017uz}, \citealp{den-Brok:2022oy}).

The MIR is an ideal waveband to probe AGN activity in obscured systems, due to the relatively low effect of dust attenuation (e.g., \citealp{2005ApJ...633..706W}, \citealp{2006ApJ...640..204A}, \citealp{Goulding:2009dq}). \citeauthor{2009A&A...502..457G} (\citeyear{2009A&A...502..457G}; see also \citealp{2015MNRAS.454..766A}) found that the nuclear MIR luminosities of AGN are strongly correlated to their X-ray luminosities. However, high-resolution (HR) observations are needed to mitigate the influence of the host galaxy, particularly for low-luminosity AGN.
In the MIR there are several forbidden lines which could be used as tracers of AGN activity.
Examples of these lines are [{\textsc Ne\,V}] 14.32 and 24.32\,\micron\ (e.g., \citealp{Satyapal:2007nx}, \citealp{Satyapal:2008uq}), [{\textsc O\,IV}] 25.89\,\micron\ (e.g., \citealp{2008ApJ...682...94M}), and [{\textsc S\,III}] 18.71 and 33.48\,\micron\  which have ionization energies of 97, 55 and 24\,eV respectively (see Table 1 in \citealp{2021ApJ...906...35S} for a list of CL ionization energies). Therefore, these high-ionization MIR emission lines have the potential to be good proxies of the intrinsic luminosity of AGN, even in obscured systems where it can be particularly complicated to estimate the AGN power (e.g., \citealp{2018ARA&A..56..625H}). It should be mentioned, however, that this might not be the case for AGN in galaxies undergoing the final stages of a merger, where [{\textsc Ne\,V}] and [{\textsc O\,IV}] have been found to be weak \citep{Yamada:2024pm}, likely because of the very large covering factor of the obscuring material \citep{Ricci:2017aa,Yamada:2024pm}.

In this work, we present the largest study to date of MIR coronal emission lines for a sample of hard X-ray selected AGN, to test whether they can be used as proxies of AGN activity. We carefully analyzed the MIR spectra of 140 sources using \textit{Spitzer}/IRS data. Our sample is selected by cross-matching the $\sim 860$ sources of the {\it Swift} BAT AGN Spectroscopic Survey (BASS\footnote{\url{www.bass-survey.com}}) with the \textit{Spitzer} archive. The BASS sample contains a wealth of multi-wavelength ancillary data and derived properties (e.g., SMBH mass, accretion rate, and column density; \citealp{Ricci:2017pm,Koss:2017wu,Koss:2022qi,2022ApJS..261....5M}), which allows us to test the dependence of CL fluxes on AGN properties, expanding previous studies of smaller samples (e.g., \citealp{2008ApJ...682...94M,2008ApJ...676..836T,2010ApJ...716.1151W,2010ApJ...709.1257T,Gruppioni:2016sy,Spinoglio:2022of,Feltre:2023lz}). The structure of the paper is as follows. Our sample, the \textit{Spitzer}/IRS and multi-wavelength data we use are presented in \S\ref{sec:Data}. The fitting methods for determining the fluxes of the forbidden emission lines [{\textsc Ne\,V}] 14.32/24.32\,\micron, [{\textsc O\,IV}] 25.89\,\micron, and [{\textsc S\,III}] 18.71/33.48\,\micron\ are provided in \S\ref{sec:Fitting}. Our analysis is described in \S\ref{sec:analyses}, while \S\ref{sec: results} is dedicated to results and discussion. Throughout this work we assume the following standard cosmological parameters: $H_{0} = 71$\,km\,s$^{-1}$\,Mpc$^{-1}$, $\Omega_{m} = 0.30$, and $\Omega_{\Lambda} = 0.70$.

\section{Sample and Data}
\label{sec:Data}

\begin{figure}
    \centering
    \includegraphics[width = \columnwidth]{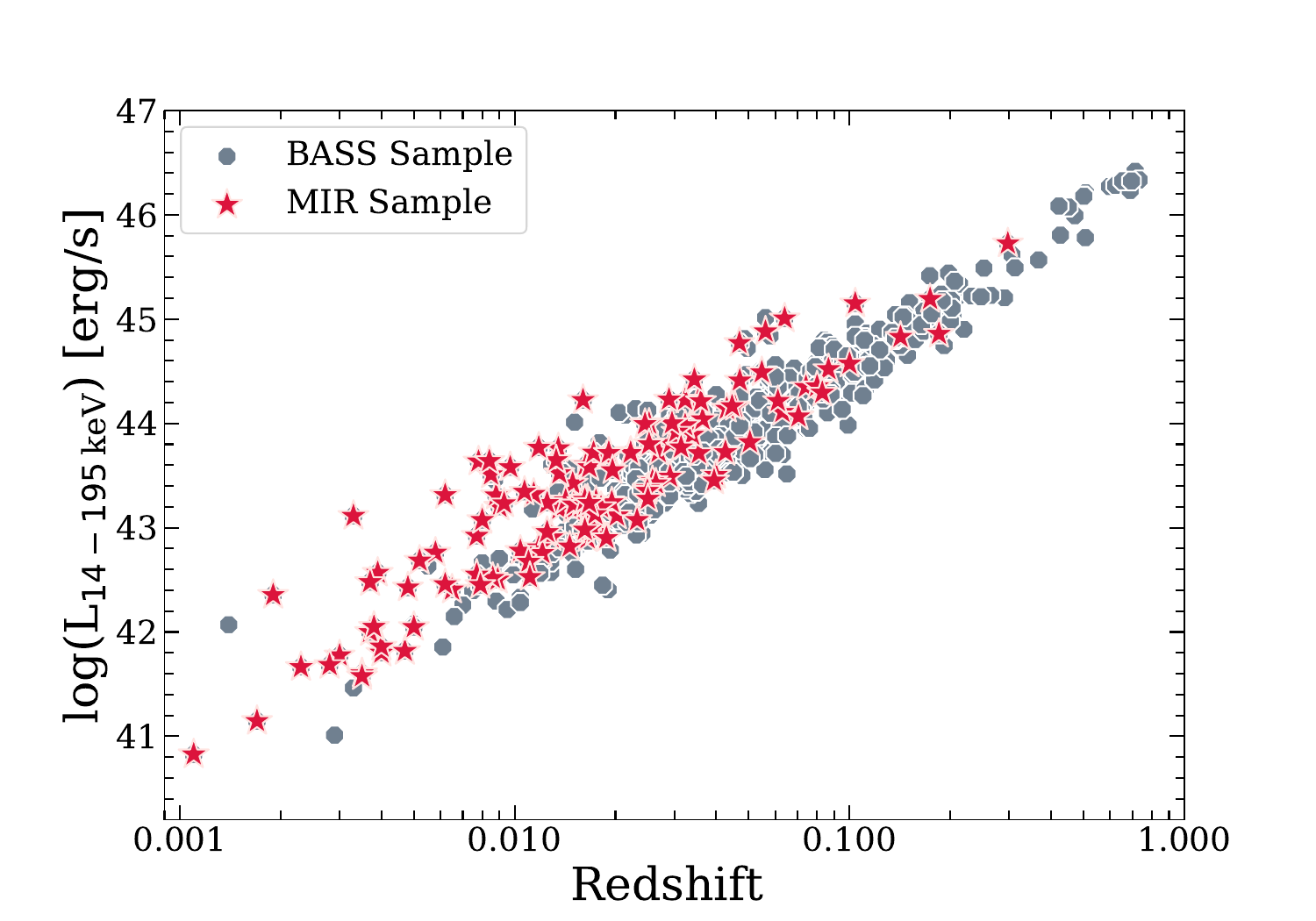}
    \caption{The 14-195\,keV X-ray luminosities as a function of redshift for the sample with MIR spectroscopy used in this work (red stars) compared to the complete BAT-selected  sample (gray points).
    The majority of the objects with \textit{Spitzer} IRS spectra have $z<0.1$, and span nearly the full range of luminosities seen in the wider BASS population.}
    \label{fig:redshift}
\end{figure}

\subsection{BASS Sample: Data \& Derived Properties}
To test the effectiveness of high-ionization emission lines as a proxy for AGN activity, we use the sample of AGN detected in the 14--195\,keV band by the Burst Alert Telescope (BAT; \citealp{Barthelmy:2005uq}) on board the {\it Neil Gehrels Swift Observatory} \citep{Gehrels:2004dq} during its first 70\,month of operations \citep{Baumgartner_2013}. This sample is highly complete, being almost unbiased by obscuration up to $N_{\rm H}\sim 10^{24}$~cm$^{-2}$ (e.g., \citealp{Ricci:2015fk}).
The sample also has a large amount of multi-wavelength data available, which includes data in the X-rays \citep{Ricci:2017pm}, optical \citep{Koss:2017wu,Koss:2022nt,2022ApJS..261....5M}, infrared \citep{Ichikawa:2017zp,Ichikawa:2019zz,Lamperti:2017uz,den-Brok:2022oy}, millimeter \citep{Koss:2021zw,Kawamuro:2022mo,Ricci:2023vy} and radio \citep{Baek:2019pp,Smith:2020ng}. This large effort has led to several studies comparing the optical and X-ray properties of AGN with their black hole masses and accretion rates \citep{Ricci:2017rn,Ricci:2022eo,Ricci:2022ke,Oh:2017di,Trakhtenbrot:2017xm,Rojas:2020or,Kakkad:2022xc}. 

The X-ray properties (i.e. intrinsic X-ray luminosities and column densities) for all AGN in the BASS sample were taken from the dedicated X-ray spectroscopic analysis of the AGN from the flux-limited 70\,month {\it Swift}/BAT catalog \citep{Ricci:2017pm}, while the black hole masses for 780 of these objects were taken from the second BASS data release \citep{Koss:2022nt}. Black hole masses ($M_{\rm BH}$) in BASS have been determined using different approaches, such as the virial method using the H$\alpha$ and H$\beta$ lines (e.g., \citealp{Shen2013}) and the $M_{\textrm{BH}}$-$\sigma_*$ relation (e.g., \citealp{2013ARA&A..51..511K}). For unobscured AGN BASS uses masses derived from broad H$\alpha$ and H$\beta$ lines. For obscured sources, BASS considers black hole masses inferred from velocity dispersion measurements \citep{Koss:2022uw}, since in these objects, the broad H$\alpha$ and H$\beta$ lines can be significantly affected by obscuration, leading to an underestimation of $M_{\rm BH}$ (e.g., \citealp{2022ApJS..261....5M, Ricci:2022eo}). The large multi-wavelength coverage of BASS sources allows us to have well-determined measurements of AGN properties such as black hole mass ($M_{\rm BH}$), Eddington ratio ($\lambda_{\rm Edd}=L_{\rm bol}/L_{\rm Edd}$), intrinsic X-ray luminosity ($L_{\rm 14-150\,keV}$), and line-of-sight column density ($N_{\rm H}$). The bolometric luminosities were calculated using a uniform bolometric correction in the 14--150\,keV band ($\kappa_{14-150} = 8.48$), which is equivalent to a 2--10\,keV bolometric
correction of $\kappa_{2-10} = 20$ \citep{Vasudevan:2009zh}
for $\Gamma=1.8$, the median photon index of nearby AGN \citep{Ricci:2017pm}. However, in \S\,\ref{sec:corr} we also consider recently calculated bolometric corrections for most of the unobscured AGN in our sample, using the recent results of \citet{Gupta:2024bp}.

\begin{figure*}
    \centering
    \includegraphics[width=0.48\textwidth]{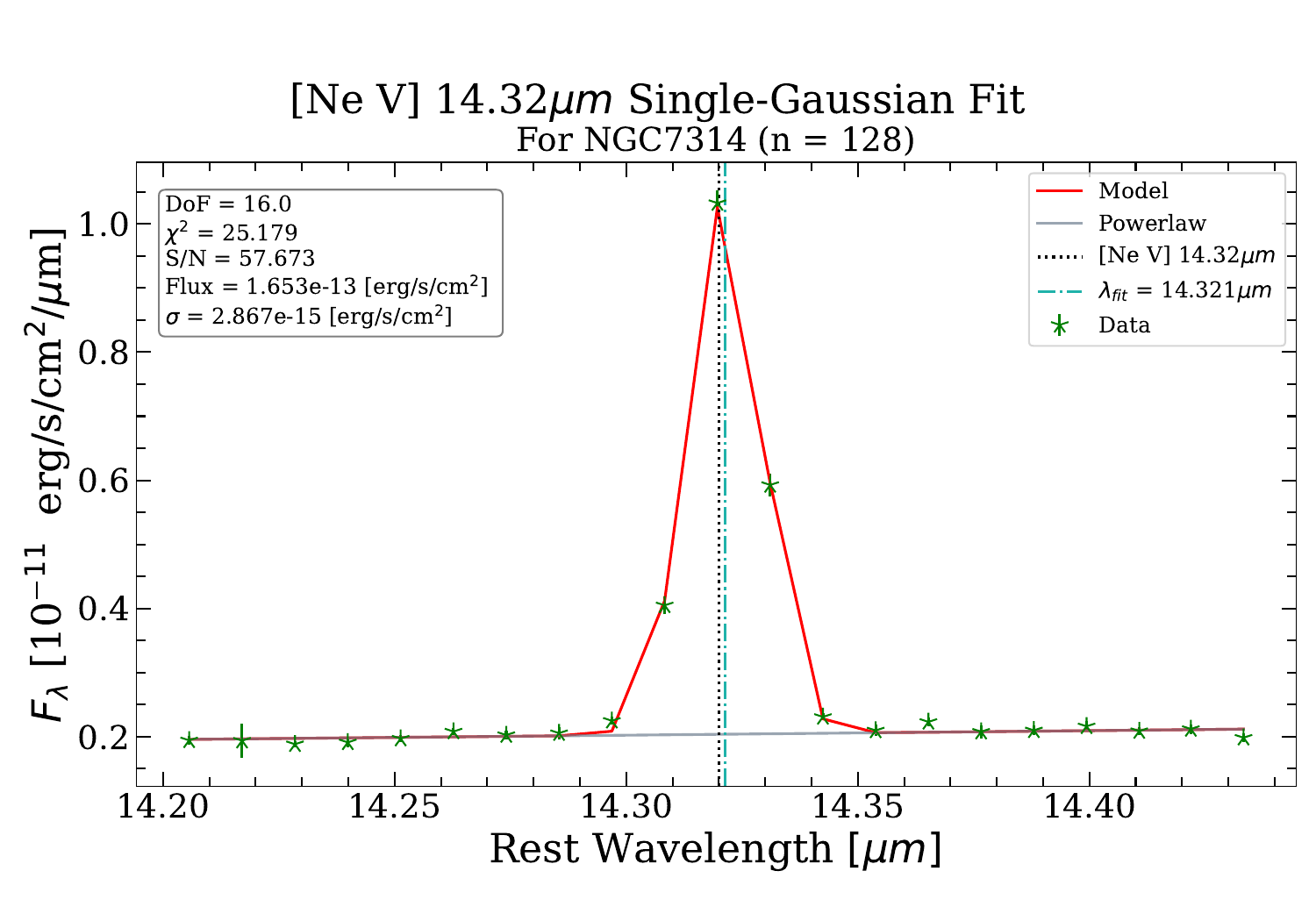} 
    \includegraphics[width=0.48\textwidth]{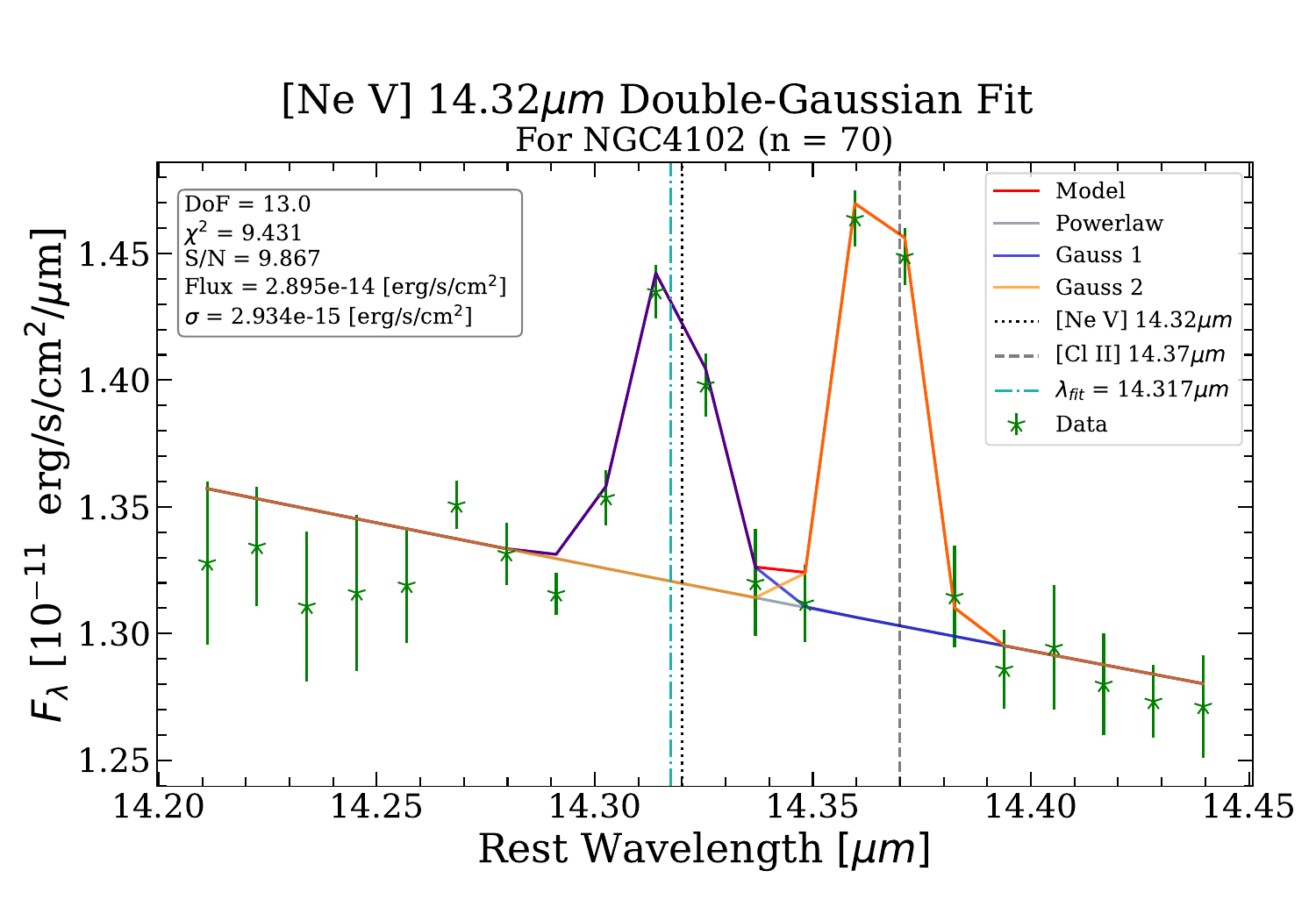} 
    \includegraphics[width=0.48\textwidth]{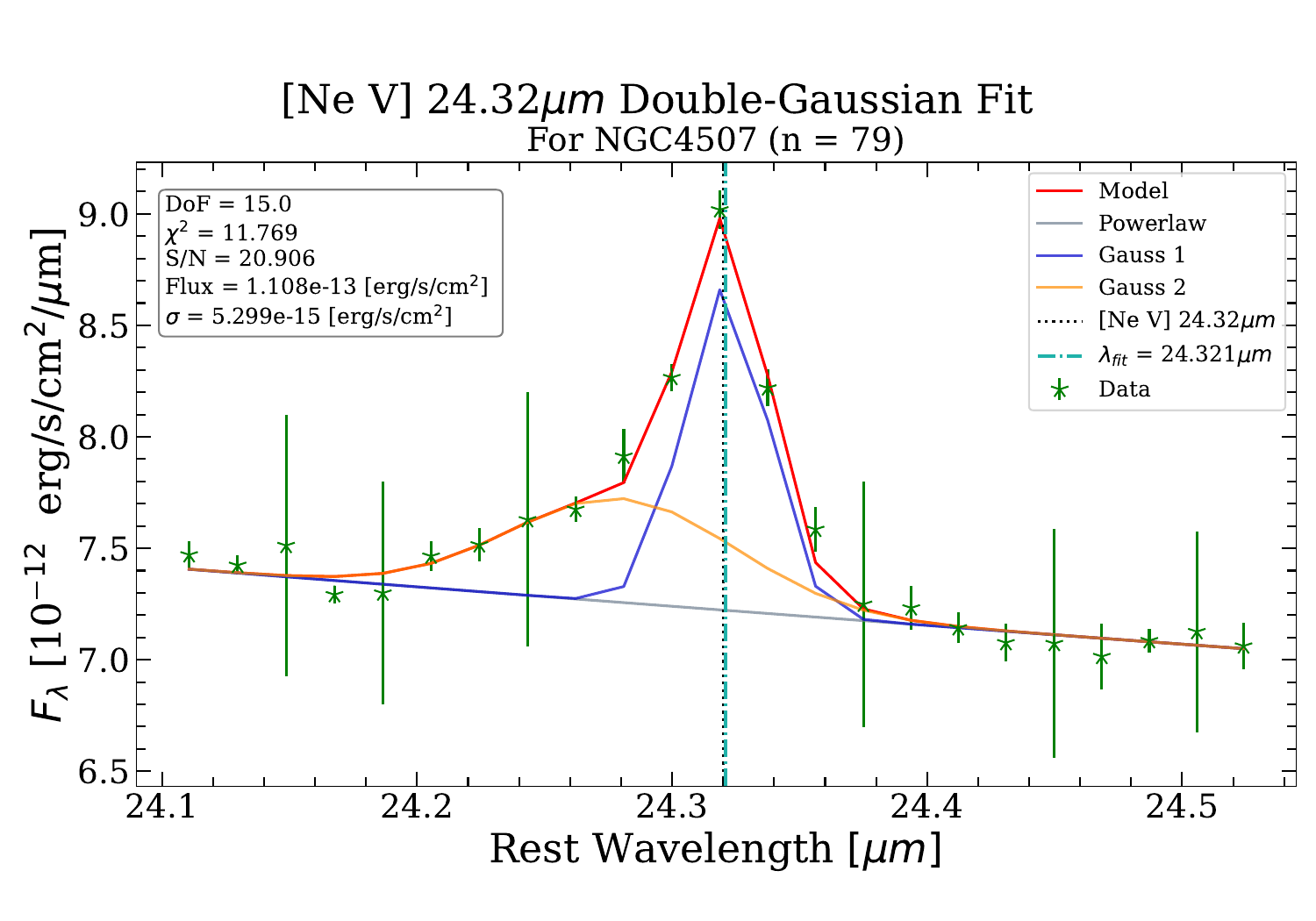} 
    \includegraphics[width=0.48\textwidth]{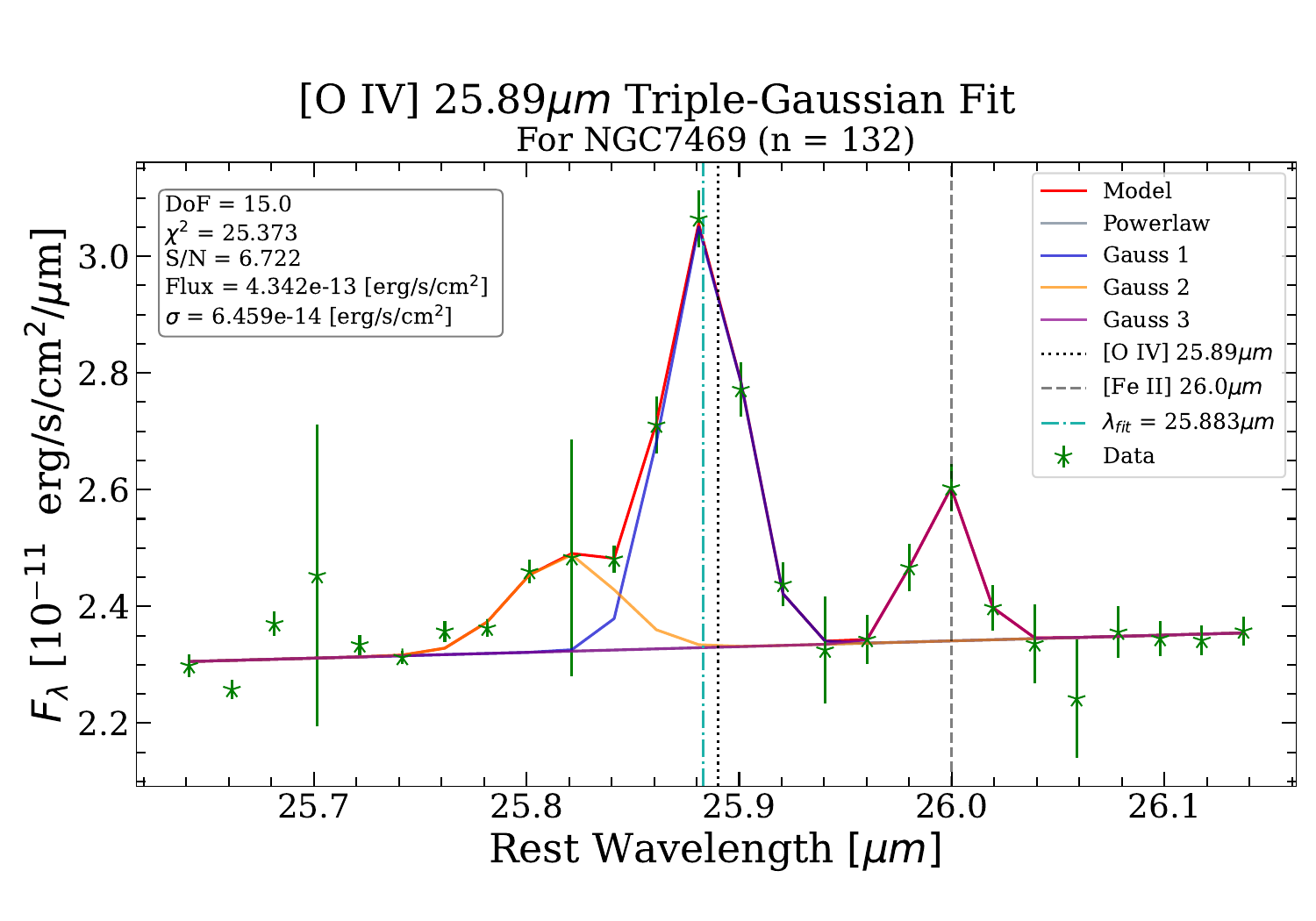}
    \caption{Examples of fits and special cases. {\it Top left panel}: Single Gaussian plus a power-law model for the [{\textsc Ne\,V}] emission line at 14.32\,\micron\ in the IR spectrum of NGC\,7314; see \S\ref{sect:linesfittingmethod}. {\it Top right panel:} Double-Gaussian plus a power-law model for the [{\textsc Ne\,V}] emission at 14.32\,\micron\ in the IR spectrum of NGC\,4102. This model included the nearby [{\textsc Cl\,II}] 14.37\,\micron\ line. The reported flux for this source is that of the first Gaussian curve only (blue line); see \S\ref{sec:nearby_lines}. {\it Bottom left panel}: Double-Gaussian plus a power-law model for the [{\textsc Ne\,V}] emission at 24.32\,\micron\ in the IR spectrum of NGC\,4507. This spectra had a blue-shifted velocity peak, which may suggest the presence of outflows. In such cases we also consider the flux of the second Gaussian, thus, the total flux (red line) for this line is the sum of the fluxes of both Gaussians; see \S\ref{sec:blue}. {\it Bottom right panel}: Triple-Gaussian plus a power-law model for the [{\textsc O\,IV}] emission at 25.89\,\micron\ in the IR spectrum of NGC\,7469. This spectrum included a blue-shifted velocity peak as well as emission at longer wavelengths. In this case, we adopt a similar method as for the blue excess by considering the fluxes of the first and second Gaussian lines as the total flux of the emission line (blue line + orange line). The reported flux determination varies depending on the specific case, see  \S\ref{sec:tripgauss}.}%
    \label{fig:fits}
\end{figure*}

\subsection{\textit{Spitzer}/IRS Spectral Data}
To measure the MIR emission line properties, we use archival spectra from the InfraRed Spectrograph (IRS) aboard the \textit{Spitzer} Space Telescope. \textit{Spitzer} IRS spectra were downloaded from the HR1 version of the \textit{Combined Atlas of Sources with Spitzer IRS Spectra} (CASSIS; \citealp{2015ApJS..218...21L}), available online.\footnote{\url{https://cassis.sirtf.com/}}
We use the `High-Resolution' R$\sim$600 spectra (measuring emission on 5-20\arcsec\ angular scales; see Table 3 of \citealp{2015ApJS..218...21L}), which typically cover the 10$-$38\,\micron\ range.\footnote{One source had a spectrum covering only the 9-18\,\micron\ range (SDSS\,J130005.35+163214.8; see Appendix~\ref{app:exc})} This is sufficient to separate commonly blended emission features such as [{\textsc O\,IV}] and [{\textsc Fe\,II}] at 25.89 and 26.00\,\micron, respectively, and [{\textsc Ne\,V}] and [{\textsc Cl\,II}] at 14.32 and 14.37\,\micron. Our final sample was obtained by cross-matching the BASS and CASSIS samples, resulting in 140 objects with median $\log(N_\textrm{H}/\rm cm^{-2})=22.12$, $\log (M_\textrm{BH}/M_{\odot})=7.68$, $\log \lambda_{\rm Edd}=-1.46$, and $\log(L_\textrm{14-150\,keV}/\rm erg\,s^{-1})=43.34$. 
In Figure\,\ref{fig:Sample_AGN_properties} we show the distributions of column density (top left panel), black hole mass (top right panel), 14--150\,keV luminosity (bottom left panel) and Eddington ratio (bottom right panel) for the whole BASS sample (dark colors) and for our subsample of AGN with \textit{Spitzer} IRS spectra available.
A Kolmogorov-Smirnov test shows that the \textit{Spitzer} IRS sample studied here is broadly representative of the wider BASS sample, although it preferentially selects lower luminosity AGN. This is also clearly shown in Figure\,\ref{fig:redshift}, and is associated to the typically lower redshifts ($z\lesssim0.1$) of AGN with \textit{Spitzer} IRS observations. About half of the objects of our sample have been studied in \citet{2010ApJ...716.1151W}, and in Appendix~\ref{app:Weaver} we report a comparison with their study.

\begin{table*}
 	\centering
	\begin{tabular}{ccccccccccccccccc}
		\hline
		\hline
		(1) & (2) & (3) & (4) & (5) &  & (6) & (7) & (8) & (9) & & (10) & (11) & (12) & (13)  \\
		\noalign{\smallskip}
		\multirow{2}{*}{Emission line}   &   \multirow{2}{*}{Fit total} &  \multirow{2}{*}{Detections}  & \multirow{2}{*}{Upper Limits}   	&  Single &    & \multicolumn{4}{c}{Double} & &   \multicolumn{4}{c}{Triple}\\
		\cline{5-5}
		\cline{7-10}
		\cline{12-15}
            & & & & Tot &    & Tot & O   & R   &   B & &   Tot   & O + R &   O + B &   B + R\\
        \hline
        [{\textsc Ne\,V}] 14.32\,\micron & 140 & 123 (88\%) & 17 (12\%) &    105 & &   17  &   4 	&    8 	&    5 & &   1    	&    1 	&    0 	&    0\\
        {[{\textsc S\,III}]} 18.71\,\micron & 136 & 128 (94\%) & 8 (6\%) & 121  & &   7  &   4 	&    0 	&    3 & &   N/A 	&    --- 	&    ---  	&    ---\\
        {[{\textsc Ne\,V}]} 24.32\,\micron & 139 & 118 (85\%) & 21 (15\%) &    109 & &   9  &   6 	&    0 	&    3  &  &  N/A &    --- 	&    ---  	&    ---\\
        {[{\textsc O\,IV}]} 25.89\,\micron & 139 & 135 (96\%) & 4 (3\%) &     77 & &   50  &   2 	&    43 	&    5 & &   8 	&    5 	&    1 	&    2\\
        {[{\textsc S\,III}]} 33.48\,\micron & 131 & 120 (92\%) & 11 (8\%) &    118 & &   2  &   1	&    0 	&    1 & &   N/A 	&    --- 	&    ---  	&    ---\\
		\hline
	\end{tabular}
	\caption{Fit statistics. Breakdown of the different models required to fit the spectra for each CL. The columns include the following information: \\
	(1) List of the five forbidden emission lines studied here. \\
	(2) Total number of spectra fitted for each line. Although we started with a full sample of 140 sources, for various reasons the spectra of some sources could not be used. This is explained in detail in  Appendix~\ref{app:exc}. \\
    (3) Number of detections for each emission line described in \S\ref{sect:linesfittingmethod}. \\
	(4) Number of non-detections/upper limits described in \S\ref{sec:ULCalc}. \\
	(5) Number of objects that only required a single Gaussian. \\
	(6-9) Number of sources that needed two Gaussian lines. \\
	(7) Sources for which the second Gaussian accounts for a blue-shifted line, which could be an outflow signature ("O"), described in \S\ref{sect:blueshifted}. We use "O" to distinguish these objects from those showing "broad" lines; see \S\ref{sec:broad}. \\
	(8) AGN in which a second Gaussian was included to take into account a nearby emission line at longer (redder) wavelengths ("R"). These components describe the fitting accommodation of [{\textsc Cl\,II}] 14.37\,\micron\ and [{\textsc Fe\,II}] 26.0\,\micron, for [{\textsc Ne\,V}] 14.32\,\micron\ and [{\textsc O\,IV}] 25.89\,\micron, respectively; see \S\ref{sec:nearby_lines}.\\
	(9) Sources that required an additional broad Gaussian line centered at the same wavelength as the expected emission line ("B"), described in \S\ref{sec:broad}.\\
	(10-13) Number of objects that required two additional Gaussian lines to fit the spectra. \\
	(11) Sources showing an blue-shifted line as well as a nearby emission line on the red side ("O+R"); see \S\ref{sec:tripgauss}. \\
	(12) Objects showing blue-shifted lines and one broad component ("O+B"); see \S\ref{sec:tripgauss}. \\
	(13) AGN that needed one broad line and one nearby emission line on the red side ("B+R"); see \S\ref{sec:tripgauss}.}
	\label{tab:fitstats}
\end{table*}

\section{Spitzer/IRS Spectral Analysis}\label{sec:Fitting}

\begin{figure}
    \includegraphics[width=0.95\columnwidth]{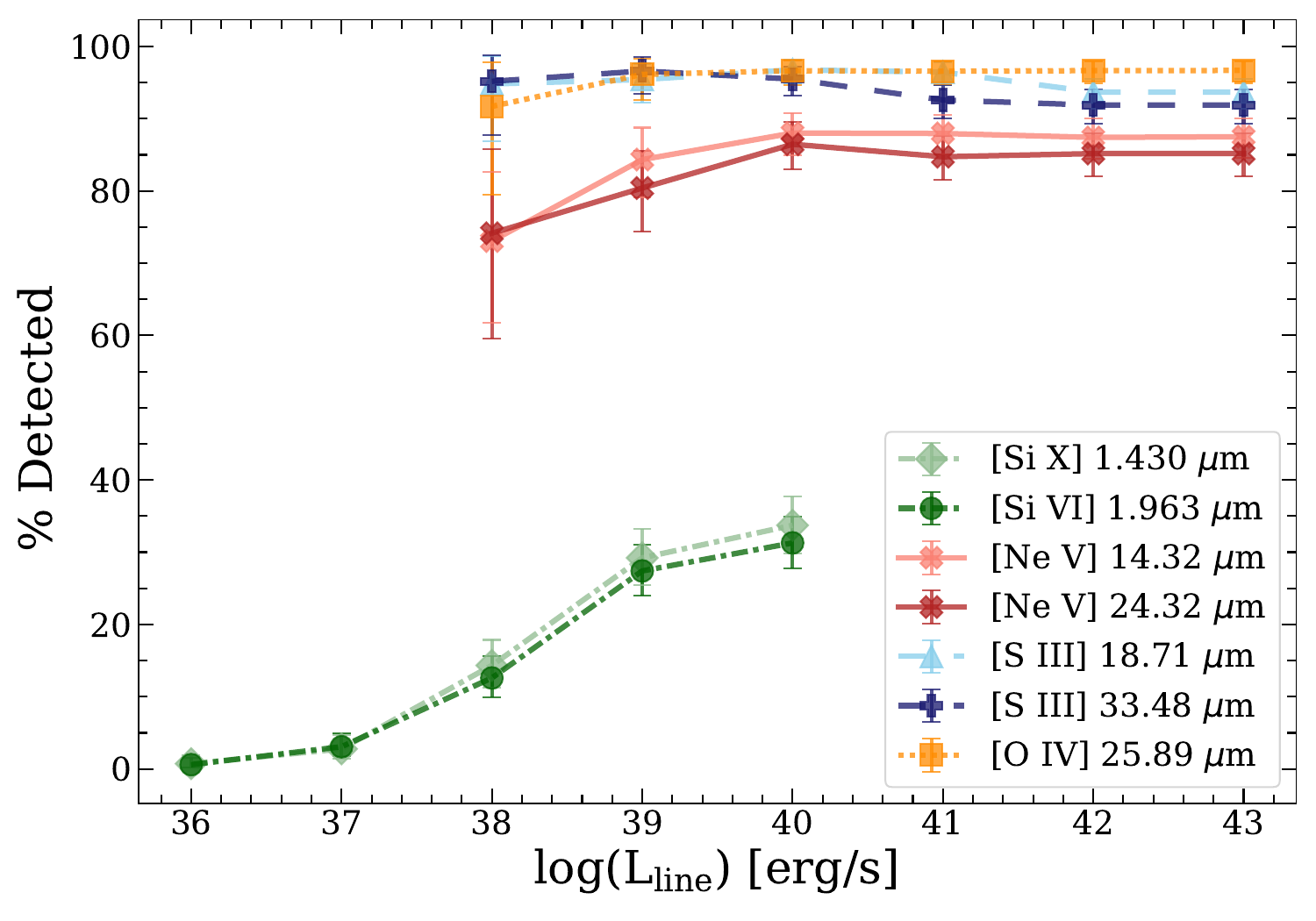}
    \includegraphics[width=0.95\columnwidth]{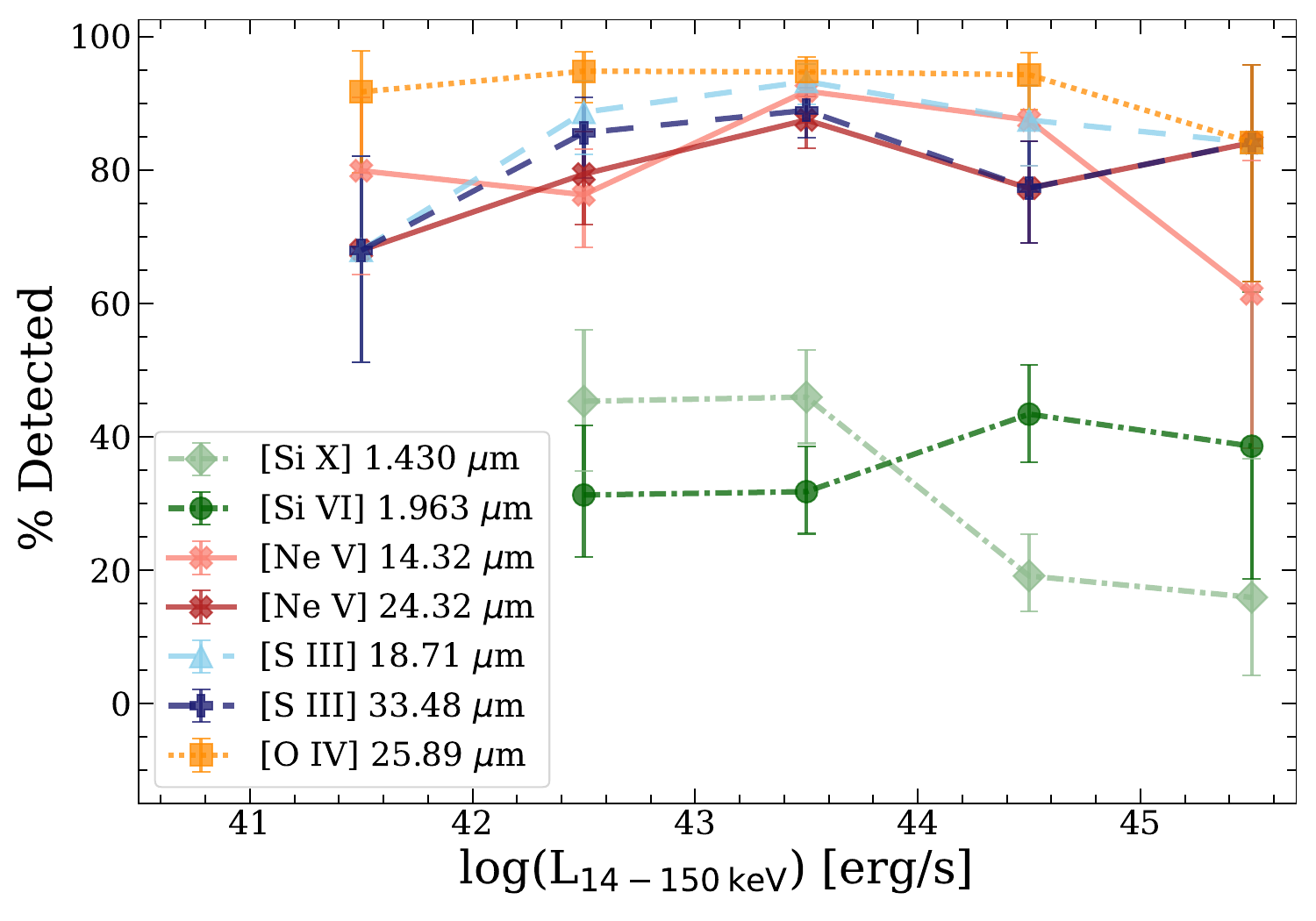}
    \includegraphics[width =0.95\columnwidth]{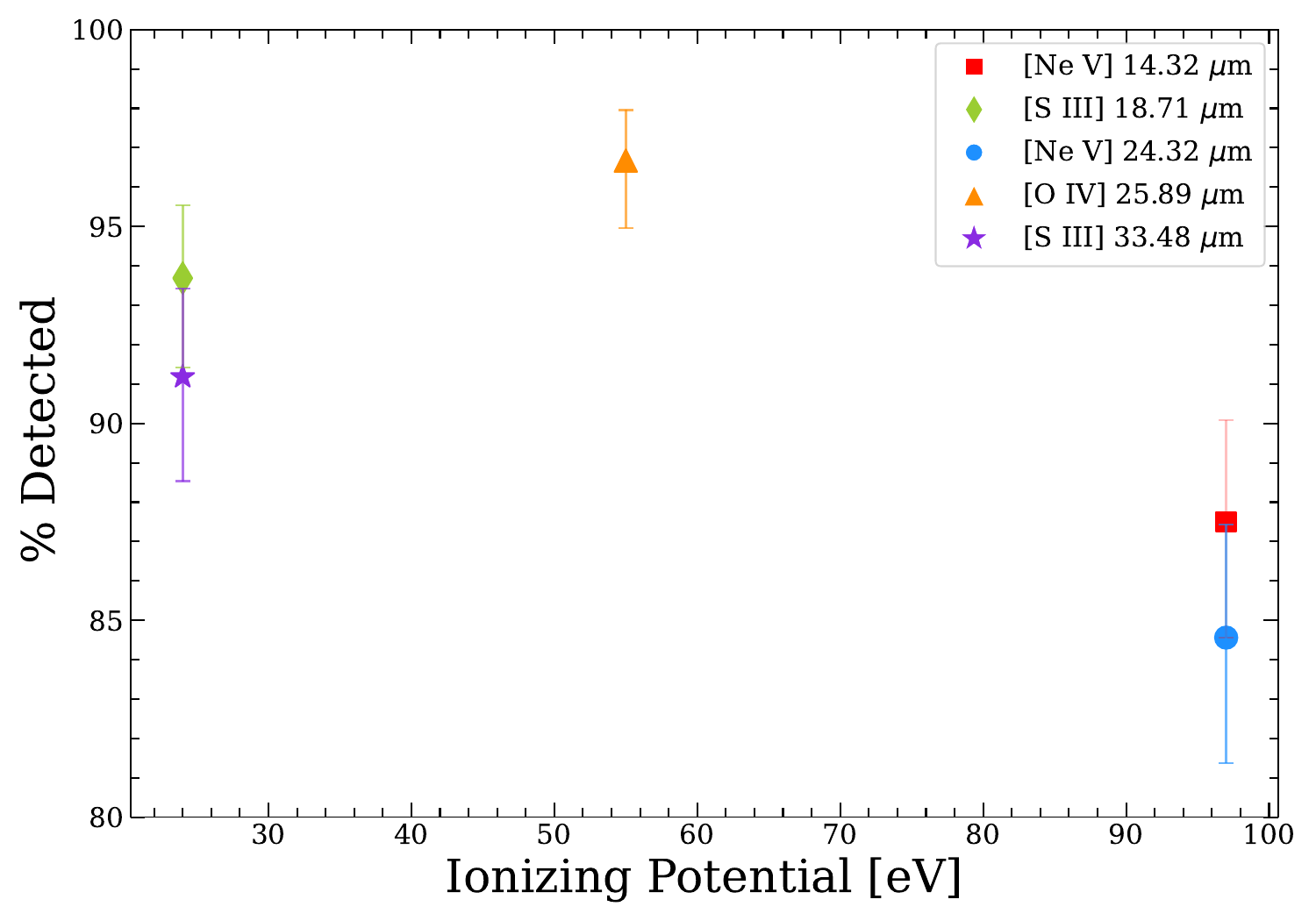}
    \caption{Comparison of the detection fraction for the MIR lines studied in this work with the NIR [{\textsc Si\,VI}] and [{\textsc Si\,X}] lines from \citet{denBrok_DR2_NIR} as a function of the logs of the line (top panel) and hard X-ray (central panel) luminosities. The MIR lines in this work have significant detections in $\gtrsim$85\% of local AGN, meaning they are almost twice as likely to be detected than their near-infrared counterparts. The bottom panel shows the detection rate versus the ionizing potential for the MIR emission lines studied here. }%
    \label{fig:DetFracPlot}
\end{figure}

\subsection{Line fitting}\label{sect:linesfittingmethod}
To measure the MIR emission line fluxes, we used the \textsc{python} \textsc{lmfit} package \citep{2014zndo.....11813N} to model the emission lines in the {\it Spitzer} spectra. Our model includes a power-law for the continuum (spanning across $\pm$0.12\,\micron\, for [{\textsc Ne\,V}] 14.32\,\micron, $\pm$0.15\,\micron\, for [{\textsc S\,III}] 18.71\,\micron, $\pm$0.22\,\micron\, for [{\textsc Ne\,V}] 24.32\,\micron, $\pm$0.25\,\micron\, for [{\textsc O\,IV}] 25.89\,\micron, and $\pm$0.25\,\micron\, for [{\textsc S\,III}] 33.48\,\micron) and Gaussian lines for each emission feature. For each line, we set the centroid wavelength to the rest-frame wavelength of the emission line with an uncertainty range of $\pm$ $\lambda/\textrm{spectral resolution}$ = $\pm$ $\lambda/600$, allowing the line amplitude, continuum parameters, and sigma to vary. The latter parameter was allowed to vary within the range of $\pm$3$\lambda/600 / 2.36$. In most cases, the best-fit centroid wavelength was consistent with the expected value (see Appendix\,\ref{app:bestlamb}).
This model was applied to five emission lines: [{\textsc Ne\,V}] 14.32/24.32\,\micron, [{\textsc S\,III}] 18.71/33.48\,\micron, and [{\textsc O\,IV}] 25.89\,\micron. To determine the error on the flux, we applied a bootstrapping method to generate, for every source and each MIR emission line, 100 simulated spectra by considering the errors on each individual wavelength bin, and then obtained the flux and corresponding continuum parameters from each individual simulated spectra. The standard deviation of the one hundred fluxes was used as the uncertainty on the reported flux. An example fit of a clear detection (a detection being when the signal-to-noise ratio was greater than or equal to 3.0, i.e. S/N $\geq$ 3) of [{\textsc Ne\,V}] 14.32\,\micron\ is presented in the top left panel of Figure\,\ref{fig:fits}.

Given the limited ($R\approx600 \approx 500\rm\,km\,s^{-1}$) resolution of \textit{Spitzer} IRS, the CLs studied in this work are essentially unresolved in the vast majority of objects in our sample.
However, while most lines were well-described by a single Gaussian (between $\sim$55\% and $\sim$90\%, see Table \ref{tab:fitstats}), in some cases a multi-component decomposition of up to three Gaussians was required to adequately model the data (see \S\ref{sec:tripgauss} and bottom right panel of Figure\,\ref{fig:fits}). 
For each object, we start with the simplest model of a single Gaussian, then progress to more complex models by adding additional Gaussian components, depending on how much the reduced $\chi^2$ value deviates from 1.0. These special cases are described in more detail in \S\ref{sec:specialcases}. 

\subsection{Upper limits}\label{sec:ULCalc}
While in the vast majority of cases we were able to detect the CLs, the emission features were not detected for a small percentage of the sample, between 4 and 21 sources (i.e., between 3\% and 15\%) across all five lines (see Table\,\ref{tab:fitstats}). For non-detections, i.e. emission line fluxes with signal-to-noise ratio S/N < 3, we determined upper limits on the flux. This was done by fixing the centroid wavelength to the expected emission line wavelength (see Appendix\,\ref{app:bestlamb}) and the width of the Gaussian to the median of the widths for detected sources, allowing only the amplitude and continuum parameters to vary in our fit. We then fit the spectra with the \textsc{lmfit} method, and used the resulting flux from this fit as the upper limit flux. 
To ensure that we provide the most conservative upper limits, we considered the larger flux between that inferred from our original fit (with all parameters free) and that obtained by fixing most parameters, as reported above. The list of sources for which only upper limits were derived is reported in Table\,\ref{tab:UL_list}.

\begin{table*}
 	\centering
	\begin{tabular}{cccccccccccc}
		\hline
		\hline
		\multirow{2}{*}{Emission line} & \multicolumn{5}{c}{L$_\textrm{14-150 keV}$ vs. L$_\textrm{Line}$}    &    & \multicolumn{5}{c}{F$_\textrm{14-150 keV}$ vs. F$_\textrm{Line}$}\\
		\cline{2-6}
		\cline{8-12}
        & Slope ($\alpha$) & Intercept ($\beta$)  & $\sigma$ [dex] & r & p-value &  & Slope ($\alpha$) & Intercept ($\beta$)   & $\sigma$ [dex] & r & p-value \\
        \hline
        {[{\textsc Ne\,V}]} 14.32\,\micron &    0.93 $\pm$ 0.05 	&    -0.3 $\pm$ 2.0 &    0.5 	&    0.7 	&    3.1e-18 	& &    0.76 $\pm$ 0.10 	&    -5.6 $\pm$ 1.0 	&0.5 	&    0.4 	&    1.5e-04\\
        {[{\textsc Ne\,V}]} 24.32\,\micron &    0.96 $\pm$ 0.04 	&    -1.4 $\pm$ 1.9 &    0.4 	&    0.8 	&    2.1e-18 	& &    0.82 $\pm$ 0.08 	&    -4.9 $\pm$ 0.9 	&0.4 	&    0.4 	&    3.3e-02\\
        {[{\textsc O\,IV}]} 25.89\,\micron &    0.92 $\pm$ 0.05 	&    0.8 $\pm$ 2.0 	&    0.5 	&    0.8 	&    4.8e-18 	& &    0.93 $\pm$ 0.10 	&    -3.2 $\pm$ 1.0 	&0.5 	&    0.5 	&    3.5e-07\\
        {[{\textsc S\,III}]} 18.71\,\micron &   0.79 $\pm$ 0.04 	&    6.3 $\pm$ 1.6 	&    0.3 	&    0.8 	&    5.3e-19 	& &    0.72 $\pm$ 0.08 	&    -5.9 $\pm$ 0.8 	&0.4 	&    0.5 	&    3.6e-06\\
        {[{\textsc S\,III}]} 33.48\,\micron &   0.79 $\pm$ 0.04 	&    6.6 $\pm$ 1.7 	&    0.2 	&    0.8 	&    2.5e-20 	& &   0.65 $\pm$ 0.07 	&    -6.4 $\pm$ 0.7 	&0.2 	&    0.6 	&    2.8e-07\\
        \hline
		\multirow{2}{*}{Emission line} & \multicolumn{5}{c}{L$_\textrm{2-10 keV}$ vs. L$_\textrm{Line}$}    &    & \multicolumn{5}{c}{F$_\textrm{2-10}$ vs. F$_\textrm{Line}$}\\
		\cline{2-6}
		\cline{8-12}
        & Slope ($\alpha$) & Intercept ($\beta$)  & $\sigma$ [dex] & r & p-value &  & Slope ($\alpha$) & Intercept ($\beta$)  & $\sigma$ [dex] & r & p-value \\
        \hline 
        {[{\textsc Ne\,V}]} 14.32\,\micron &    0.63 $\pm$ 0.06  	&    13.2 $\pm$ 2.4 	&    0.8 	&    0.5 	&    1.2e-05 	& &   0.76 $\pm$ 0.08 	&    -5.3 $\pm$0.9 	&0.5 	&    0.5 	&    1.1e-05\\
        {[{\textsc Ne\,V}]} 24.32\,\micron &    0.69 $\pm$ 0.05  	&    10.8 $\pm$ 2.4 	&    0.7 	&    0.5 	&    7.8e-03 	& &    0.73 $\pm$ 0.07 	&    -5.6 $\pm$0.8 	&0.4 	&    0.6 	&    7.1e-07\\
        {[{\textsc O\,IV}]} 25.89\,\micron &    0.62 $\pm$ 0.05	&    14.1 $\pm$ 2.3 	&    0.7 	&    0.5 	&    3.9e-06 	& &    0.80 $\pm$ 0.08 	&    -4.3 $\pm$0.9 	&0.5 	&    0.5 	&    2.1e-05\\
        {[{\textsc S\,III}]} 18.71\,\micron &   0.52 $\pm$ 0.05 	&    17.9 $\pm$ 1.9 	&    0.5 	&    0.5 	&    3.3e-04 	& &    0.61 $\pm$ 0.07 	&    -6.9 $\pm$0.8 	&0.4 	&    0.5 	&    1.3e-04\\
        {[{\textsc S\,III}]} 33.48\,\micron &    0.53 $\pm$ 0.04 	&    17.8 $\pm$ 1.9 	&    0.4 	&    0.5 	&    3.2e-04 	& &    0.48 $\pm$ 0.06 	&    -7.9 $\pm$0.7 	&0.3 	&    0.5 	&    1.1e-06\\
        \hline
	\end{tabular}
	\caption{Table of linear regression fit data corresponding to the fits shown in Figure \ref{flux_and_lum_NeV}, in the form: $\textrm{Y}=\alpha*\textrm{X} + \beta$. Here, Y and X represent the log of  line flux/luminosities and the log of  X-ray flux/luminosities, respectively, where we included the linear regression data with the log of the 2-10 keV X-ray luminosity. $\alpha$ and $\beta$ represent the slope and y-intercept of the linear fits, respectively. Using \textsc{pymccorrelation}, we determined the Spearman's Rank correlation coefficient, r, and the corresponding p-value. }
    \label{tab:flux_lum_tbl}
\end{table*}

\begin{table*}
 	\centering
	\begin{tabular}{cccccc}
		\hline
        \hline
		\multirow{2}{*}{Emission line} & \multicolumn{5}{c}{L$_\textrm{Bol}^*$ vs. L$_\textrm{Line}$}\\
		\cline{2-6}
        & Slope ($\alpha$) & Intercept ($\beta$) & $\sigma$ [dex] & r & p-value\\
        \hline
        {[{\textsc Ne\,V}]} 14.32\,\micron &     1.00 $\pm$ 0.07 	&    -3.8 $\pm$ 3.2 	&    0.3 	&    0.8 	&    1.9e-11\\
        {[{\textsc Ne\,V}]} 24.32\,\micron &    1.09 $\pm$ 0.08	&    -7.8 $\pm$ 3.7 	&    0.3 	&    0.8 	&    3.6e-10\\
        {[{\textsc O\,IV}]} 25.89\,\micron &     0.93 $\pm$ 0.06	&    -0.3 $\pm$ 2.7 	&    0.2 	&    0.9 	&    1.1e-12\\
        {[{\textsc S\,III}]} 18.71\,\micron &   0.82 $\pm$ 0.07 	&    4.0 $\pm$ 3.1 	    &    0.3 	&    0.8 	&    1.3e-07\\
        {[{\textsc S\,III}]} 33.48\,\micron &   0.92 $\pm$ 0.09 	&    -0.2 $\pm$ 3.8 	&    0.2 	&    0.8 	&    1.6e-07\\
        \hline
	\end{tabular}
	\caption{Table of linear regression fit data corresponding to the fits shown in Figures \ref{fig:lbol} in the form: $\textrm{Y}=\alpha*\textrm{X} + \beta$. Here, Y and X represent the logs of the line luminosity and bolometric luminosity (from \citealp{Gupta:2024bp}), respectively. $\alpha$ and $\beta$ represent the slope and y-intercept of the linear fits, respectively. Using \textsc{pymccorrelation}, we determined Spearman's Rank correlation coefficient ($r$) and the corresponding p-value.}
    \label{tab:lineVSbol}
\end{table*}

\subsection{Special cases}
\label{sec:specialcases}
In some special cases, we added a second or even a third Gaussian component. Some spectra that had [{\textsc Ne\,V}] 14.32 \micron\ emission were influenced by the emission of [{\textsc Cl\,II}] at 14.37\micron\, while at longer wavelengths, many sources showed [{\textsc Fe\,II}] 26.0\,\micron\ emission close to the [{\textsc O\,IV}] 25.89\,\micron\ line. A handful of sources showed blue-shifted velocity peaks, suggesting the presence of outflowing gas. In some cases, the source showed signatures of both a nearby emission line and a blue-shifted velocity peak. Some other spectra showed broad emission lines, where one Gaussian at the center was insufficient for the model to accurately fit the data, and thus a second Gaussian at the same center wavelength was added. These cases indeed required more attention to detail in the fitting process and will be discussed in the following sections; a detailed breakdown of the number of such cases for each CL can be found in Table \ref{tab:fitstats}.

\subsubsection{Nearby emission lines}
\label{sec:nearby_lines}
For many sources, a close-by, but well-distinguished, emission line is present at redder wavelengths, which would greatly impact our fit, particularly the slope of the continuum, making it difficult to accurately assess the flux of the CLs. For the [{\textsc Ne\,V}] 14.32\,\micron\ and the [{\textsc O\,IV}] 25.89\,\micron\ lines, a second Gaussian was added to the model to account for the emission of [{\textsc Cl\,II}] 14.37\,\micron\ (8 sources or $\sim$6\%) and [{\textsc Fe\,II}] 26.0\,\micron\ (43 sources or $\sim$31\%), respectively. We refer to Table\,\ref{tab:fitstats} for the number of sources that had a nearby emission line, denoted under "R" for \textit{Red}.
For these cases we took the same steps in calculating the flux and its uncertainty as for the \textit{single}-Gaussian fits (see \S\ref{sect:linesfittingmethod}), excluding the flux and error of the second Gaussian for the overall reported flux/error. An example of a Double-Gaussian fit for the [{\textsc Ne\,V}] emission at 14.32\,\micron\ in the IR spectrum of NGC\,4102 is shown in the top right panel of Figure\,\ref{fig:fits}.

\begin{figure*}
    \centering
    \includegraphics[scale = 0.216]{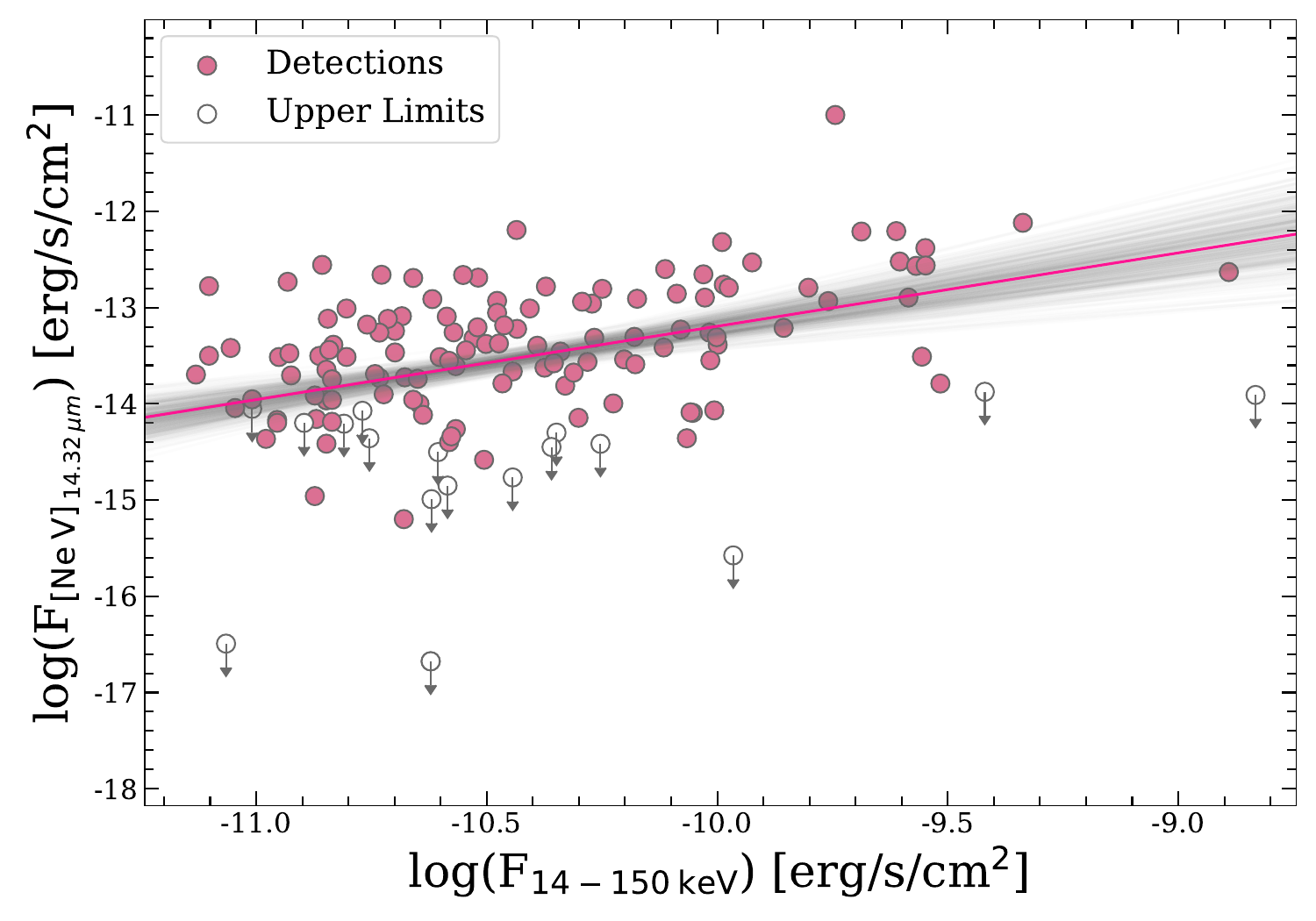} 
   \includegraphics[scale = 0.216]{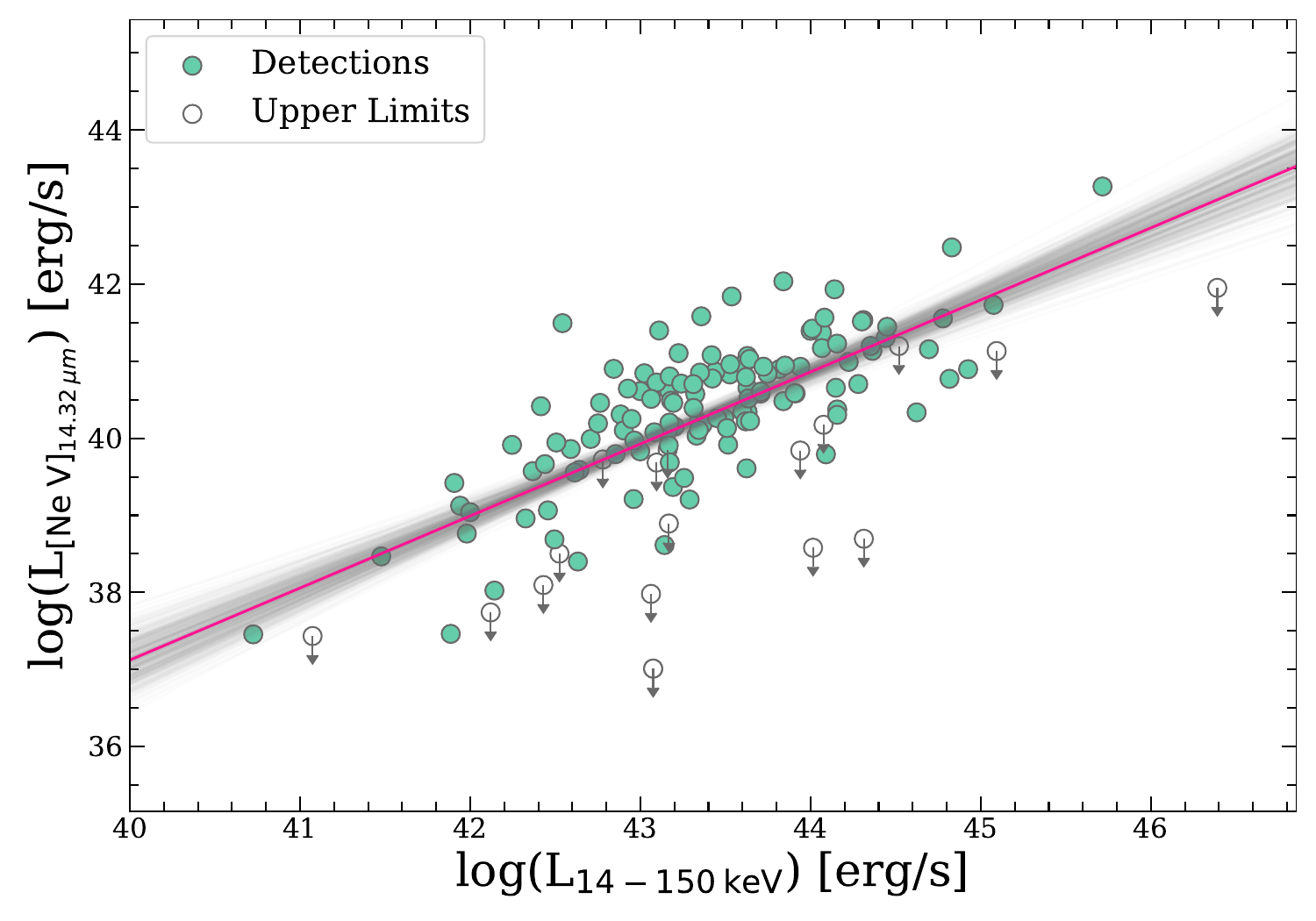} 
    \includegraphics[scale = 0.216]{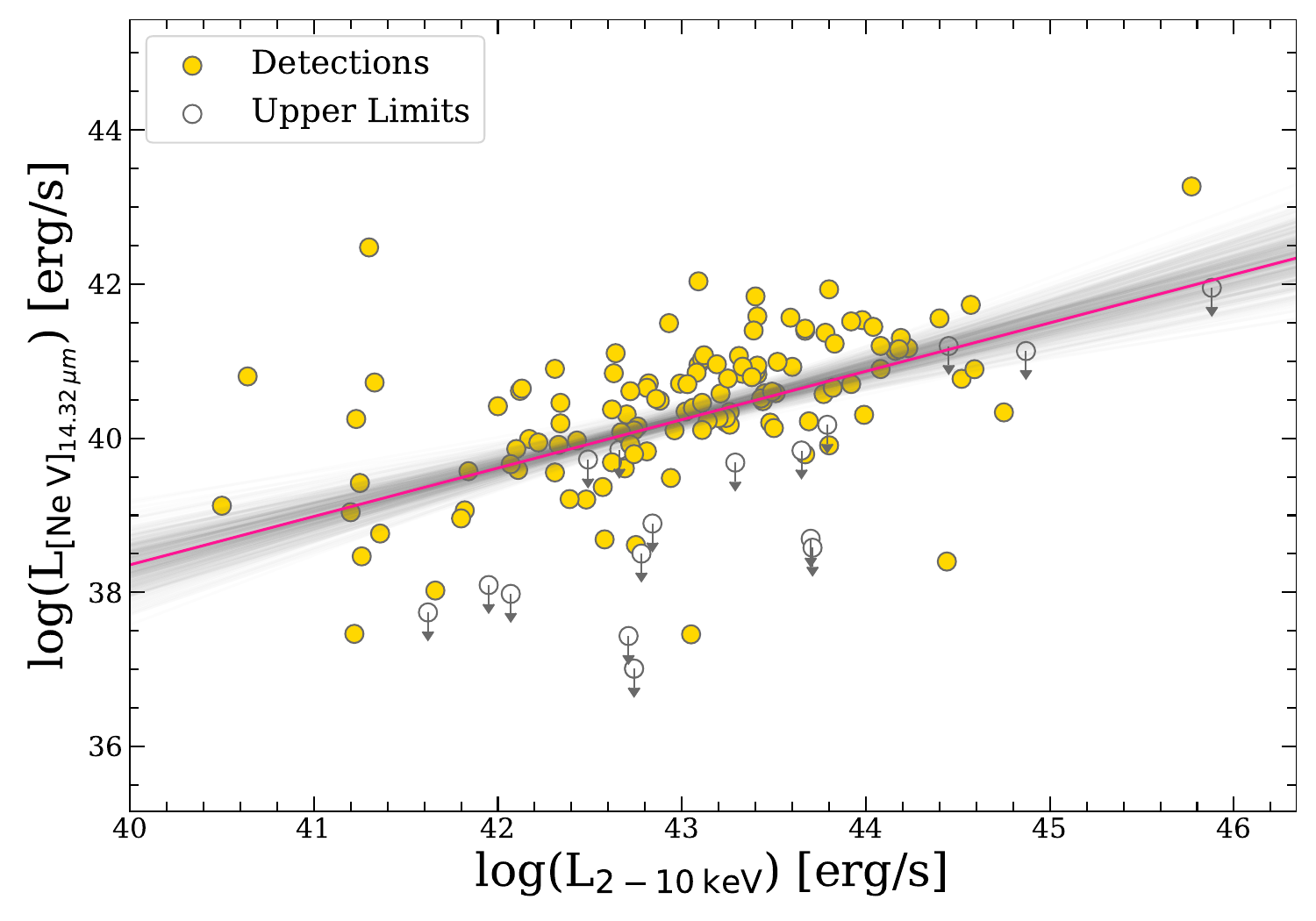} 
    \includegraphics[scale = 0.216]{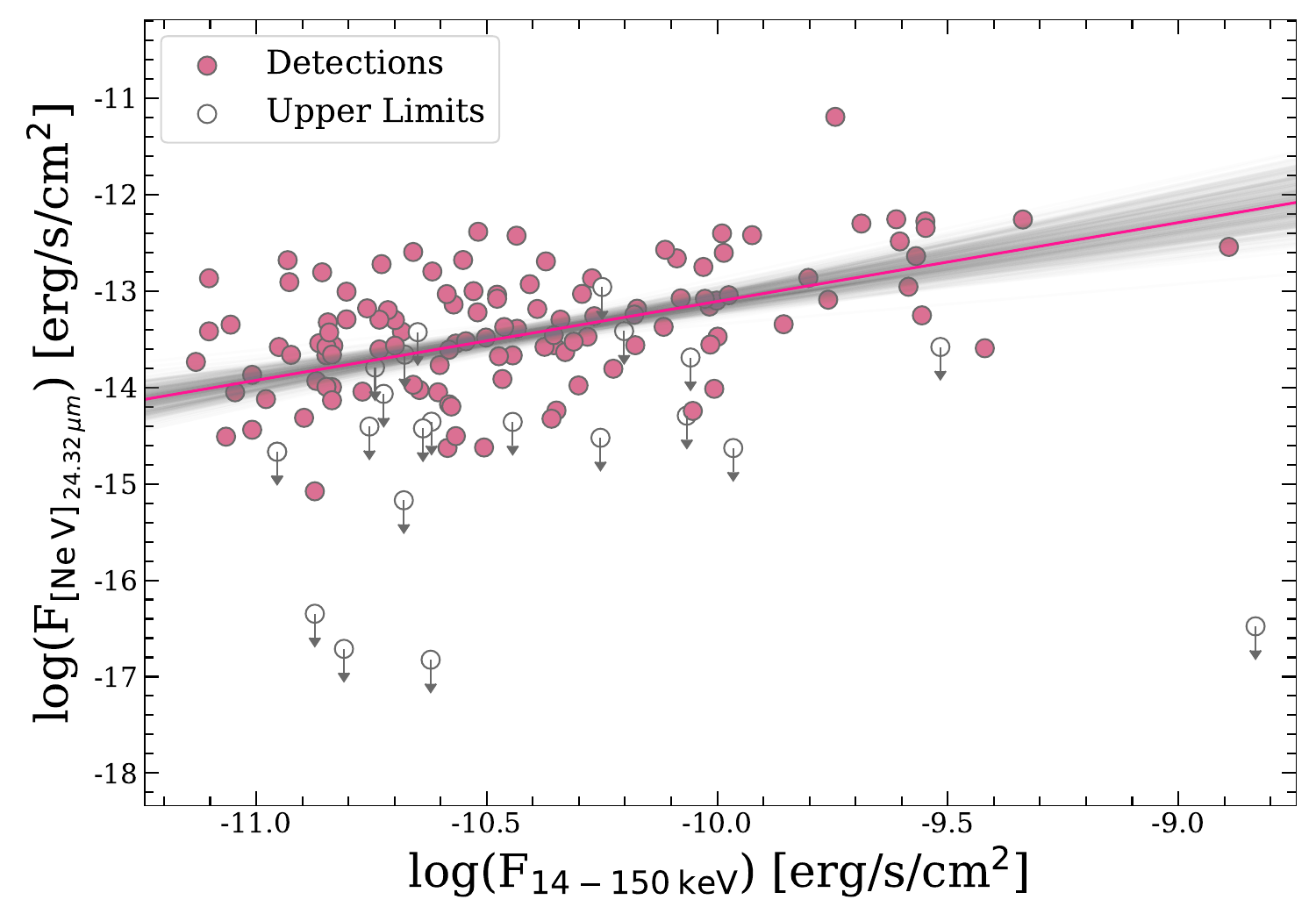} 
    \includegraphics[scale = 0.216]{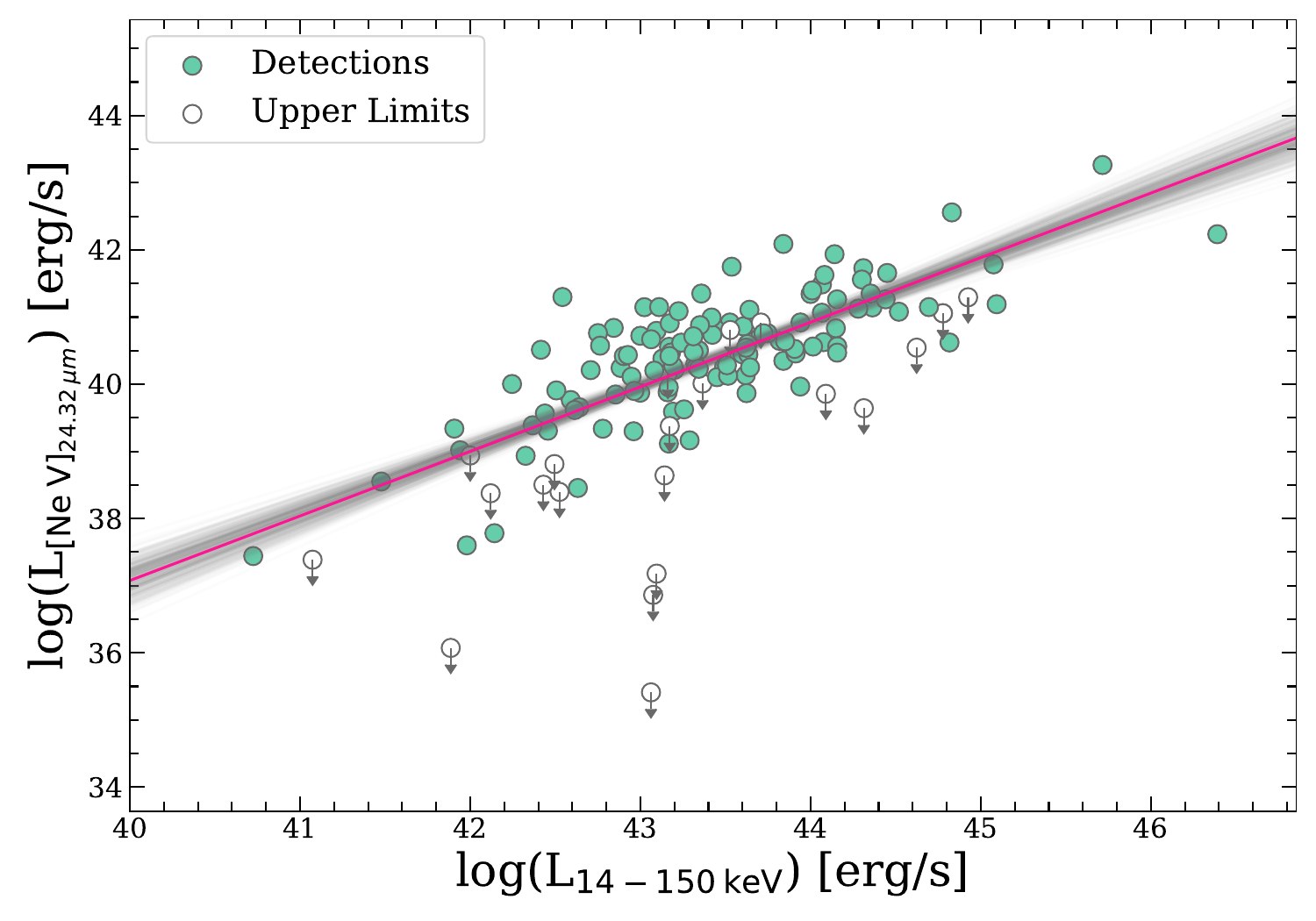} 
    \includegraphics[scale = 0.216]{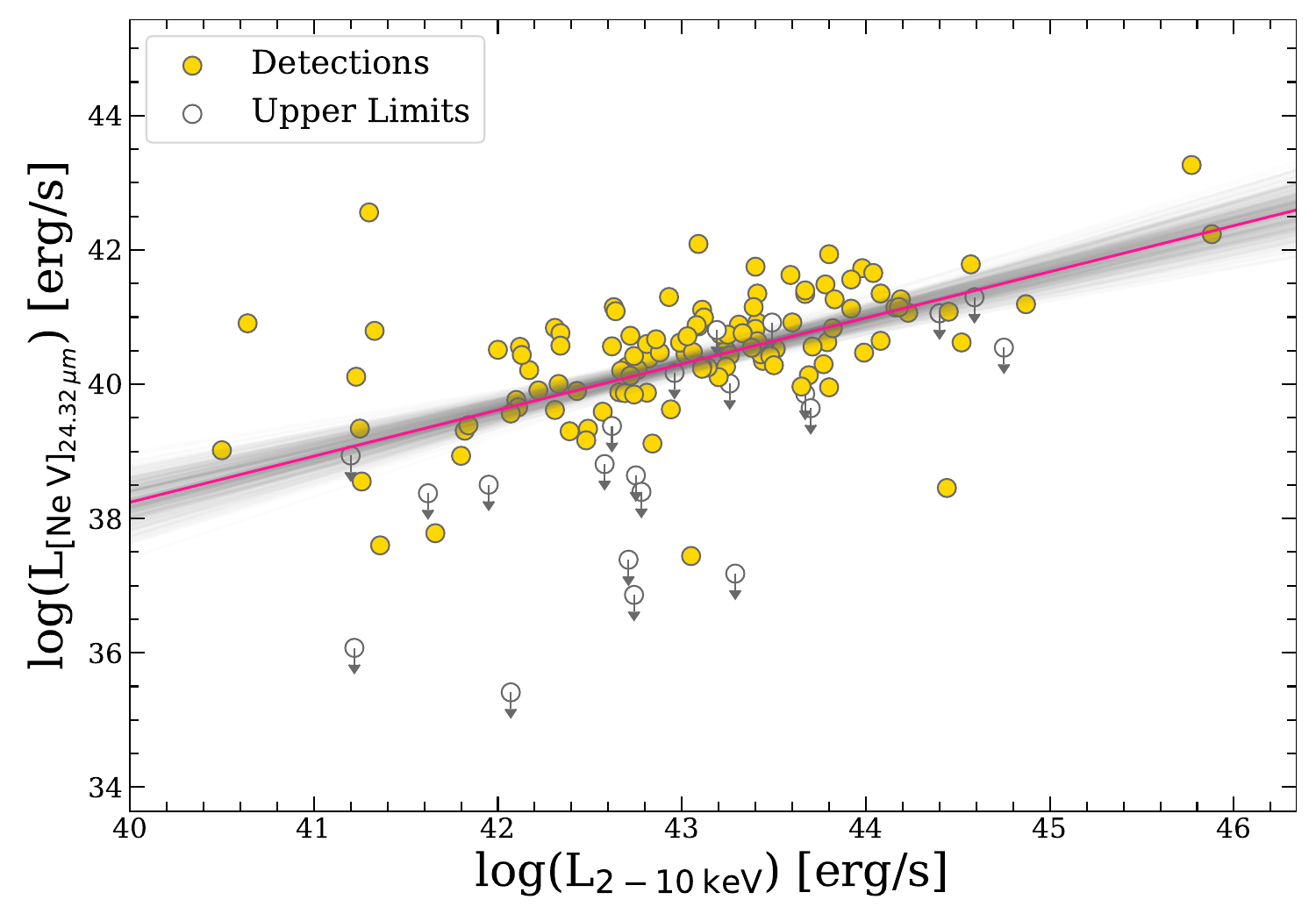} 
    \includegraphics[scale = 0.216]{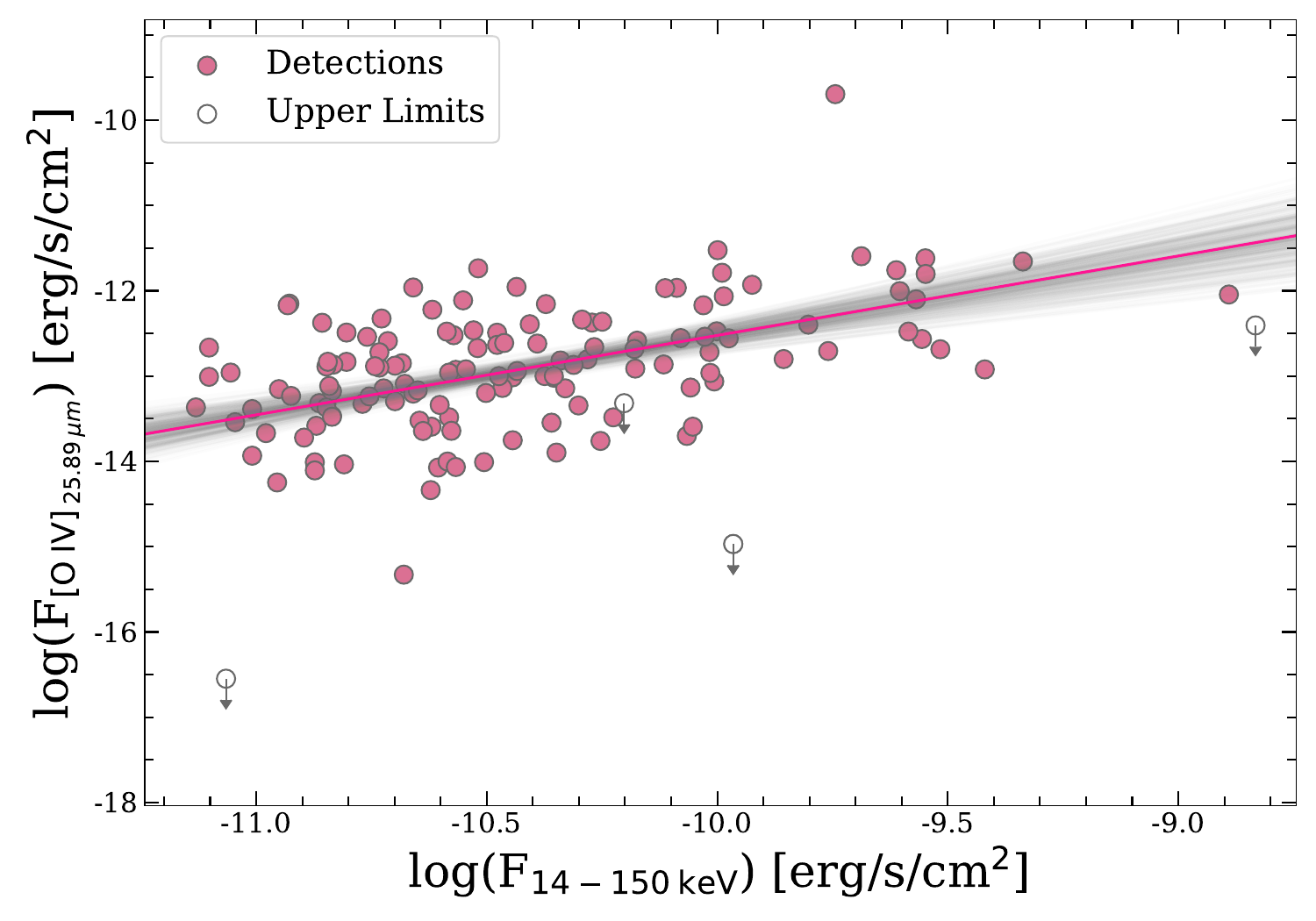} 
   \includegraphics[scale = 0.216]{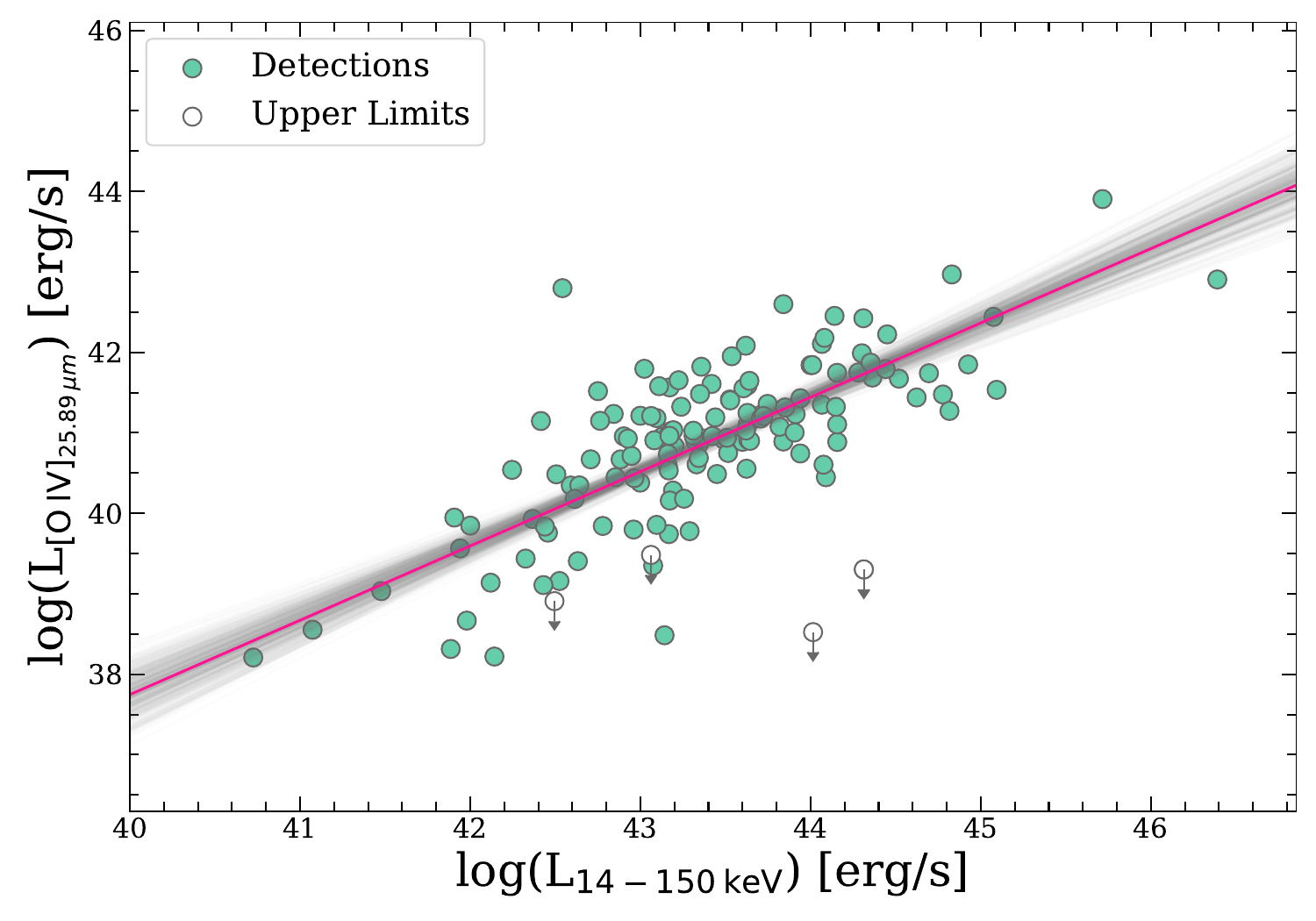} 
    \includegraphics[scale = 0.216]{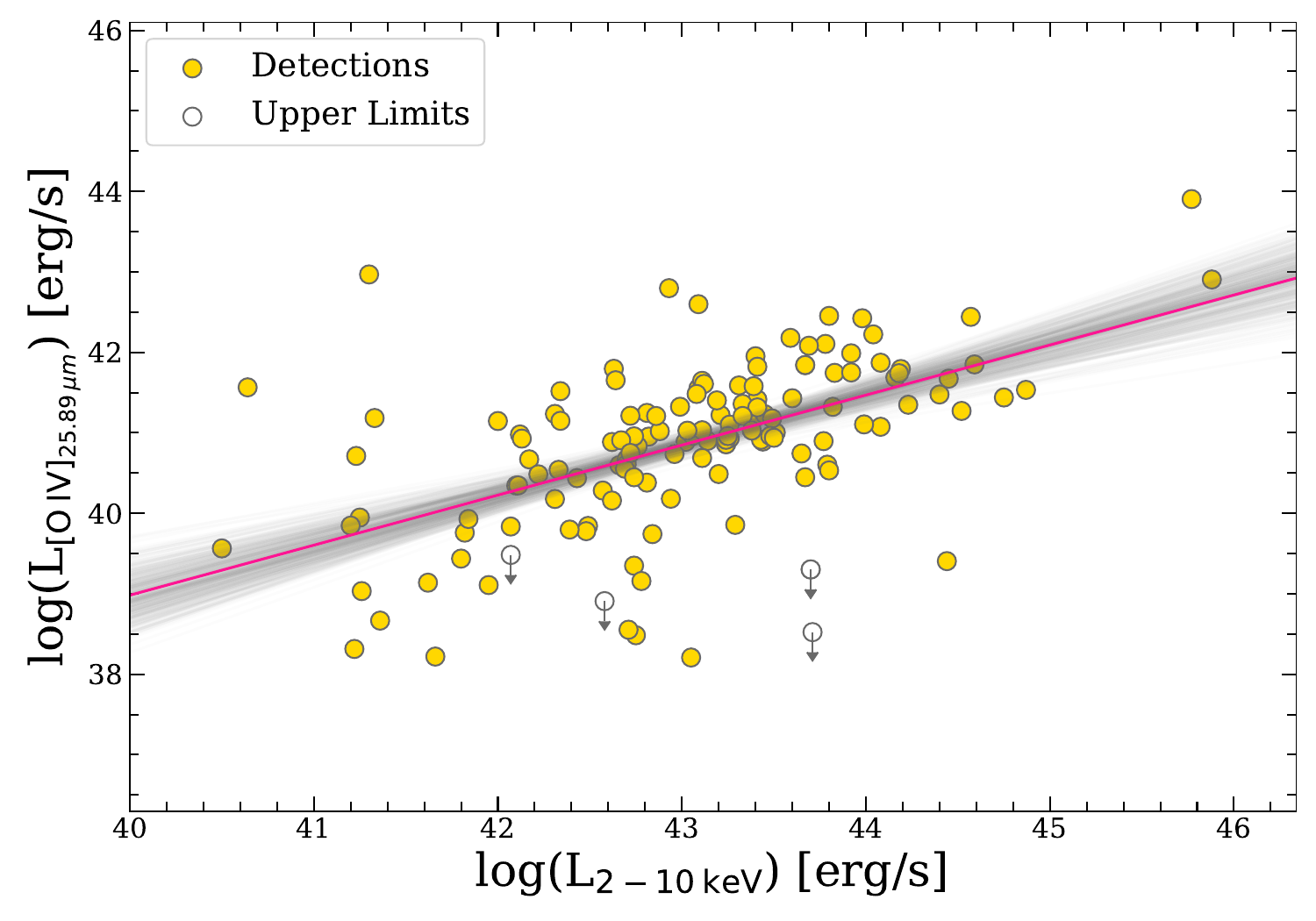} 
    \includegraphics[scale = 0.216]{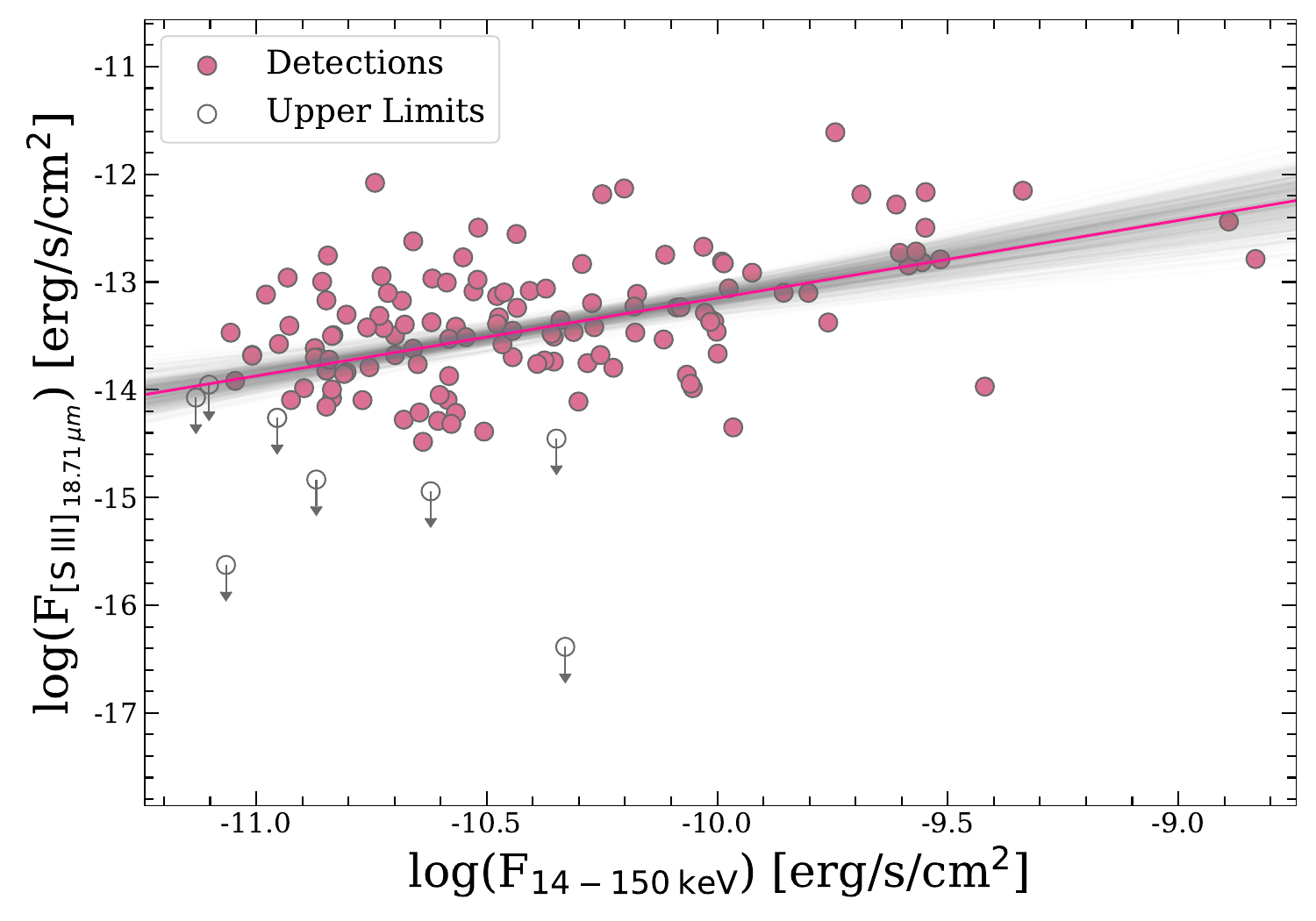} 
   \includegraphics[scale = 0.216]{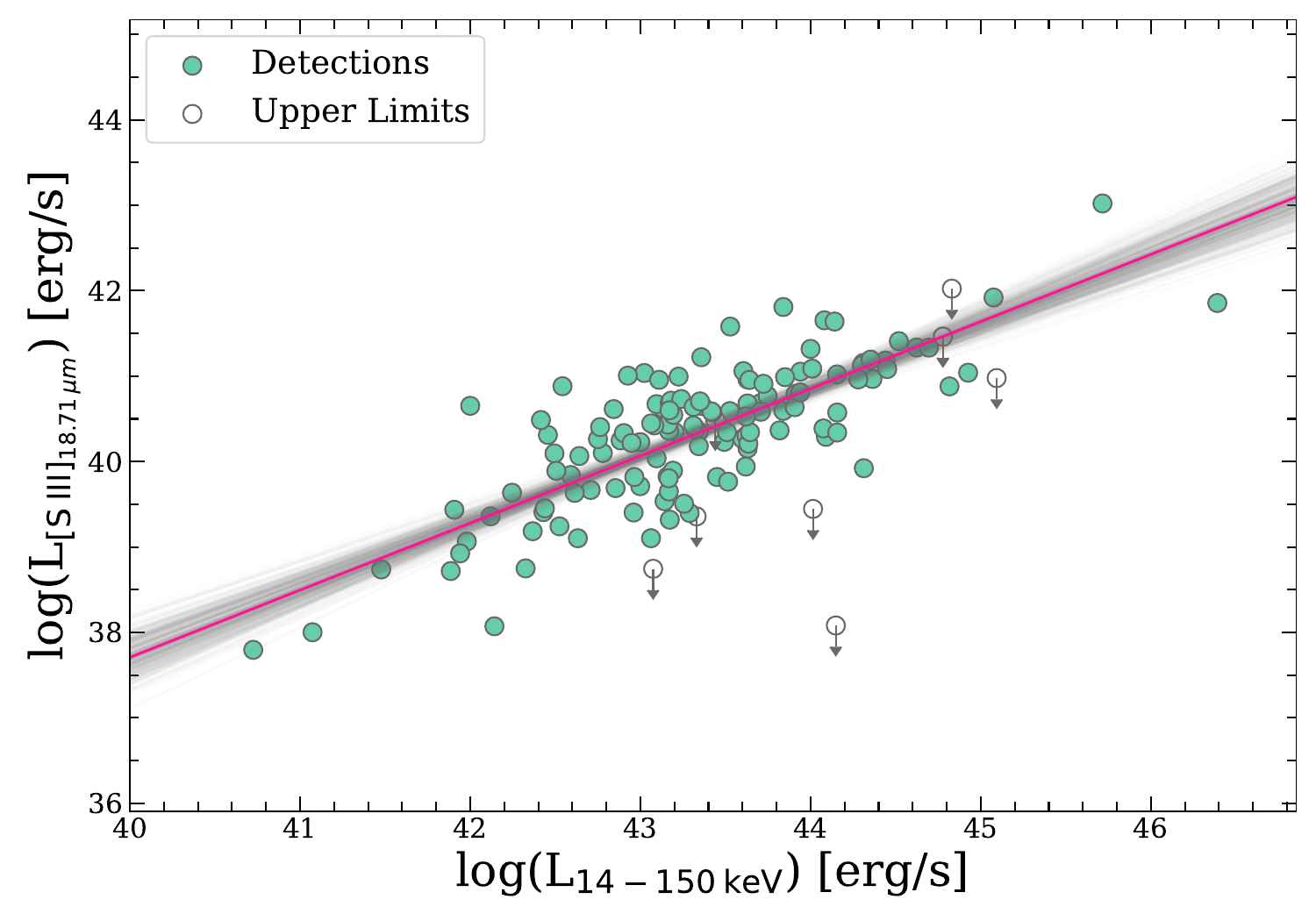} 
    \includegraphics[scale = 0.216]{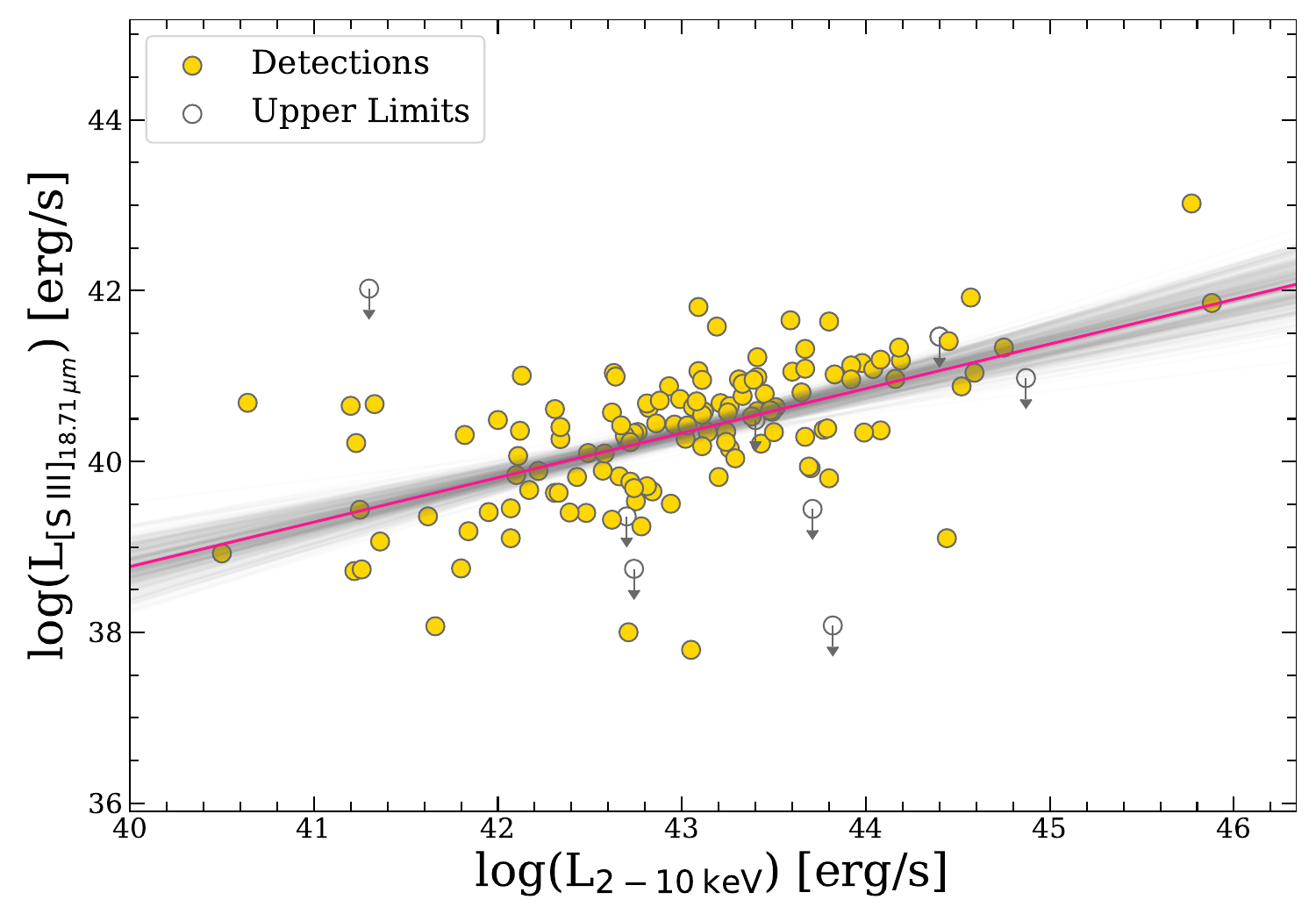} 
    \includegraphics[scale = 0.216]{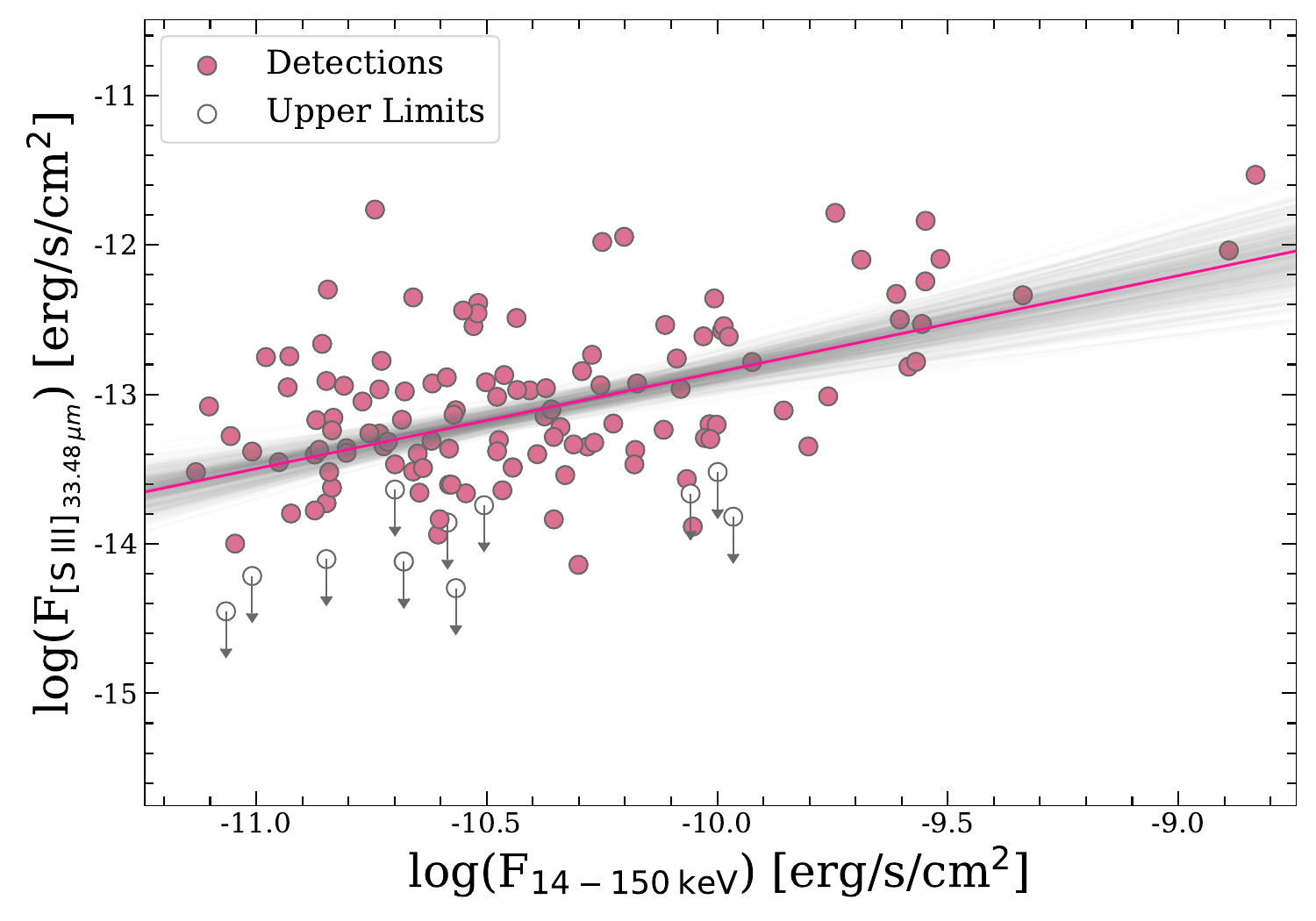} 
    \includegraphics[scale = 0.216]{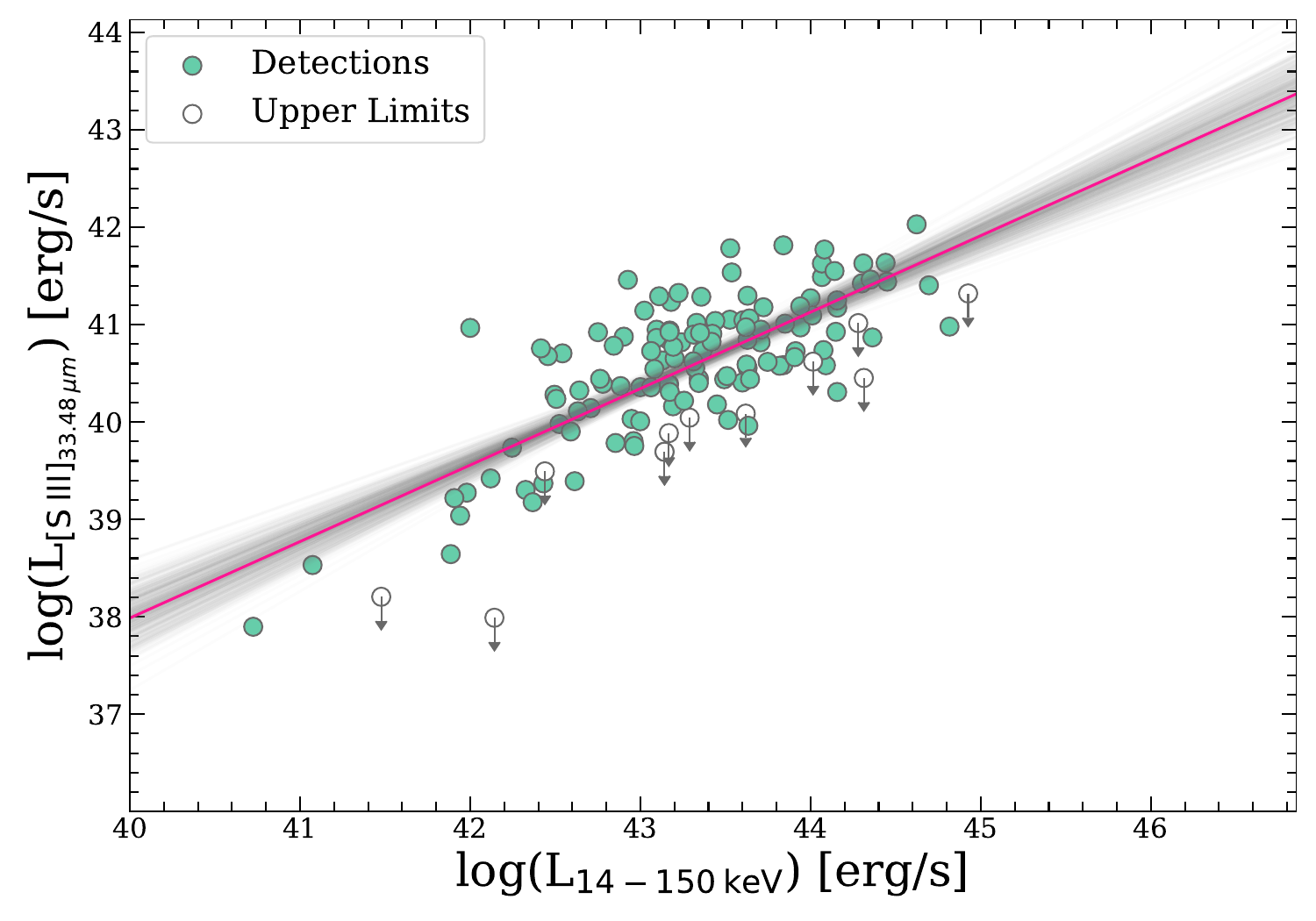} 
    \includegraphics[scale = 0.216]{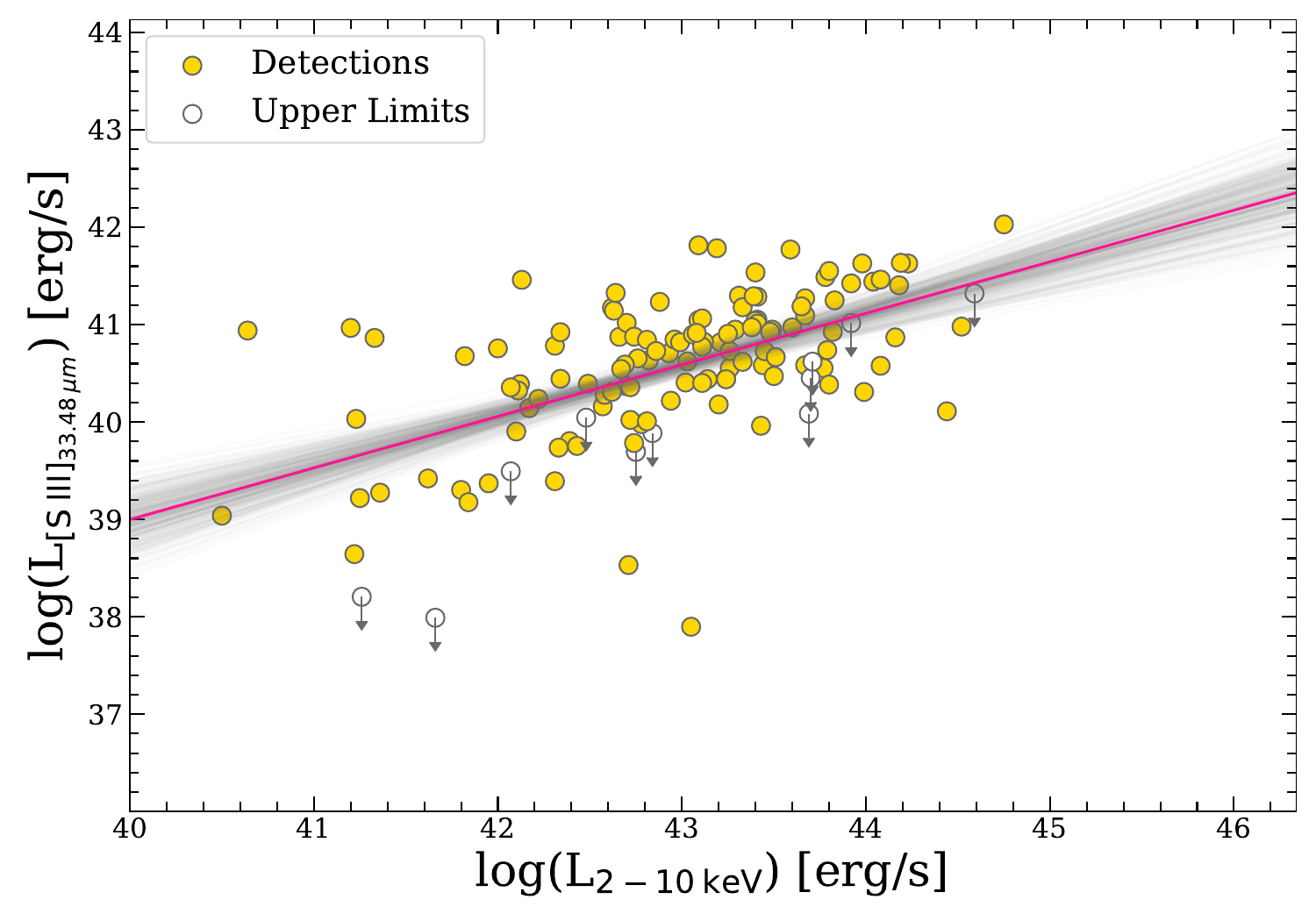} 
    \caption{Log of the emission line flux vs. log of the X-ray flux (left panels) and log of emission line luminosity vs. log of X-ray luminosity (central and right panels) for [{\textsc Ne\,V}] 14.32/24.32\,\micron,  [{\textsc O\,IV}] 25.89\,\micron\ and [{\textsc S\,III}] 18.71/33.48\,\micron. A linear regression fit in the form Y=$\alpha\cdot$X + $\beta$ was performed on the data, where Y and X represent the log of line flux/luminosities and the log of X-ray flux/luminosities, respectively; $\alpha$ and $\beta$ represent the slope and y-intercept of the linear fits, respectively. The linear fit values can be found in Table\,\ref{tab:flux_lum_tbl}. The confidence intervals are also shown on the plots.}
    \label{flux_and_lum_NeV}
\end{figure*}

\subsubsection{Blue-shifted lines}\label{sect:blueshifted}
\label{sec:blue}
A small number of sources, ranging from one to six AGN (i.e., between 0.8\% and 4\%), display broad emission lines with an asymmetric profile extending to bluer wavelengths, which may be associated with outflowing gas. 
For these objects, we used two Gaussian lines to account for the blueshifted emission and included the flux from the second Gaussian when calculating the total flux of the line. Table\,\ref{tab:fitstats} contains the number of sources exhibiting signatures of a blue-shifted line, indicated under "O" for \textit{Outflow}. We use "O" to distinguish these objects from those showing a \textit{broad} emission line, which are denoted as "B" (see \S\ref{sec:broad}). An example of a double-Gaussian fit for a  [{\textsc Ne\,V}]  24.32\,\micron\ source with a blue-shifted line is shown in the bottom left panel of Figure\,\ref{fig:fits}. The characteristic width of the Gaussians of these blue-shifted/broad lines range from 0.003 to 0.030\,\micron\ with a median of 0.010\,\micron\ for  [{\textsc Ne\,V}] 14.32\,\micron, from 0.002 to 0.037\,\micron\ with a median of 0.013\,\micron\ for [{\textsc S\,III}] 18.71\,\micron, from 0.003 to 0.046\,\micron\ with a median of 0.017\,\micron\ for  [{\textsc Ne\,V}]  24.32\,\micron, from 0.003 to 0.055\,\micron\ with a median of 0.018\,\micron\ for [{\textsc O\,IV}] 25.89\,\micron, and from 0.003 to 0.071\,\micron\ with a median of 0.020\,\micron\ for [{\textsc S\,III}] 33.48\,\micron (Figure \ref{fig:bestlam}). If a line was found to have also a broad and/or a blue-shifted component, then the respective characteristic width of the second (and third, if applicable) Gaussian was included in the histogram as an additional data point. See the next sections \S\ref{sec:broad} and \S\ref{sec:tripgauss} for a further discussion of these special cases.

\subsubsection{Broad emission lines}
\label{sec:broad}
In a few cases, ranging from one to five sources (i.e., between 1\% and 4\%), the emission lines were broad, and a single Gaussian could not produce a good fit, so a second Gaussian line was added at the same wavelength as the desired emission line. We selected these sources by visually inspecting the single-Gaussian fits, testing the double Gaussian model fit, and then comparing the values of the $\chi^2$ for the two cases. The total line flux and uncertainty was determined in the same manner as for the sources with blueshifted emission (\S\ref{sect:blueshifted}). In Table \ref{tab:fitstats}, these are denoted under "B" for \textit{broad}. It is possible that these broad lines could be interpreted as "symmetric outflows," with simultaneously blue-shifted and red-shifted outflowing gas (e.g., \citealp{2017A&A...603A..99P} and \citealp{Rojas:2020or}). 

\subsubsection{Three Gaussians}
\label{sec:tripgauss}
Only [{\textsc Ne\,V}] 14.32\,\micron\, and [{\textsc O\,IV}] 25.89\,\micron\, required a model with three Gaussian lines to account for any extra features. A few combinations were needed for such cases: an "outflow" signature, or blue-shifted line, with a nearby emission line at longer wavelengths (one source for [{\textsc Ne\,V}] 14.32\,\micron\, and five sources for [{\textsc O\,IV}] 25.89\,\micron); a blue-shifted line with a broad line (1 source for [{\textsc O\,IV}] 25.89\,\micron); broad line with a nearby emission at longer wavelengths (two sources for [{\textsc O\,IV}] 25.89\,\micron). These are denoted as "O+R", "O+B", "B+R", respectively, in Table \ref{tab:fitstats}. Again, the spectra that required this model were selected by visual inspection and by comparing the $\chi^2$ after the new model fit.
The total line flux was calculated according to the specific case. If there was a blue-shifted line and/or if the line was broad, then the corresponding flux was added to the total flux. On the other hand, if a Gaussian was added to account for any redder, separate, emission line, then the flux and uncertainty of the additional line was not considered. An example of a triple Gaussian fit, specifically for a source showing a blue-shifted peak signature along with a nearby emission line, is shown in the bottom right panel of Figure\,\ref{fig:fits}.

\section{Results and discussion}
\label{sec:analyses}
Here we present and discuss the main results of our work. In \S\ref{sec:detrates} we illustrate the detection rates of the MIR CLs studied here, and compare them with the NIR [{\textsc Si\,VI}] 1.96$\mu$m and [{\textsc Si\,X}] 1.43$\mu$m CLs measured by \citet{denBrok_DR2_NIR}. In \S\ref{sec:corr} we discuss the correlation between the CL and the X-ray flux, and how their flux ratios relate to the AGN properties (i.e., $N_{\rm H}$, $M_{\rm BH}$, $L_{14-150}$ and $\lambda_{\rm Edd}$). The fluxes of all lines studied here are reported in Table\,\ref{tab:Spitzer_tbl1} of
Appendix\,\ref{app:fluxes}.

\subsection{Detection rates}
\label{sec:detrates}
We find that each of the five MIR emission lines studied in this work have very high detection rates (85-95\%), as given in Table \ref{tab:fitstats}. The best-fit centroid wavelengths of the lines were consistent with the expected values for each line (see Appendix\,\ref{app:bestlamb}).
In Figure\,\ref{fig:DetFracPlot} we compare the detection rate of the MIR CLs with those of  [{\textsc Si\,VI}] and [{\textsc Si\,X}] lines in the NIR provided by \citet{denBrok_DR2_NIR} for the {\it Swift}/BAT sample. These two emission lines have very high ionization energies of 167 and 351\,eV respectively, and are found at rest-frame wavelengths of 1.963 and 1.430\micron. Given their high ionizing potentials, they could serve as a useful comparison with the MIR CLs.

The detection rates for all CLs were determined as follows. For each source, we evaluated whether the emission line luminosity error was below a defined threshold luminosity and counted the sources with errors below this value. Sources with a luminosity error exceeding the threshold were excluded from that bin, as their sensitivity was insufficient for detection at the threshold luminosity. Next, for each threshold luminosity bin, we counted how many sources met the detection criterion (S/N$>$3). The higher-luminosity bins contain most of the sources, as all observations are sensitive enough to detect a $10^{43} \rm\,erg\,s^{-1}$ line. The detection fraction and its uncertainty were calculated using the \textsc{scipy} Python package by applying the inverse of the regularized incomplete beta function. This process was repeated for each MIR line, the two NIR lines, and their corresponding X-ray luminosity in the 14--150 keV range.
Here and throughout the paper, fractions and uncertainties were estimated following \citet{Cameron:2011cl}. All the uncertainties represent the 16th and 84th quantiles of a binomial distribution. We performed this test for the NIR lines [{\textsc Si\,VI}] 1.96$\mu$m and [{\textsc Si\,X}] 1.43$\mu$m, as well as for our MIR lines. The top and central panels of Figure\,\ref{fig:DetFracPlot} show that the MIR CLs studied here are typically detected nearly twice as frequently as the Silicon lines in the NIR, and thus serve as a better proxy for AGN activity.

In the bottom panel of Figure\,\ref{fig:DetFracPlot} we show the detection fraction of the five CLs studied here as a function of the ionizing potential. While we do not see a large change in the detection fractions ($\sim 10\%$ at most) and do not find evidence of a clear trend, we see that the [{\textsc Ne\,V}]  lines, which have the highest ionization potential, are also those with the lowest detection fraction.
In order to understand the origin of the non detections, we studied whether the AGN for which coronal lines were not detected have different properties than those in which they are detected. This was done by performing Kolmogorov-Smirnov tests for these two samples to test whether their black hole mass, Eddington ratio, column density and X-ray luminosities are different. Our test showed that, for none of the CLs studied here, the two subsamples presented significant differences.

\begin{figure*}
    \centering
    \includegraphics[scale = 0.33]{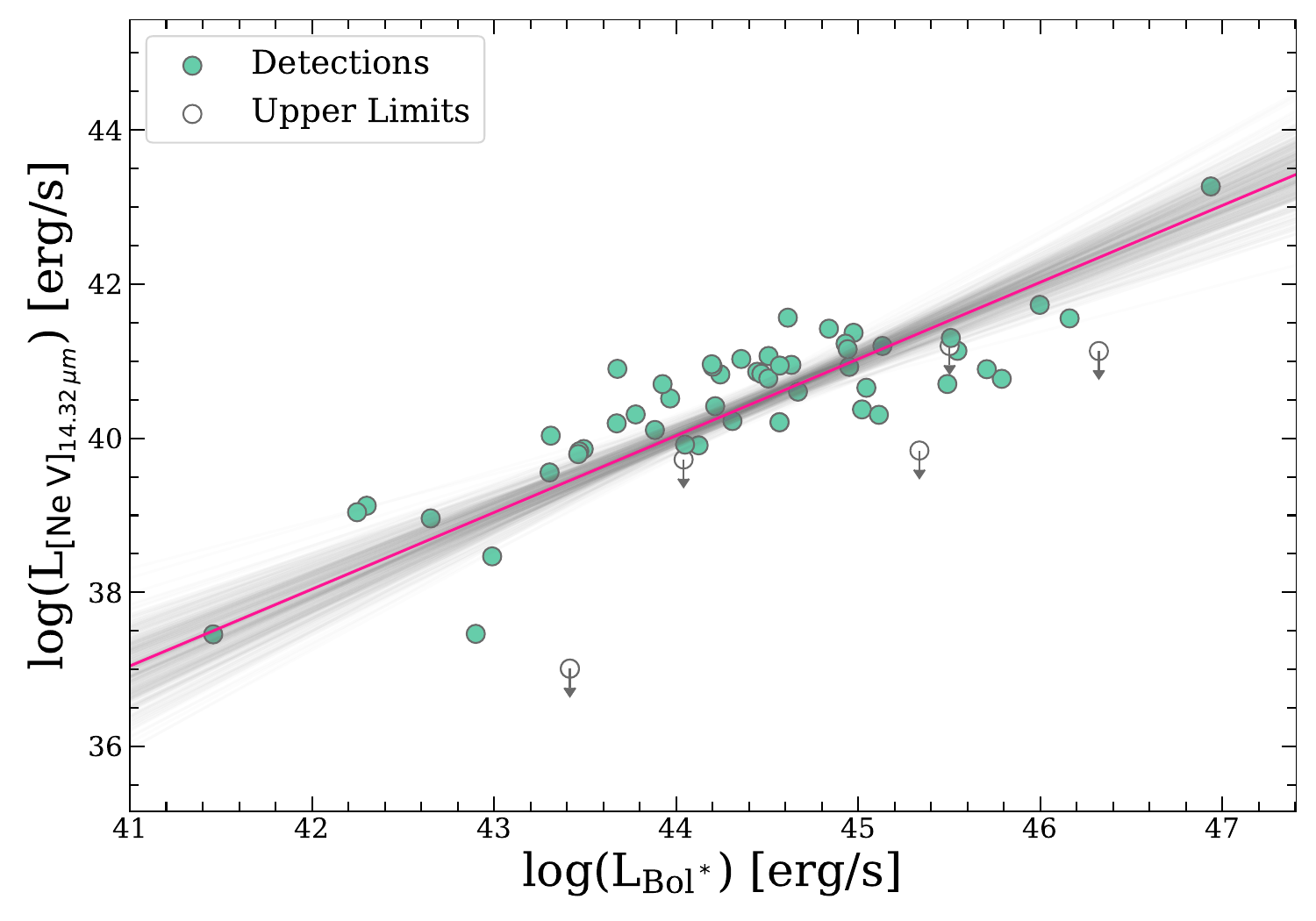}
    \includegraphics[scale = 0.33]{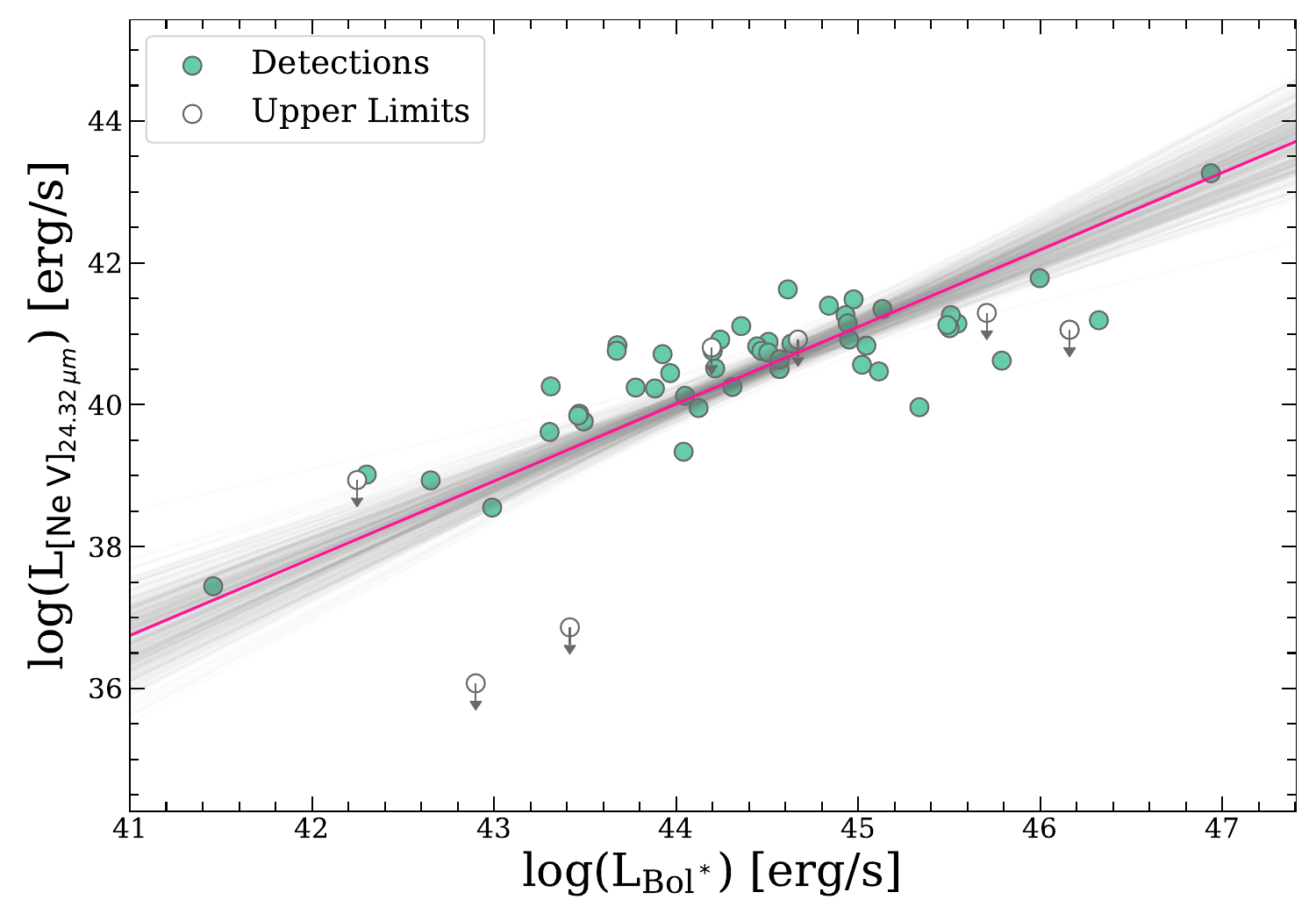}
    \includegraphics[scale = 0.33]{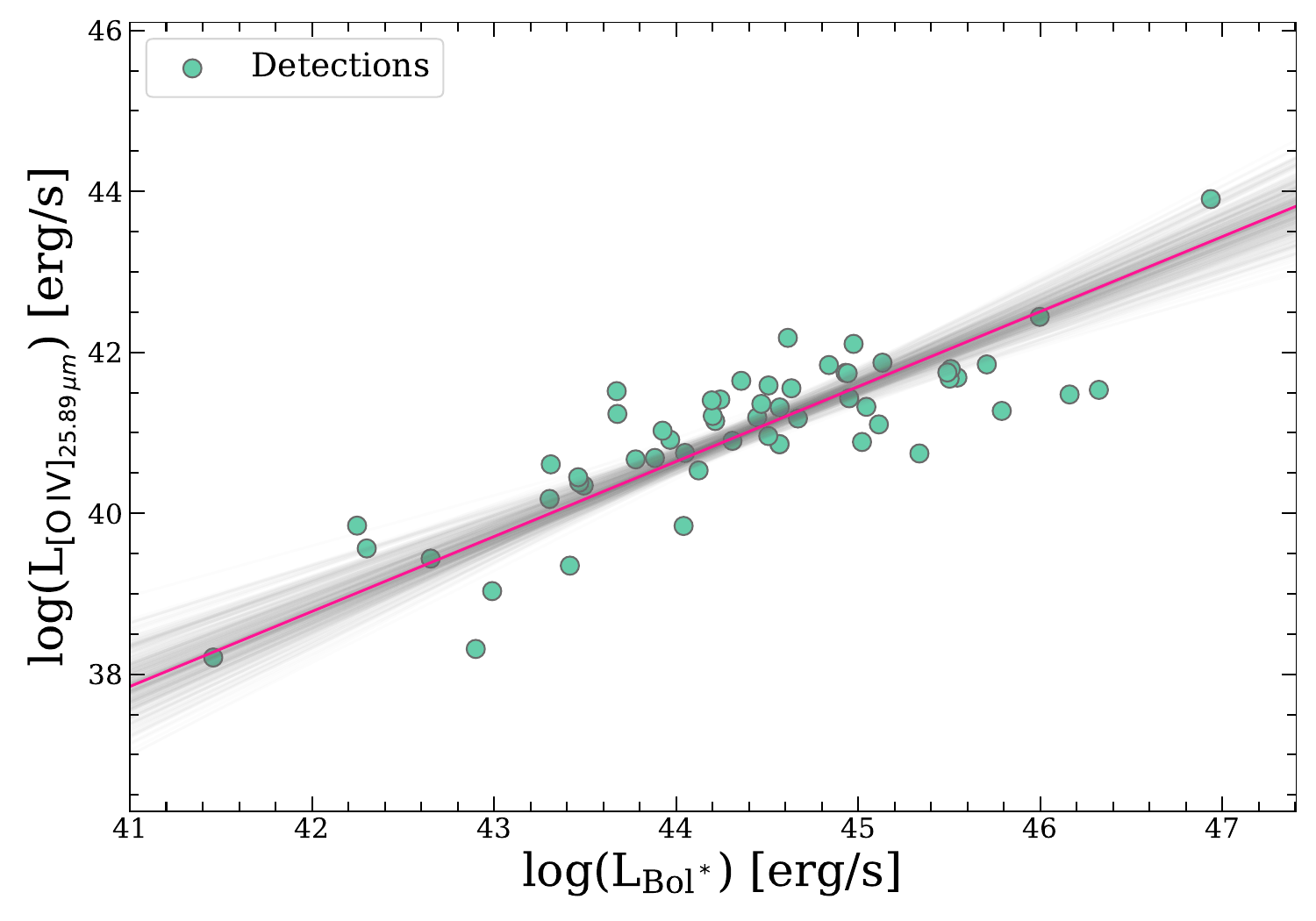}
    \includegraphics[scale = 0.33]{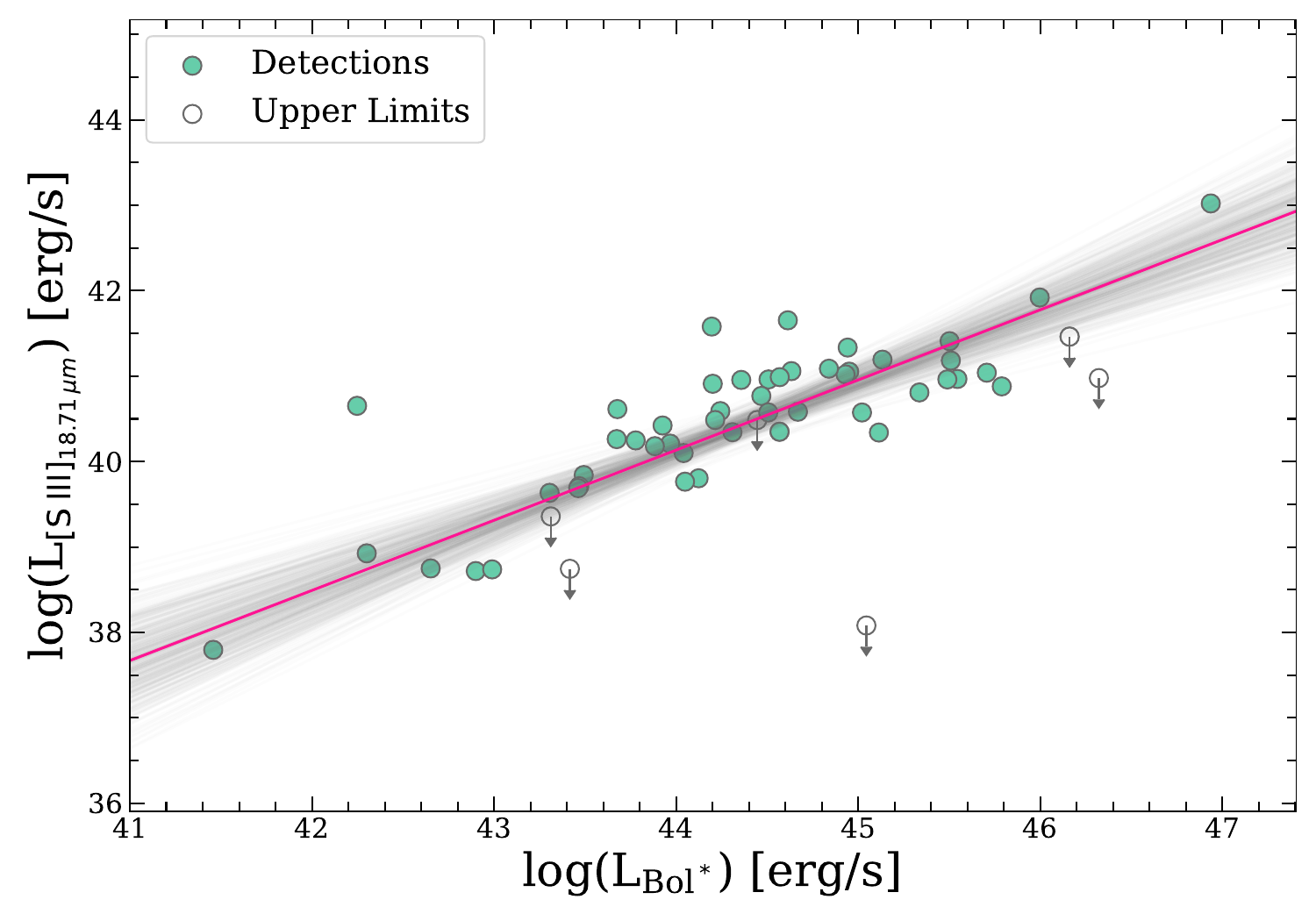}
    \includegraphics[scale = 0.33]{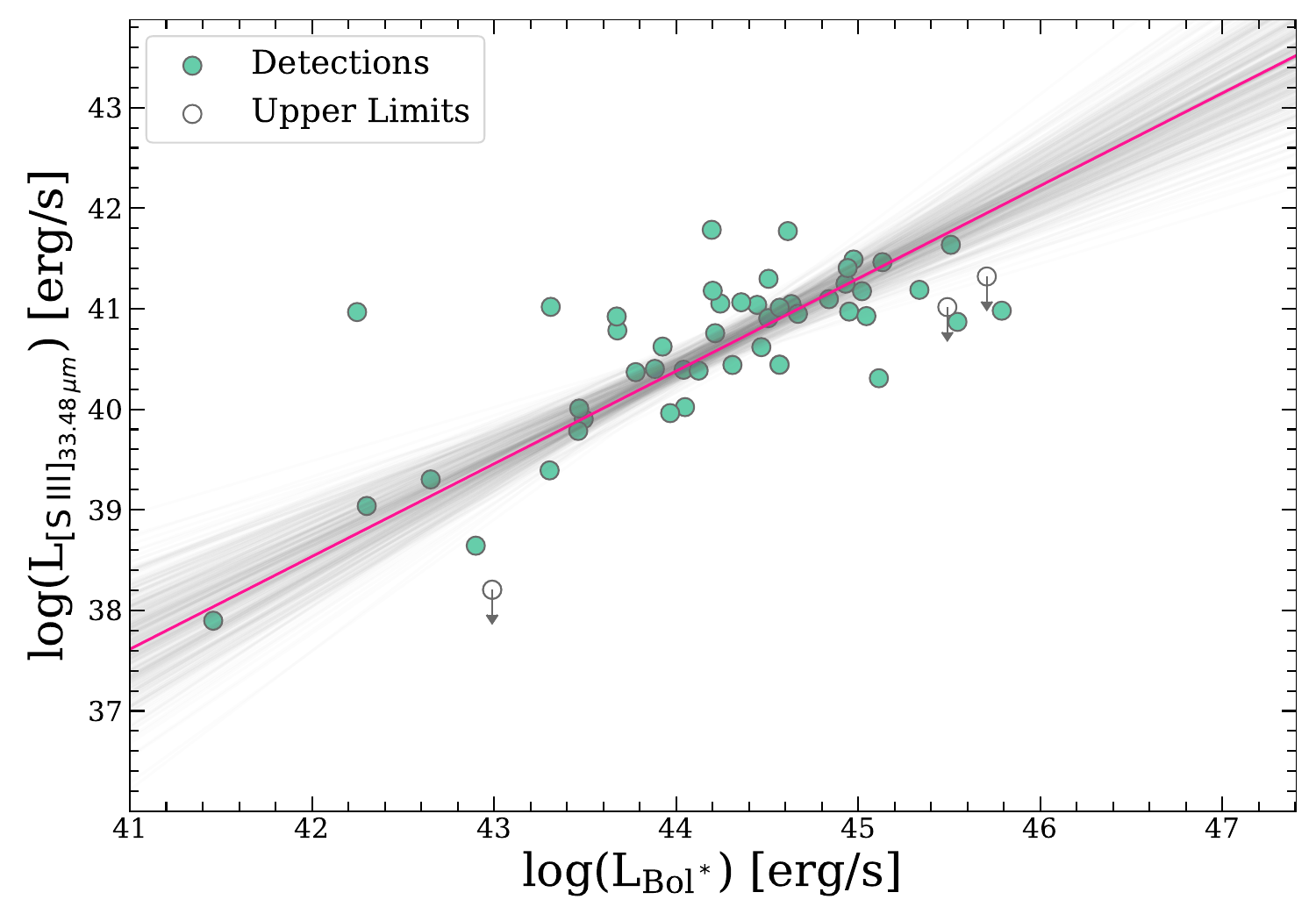}
    \caption{Log of emission line luminosities vs. log of bolometric luminosities (obtained from the spectral energy distribution analysis of \citet{Gupta:2024bp}, represented by L$_{\textrm{Bol}^*}$) for the lines studied here. We performed a linear regression fit in the form Y=$\alpha\cdot$X + $\beta$, where Y and X represent the logs of the line and bolometric luminosities, respectively; $\alpha$ and $\beta$ represent the slope and y-intercept of the linear fits, respectively. The linear fit values can be found in Table\,\ref{tab:lineVSbol}, with errors. The confidence intervals on the best fit are shown as shaded areas.}
    \label{fig:lbol}
\end{figure*}

\subsection{The relation between MIR CLs and AGN emission}
\label{sec:corr}
We studied the relations between the coronal emission line flux and the X-ray flux in the 2--10\,keV and 14--150\,keV bands for each emission line, along with the relation between the emission line luminosity vs. the X-ray luminosity, shown in Figure \ref{flux_and_lum_NeV}. 
The Python package \textsc{Linmix} was used for censoring, to take upper limits into consideration, for the overall linear regression fit in the form
\begin{equation}
\textrm{Y}=\alpha\cdot\textrm{X} + \beta.
\end{equation}
Here Y and  X represent the log of line flux/luminosities and the log of the X-ray flux/luminosities, respectively, while $\alpha$ and $\beta$ represent the slope and y-intercept of the linear fits, respectively. The \textsc{Linmix} package provides a large array of slopes and y-intercept. We use the median slope and y-intercept as the overall line of best-fit parameters. These fit parameters and their uncertainties are listed in Table\,\ref{tab:flux_lum_tbl}, along with some correlation coefficients. We also use \textsc{Linmix} to determine the intrinsic scatter, taking the median value from the \textsc{Linmix} array, and the \textsc{pymccorrelation} \textsc{python} package (\citealp{2020ApJ...893..149P}) to obtain the Spearman's Rank correlation coefficient and the corresponding p-value (\citealp{2014arXiv1411.3816C}). For the latter two coefficients, the package provided the 16th, 54th, and 84th percentiles, where we only show the 84th percentile values. We also determined confidence intervals and included these in the plots as well, for visual purposes. We find a clear positive correlation between the log of emission line luminosity and the log of X-ray luminosity, although with rather large scatters (typically 0.4 - 0.5\,dex; see Table\,\ref{tab:flux_lum_tbl}). To test for any redshift-related effect on the correlations, we performed a test considering only sources within 100\,Mpc. We found that the intrinsic scatter for this subsample was consistent with that obtained for our full sample. The slopes obtained are generally steeper than those reported for the entire sample.

We also tested how the CLs studied here are correlated with the bolometric luminosity. This was done by using the results of the recent work of \cite{Gupta:2024bp}, who studied the spectral energy distribution of $\sim 240$ unobscured ($N_{\rm H} < 10^{22}\,{\rm cm^{-2}}$) nearby AGN from the \textit{Swift}/BAT catalogue. In their work, they used simultaneous optical/UV photometric data from \textit{Swift}/UVOT and X-ray spectral data from \textit{Swift}/XRT, using \textsc{GALFIT} to correct the source magnitudes for host galaxy contamination in the optical/UV and determine the intrinsic AGN fluxes. As illustrated in Figure\,\ref{fig:lbol}, the log of the CL luminosities are tightly correlated with the bolometric output of the AGN in our sample. Interestingly, the typical scatter in these relations is lower  (typically 0.2 -- 0.3\,dex, see Table\,\ref{tab:lineVSbol}) than that obtained for the X-ray emission. This could be due to the fact that the luminosities obtained by the careful spectral energy distribution study of \cite{Gupta:2024bp} are more sensitive in the region that produces the ionizing flux responsible for the MIR CLs studied here. The difference in scatter and slope observed when comparing coronal lines to X-ray and bolometric luminosities confirms the idea that X-ray bolometric corrections are not constant (e.g.,\citealp{Vasudevan:2009zh}).

\subsection{Coronal lines and host galaxy/AGN properties}
\label{sec:fluxrat}
\begin{figure*}
    \centering
    \includegraphics[scale = 0.3]{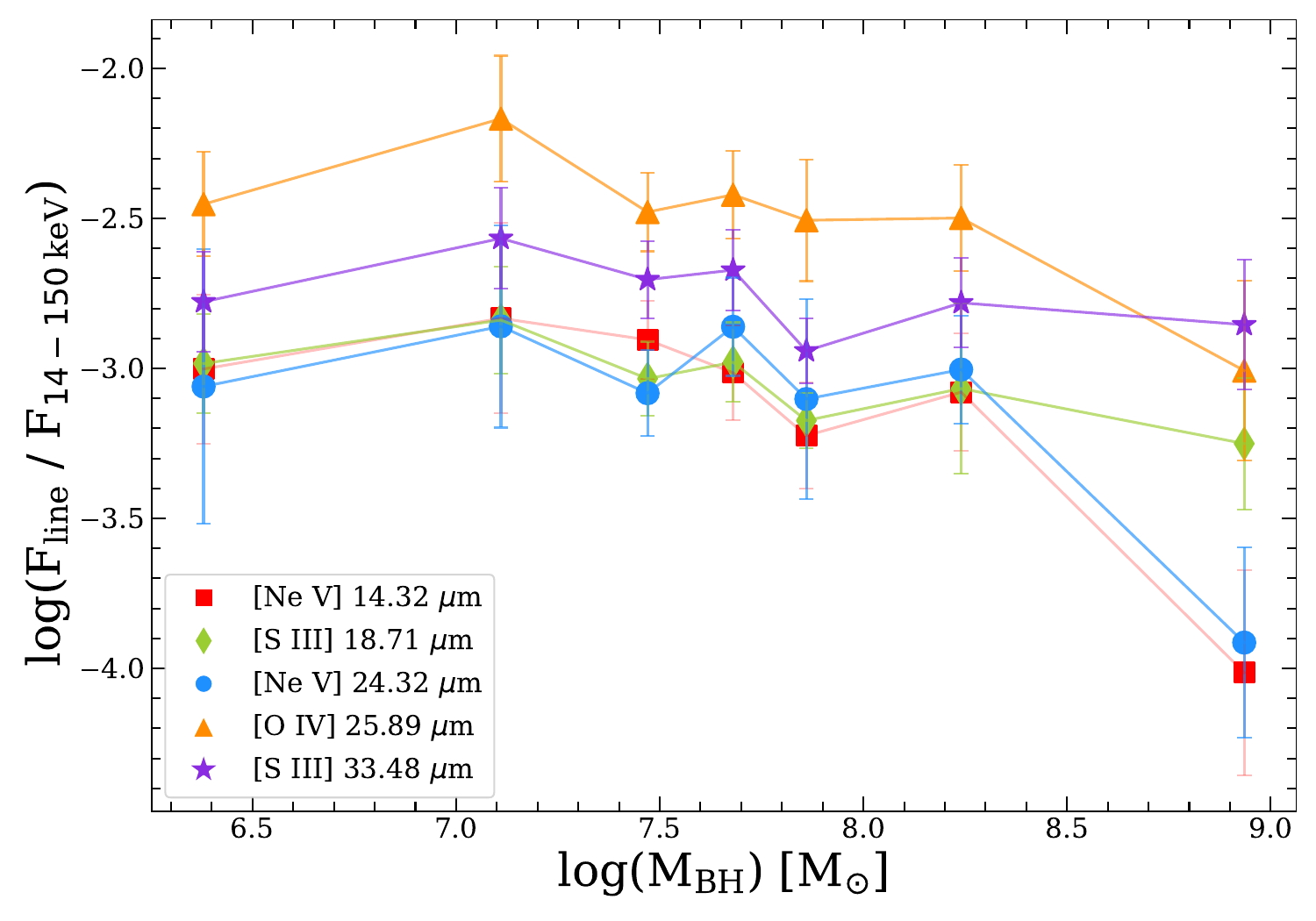} 
   \includegraphics[scale = 0.3]{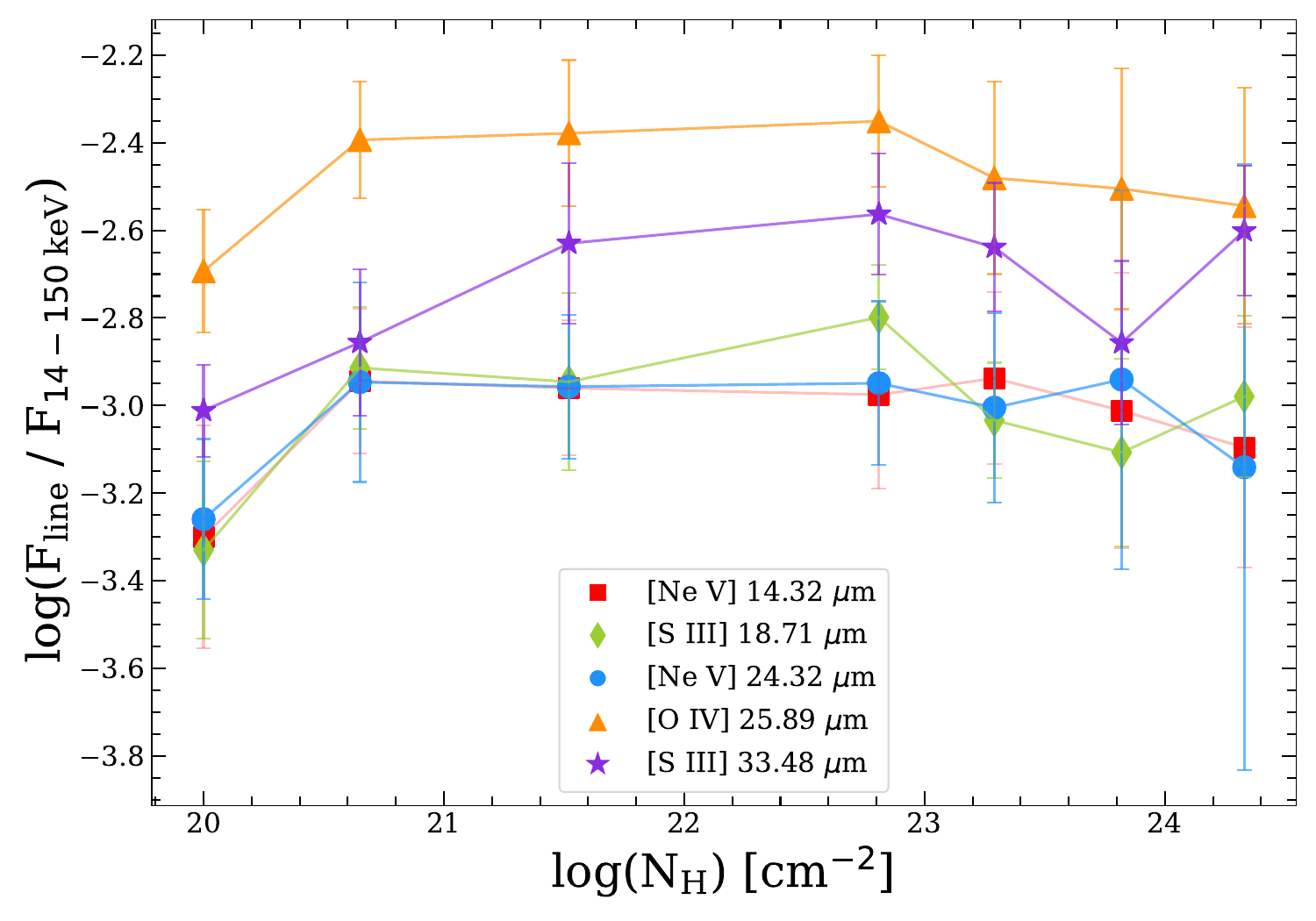} 
   \includegraphics[scale = 0.3]{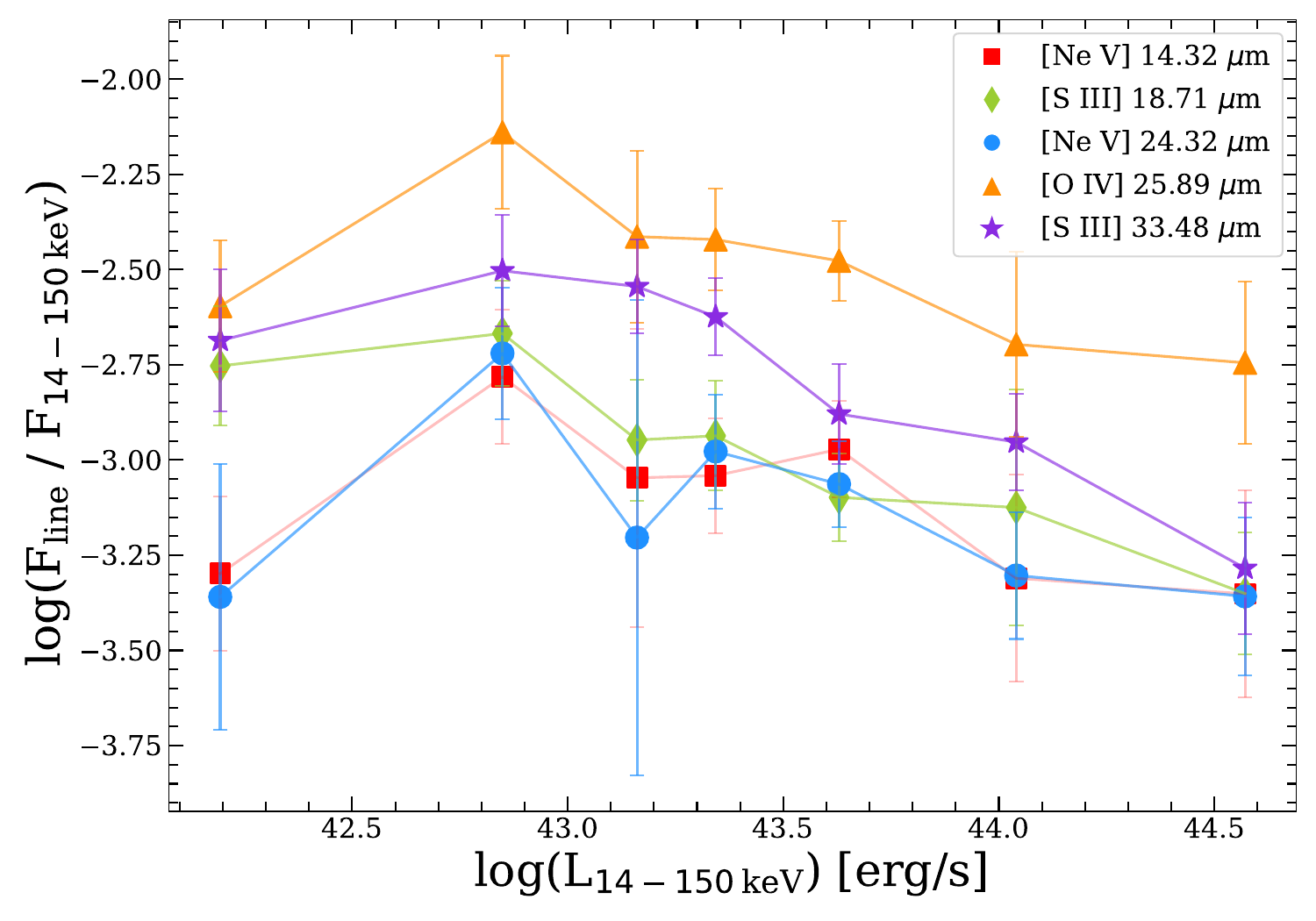} 
    \includegraphics[scale = 0.3]{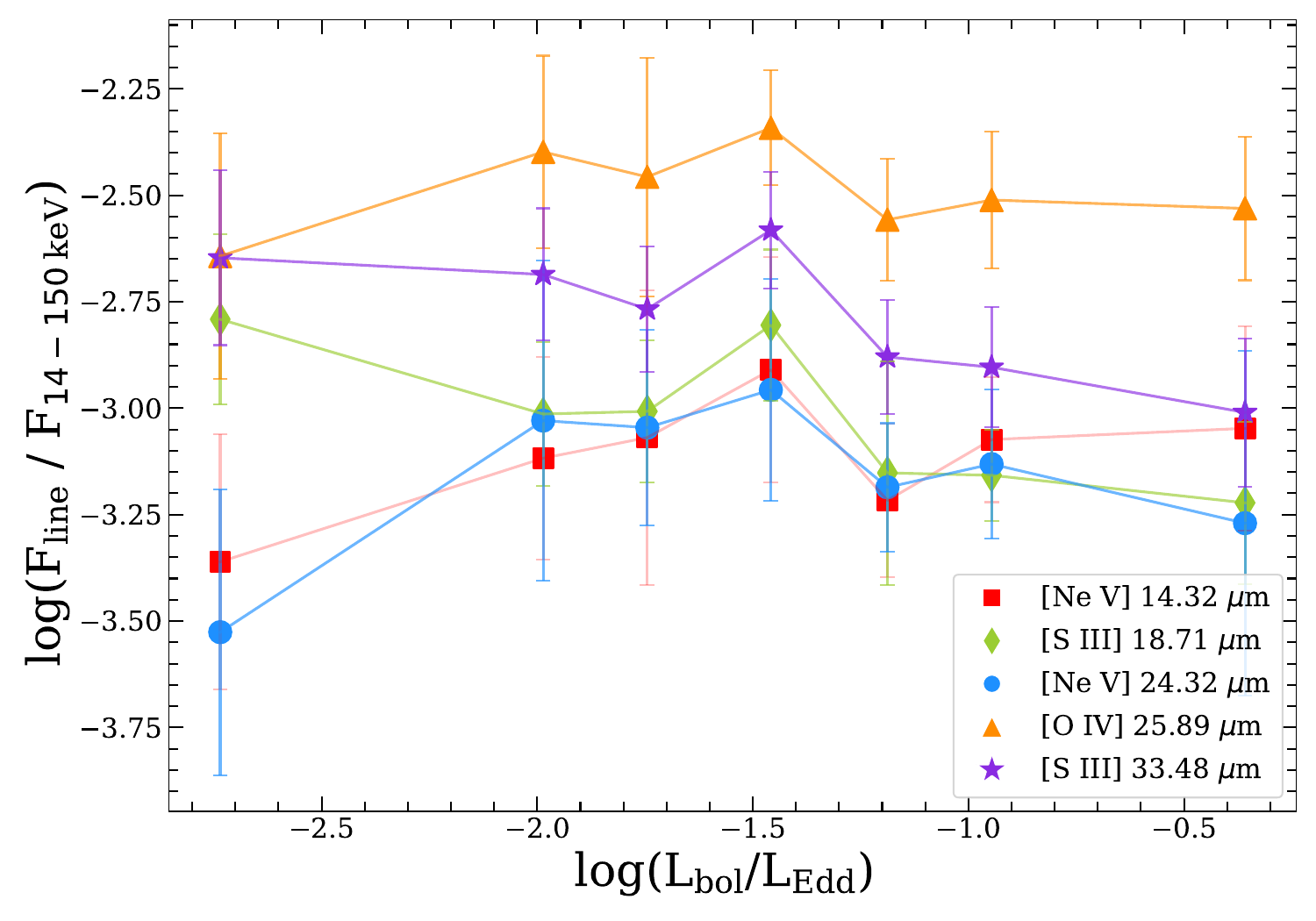}
    \includegraphics[scale = 0.3]{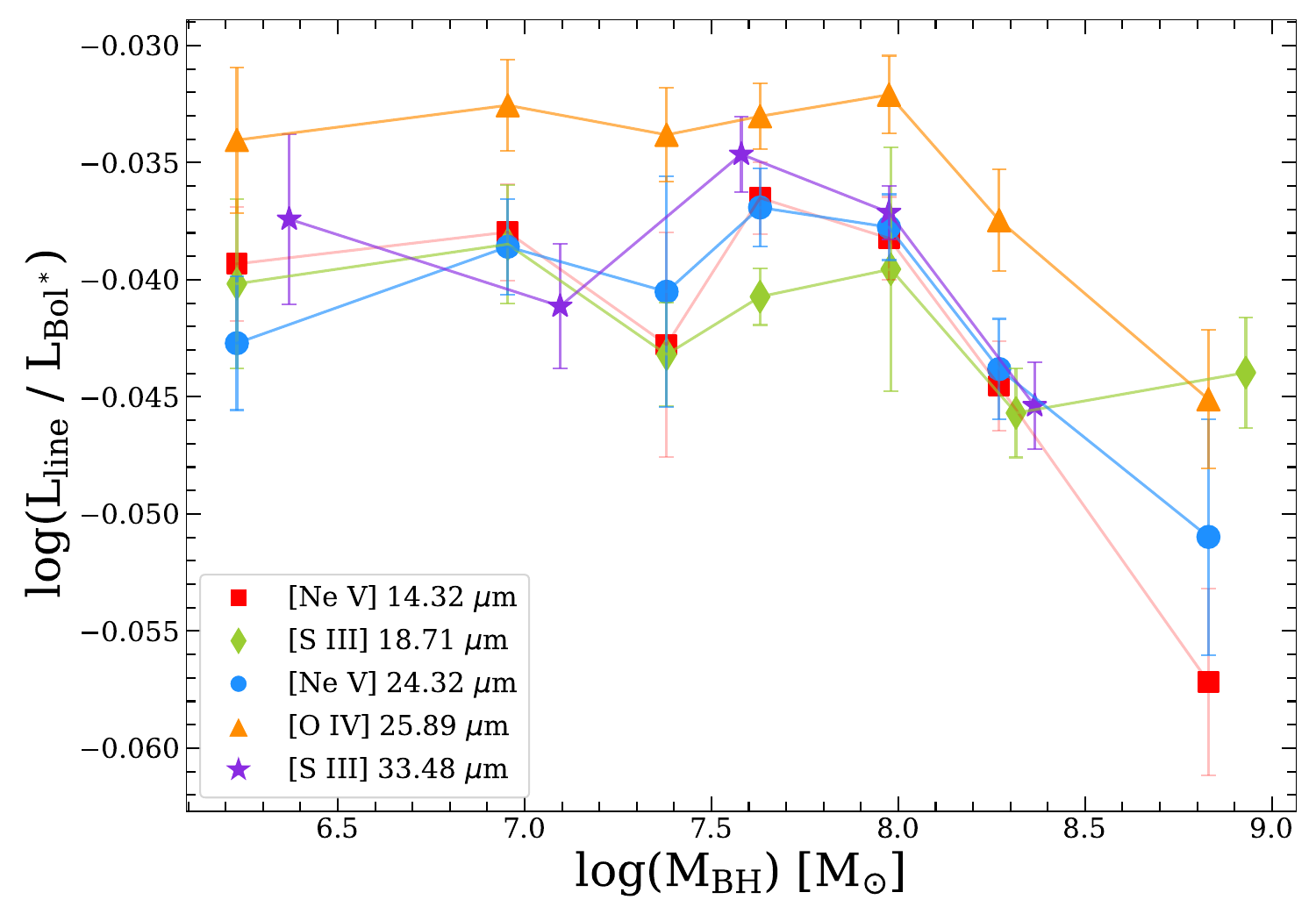} 
   \includegraphics[scale = 0.3]{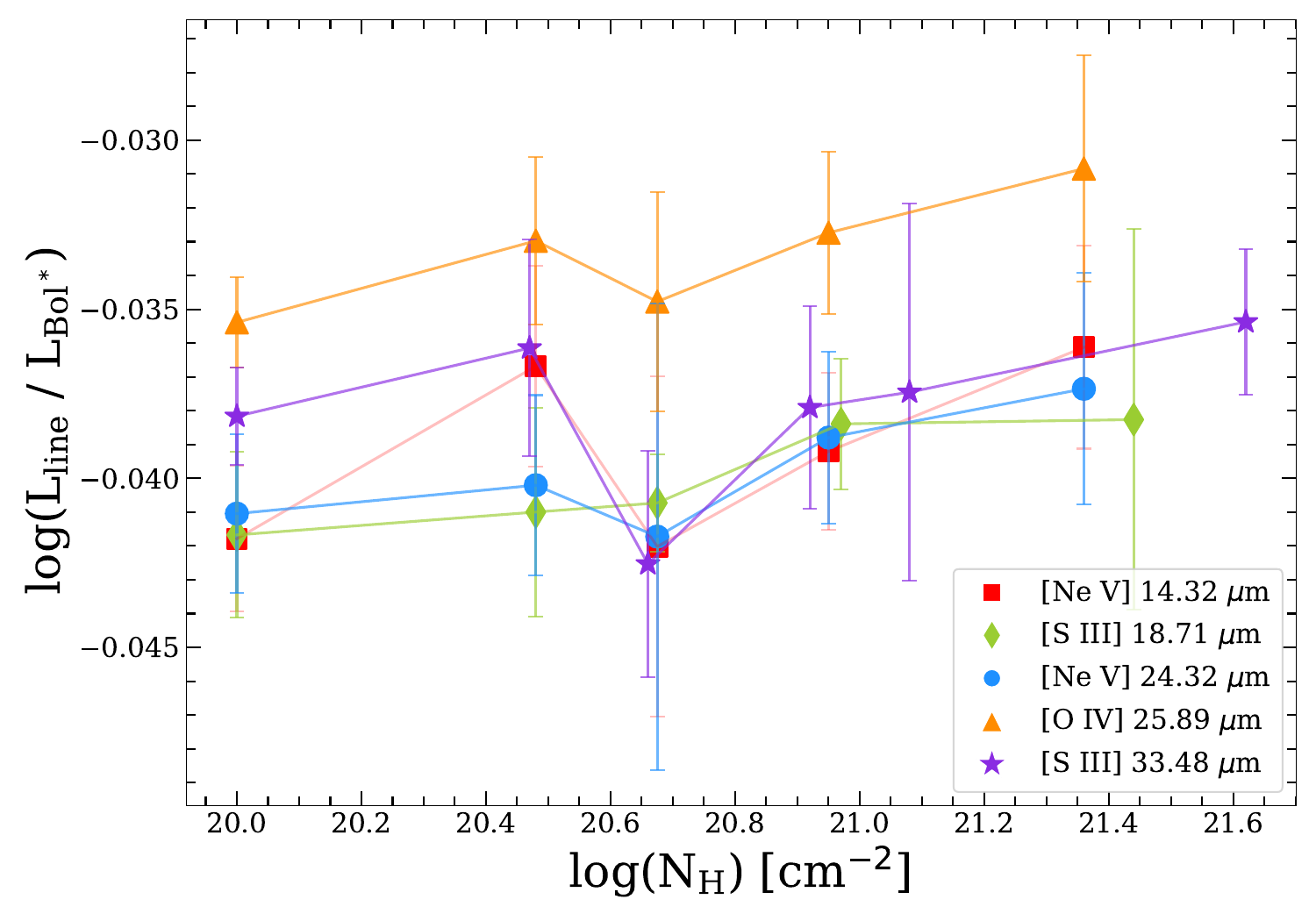} 
   \includegraphics[scale = 0.3]{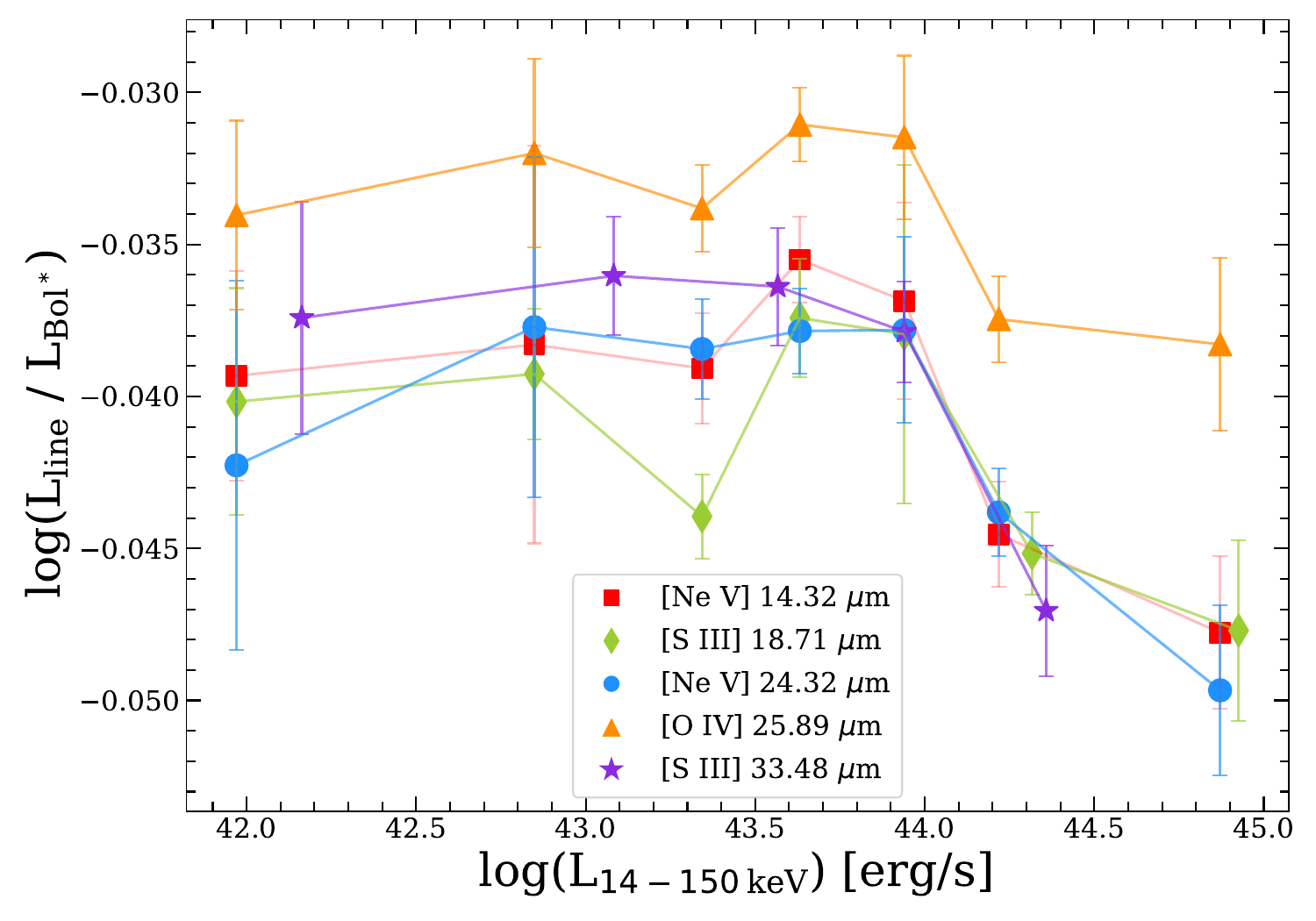} 
    \includegraphics[scale = 0.3]{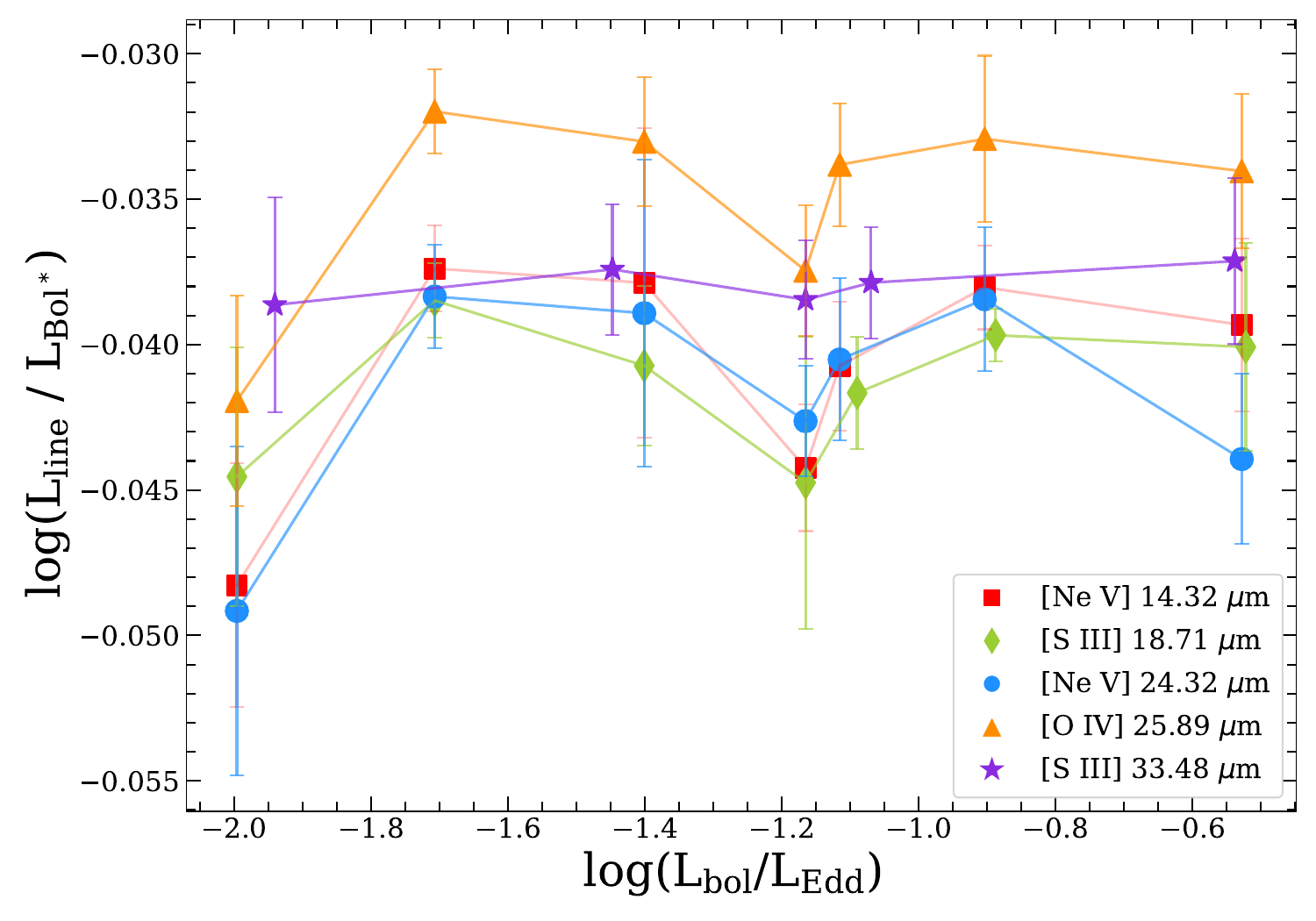}
    \caption{Median ratio between the flux/luminosity of the lines and the 14--150\,keV (top four panels) and bolometric (bottom four panels) flux/luminosity for all five emission lines studied here, versus black hole mass, column density, Eddington ratio and X-ray luminosity. The red squares represent [{\textsc Ne\,V}] 14.32\,\micron, the green diamonds represent [{\textsc S\,III}] 18.71\,\micron, the blue circles represent [{\textsc Ne\,V}] 24.32\,\micron, the orange triangles represent [{\textsc O\,IV}] 25.89\,\micron, and the purple stars represent [{\textsc S\,III}] 33.48\,\micron. The range in $N_{\rm H}$ is very narrow since the bolometric luminosities in \cite{Gupta:2024bp} were obtained for a sample of unobscured AGN. }
    \label{fig:fluxratio}%
\end{figure*}

To test whether the correlations reported in \S\ref{sec:corr} and Table\,\ref{tab:flux_lum_tbl} can be safely used through a wide range of AGN parameters, we studied how the ratio between the CL and X-ray fluxes changes with the most fundamental AGN properties. This was done for black hole mass, column density, X-ray luminosity, and Eddington ratio. We used survival analysis to take the upper limits into account and to find the median log(flux ratio) values for seven bins of the different properties. We performed this analysis for all five lines and, as shown in the four top panels of Figure \ref{fig:fluxratio}, there seems to be \textit{no} obvious correlation between the X-ray flux to emission-line flux ratios and the AGN properties, with the possible exception of a tentative decrease of [{\textsc Ne\,V}]\,24.32\,$\mu$m for the highest column densities. We also checked the same ratios but using the bolometric luminosities \citep{Gupta:2024bp} instead of the 14--150\,keV ones (four bottom panels of Figure \ref{fig:fluxratio}), and find again no clear trend, with the possible exception of a decrease of the ratio with black hole mass above $\sim 10^{7.5}\rm M_{\odot}$ for [{\textsc Ne\,V}] 14.32\,\micron, [{\textsc Ne\,V}] 24.32\,\micron, [{\textsc O\,IV}] 25.89\,\micron, and [{\textsc S\,III}] 33.48\,\micron. 

We also tested whether the ratio between lines and X-ray and bolometric flux correlates with the host galaxy inclination. In order to determine the inclination from the observed ratio of the semiminor to semimajor axes of the galaxy, we use 99 measured values from \citet{Koss11} and \citet{Koss:2021zw}. For the remaining sources, we follow the prescription outlined in \citet{Koss:2021zw}, adopting a flattening parameter (\(q_0\)) value of 0.2 for disk galaxies and 0.34 for early-type galaxies to convert the observed axis ratio into intrinsic inclination. For these 34 sources, we use inclination measurements from Pan-STARRS or the 2MASS Extended Source Catalog \citep{Jarrett00}. In seven cases, typically the most distant and luminous sources, the imaging is completely dominated by the PSF emission, and the host galaxy inclination cannot be reliably measured. We did not find any significant correlation between the inclination angles and the ratio of the coronal line and X-ray or bolometric fluxes. We also investigated whether the non detection of the coronal lines could be associated to a particular range of host galaxy inclinations. A Kolmogorov-Smirnov test showed that, for all coronal lines studied here, there is no significant difference in the distribution of inclination angles between detections and non detections.

Our results indicate that the MIR forbidden emission lines are generally unbiased to varying AGN properties, further emphasizing how they can be useful proxy for AGN activity. Studies with larger samples might allow us to better understand the possible relation of the ratio between CL and AGN bolometric luminosities with black hole mass. All correlations reported here could be particularly useful for future observations with \textit{JWST} in estimating the X-ray and bolometric luminosity of sources, once the MIR line fluxes have been determined.

\subsection{MIR coronal and Silicon lines}
\label{sec:MIRvsSilicon}
\begin{figure}
    \includegraphics[scale = 0.33]{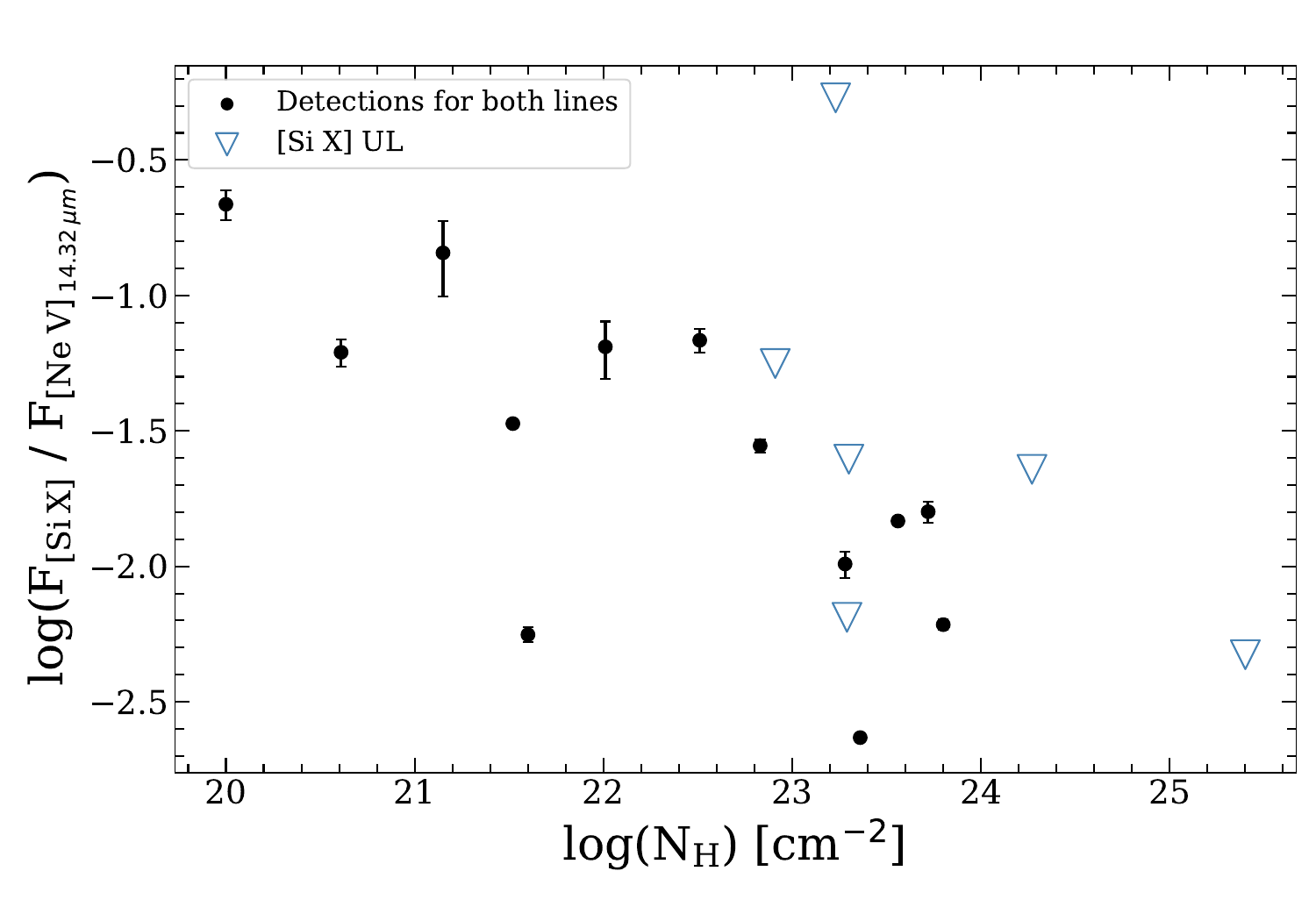} 
    \includegraphics[scale = 0.33]{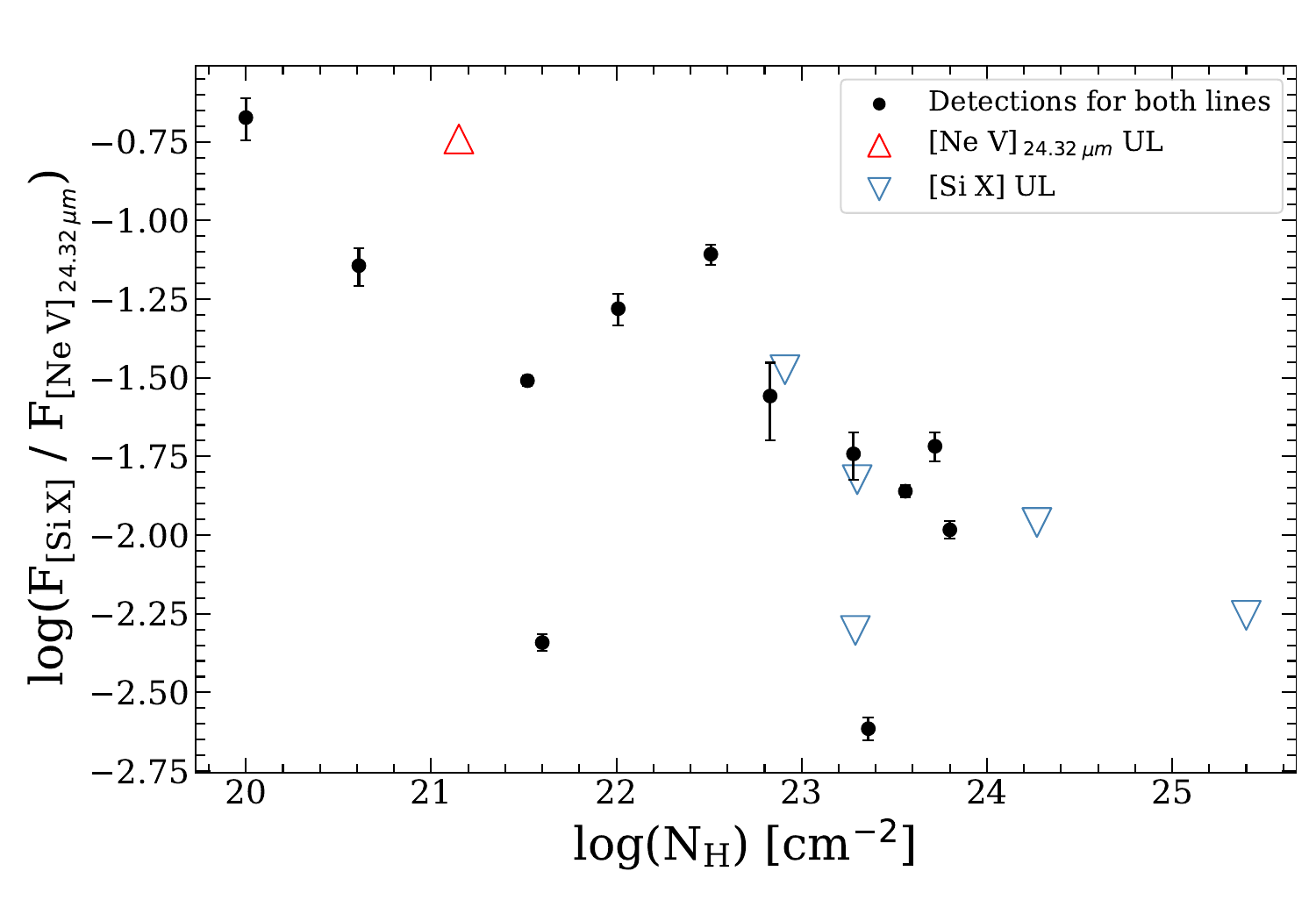} 
    \caption{The log of [{\textsc Si\,X}] 1.43$\mu$m line flux ratio vs. column density with [{\textsc Ne\,V}] 14.32\,\micron, top, and [{\textsc Ne\,V}] 24.32\,\micron, bottom. The downward triangles represent the upper limits for the Silicon sample, while the upward triangles represent an upper limit for [{\textsc Ne\,V}] 14.32\,\micron.}
    \label{fig:SiRatio}
\end{figure}

We also compared the line fluxes of coronal MIR and Silicon lines. We took the sources from \citet{denBrok_DR2_NIR} that overlapped with our sample (22 objects) and calculated the log of the ratio between silicon and MIR CL flux. We then compared this ratio to the different AGN properties. While there is no clear correlation with black hole mass, X-ray luminosity, or Eddington ratio, when this ratio was compared to column density, the line ratios tended to decrease as N$_{\textrm{H}}$ increases. Interestingly, this trend was only present for the [{\textsc Si\,X}] 1.43$\mu$m lines, and could be associated to extinction at these shorter wavelengths. This declining trend is shown in Figure \ref{fig:SiRatio}, where we show the log of the [{\textsc Si\,X}] line flux ratio with [{\textsc Ne\,V}] at 14.32\,\micron\, and 24.32\,\micron.

\subsection{Comparison with recent studies}

The 140 \textit{Swift} BAT selected objects used in this work are some of the brightest AGN in the sky, and hence have featured in a range of previous studies.
\citet{2010ApJ...716.1151W} had previously studied the MIR spectra of 79 sources selected from an earlier version of the \textit{Swift}/BAT catalogue.
In Appendix~\ref{app:Weaver}, we compare our [{\textsc Ne\,V}] and [{\textsc O\,IV}] measurements with \citet{2010ApJ...716.1151W}, we find good agreement in the 75 sources that overlap with our sample.

Instead of using the ultra-hard X-ray flux, another popular method to identify both obscured and unobscured AGN is via their MIR continuum flux, which probes the thermal emission from warm dust heated by the AGN.
The most comprehensive studies using this method have been through the 12\,\micron\ selected sample of 112 AGN \citep{2008ApJ...676..836T, 2010ApJ...709.1257T}. These studies have found that the [{\textsc Ne\,V}] and [{\textsc O\,IV}] MIR lines are good tracers of the AGN emission, by comparing with both their optical emission properties \citep[e.g.][]{Feltre:2023lz} and X-ray properties \citep{Spinoglio:2022of}.
 A total of 100 of these 12\,\micron-selected AGN have both \textit{Spitzer}-IRS and \textit{Swift}-BAT data, meaning they appear in our sample. These 100 objects were recently studied in detail by \citet{Spinoglio:2022of}, who found that the high-ionization MIR [{\textsc Ne\,V}] and [{\textsc O\,IV}] lines correlate with other AGN bolometric indicators, and that there is no evidence for systematic differences in these correlations among the various AGN populations, including types 1 and 2 and Compton-thick and Compton-thin AGN, consistent with the conclusions of this work.
Furthermore, Eqs.~1, 2, and 3 of \citet{Spinoglio:2022of} give the observed correlations of 
\begin{equation}
    L_\textrm{bol}
    \propto L_{[{\rm \textsc O\,IV}]\,25.9}^{0.88\pm0.06}
    \propto L_{[{\rm \textsc Ne\,V}]\,14.3}^{0.90\pm0.08}
    \propto L_{[{\rm \textsc Ne\,V}]\,24.3}^{1.06\pm0.10}
\end{equation}
which are consistent with the correlations we observe between the MIR emission line luminosities and the 14-150\,keV luminosities in Table~\ref{tab:flux_lum_tbl}. The bolometric luminosities in \citet{Spinoglio:2022of} were calculated using the bolometric corrections of  \cite{Lusso:2012jc}, while in our work we used the bolometric luminosities obtained for each source by a careful analysis of their simultaneous optical/UV/X-ray spectral energy distribution. Our study represents a significant increase in sample size compared to previous studies, as this is the largest study to date of Swift/BAT AGN.

\section{Summary \& Conclusions}
\label{sec: results}

In this paper we have studied the relation between MIR forbidden emission lines and the X-ray emission from AGN. By using the MIR spectra from \textit{Spitzer}/IRS, we carefully modelled the [{\textsc Ne\,V}] 14.32/24.32\,\micron, [{\textsc S\,III}] 18.71/33.48\,\micron, and [{\textsc O\,IV}] 25.89\,\micron\ CLs (see Figure\,\ref{fig:fits} and Table\,\ref{tab:fitstats}) for a sample of $\sim140$ {\it Swift}/BAT-selected AGN for which black-hole mass, column density, X-ray luminosity, and Eddington ratio properties are available. Our main results are:

\begin{itemize}
    \item We found a very high detection rate for coronal emission lines in our sample, with at least 85\% of the objects showing each of the line we studied (see Figure\,\ref{fig:DetFracPlot} and Table\,\ref{tab:fitstats} for details).
    \item The MIR coronal lines luminosities correlate well with the 14--150\,keV luminosities, with a typical scatter of 0.4--0.5\,dex confirming the idea that MIR coronal lines are good proxies for the accretion power (see Figure\,\ref{flux_and_lum_NeV}), at least the in the range of X-ray luminosities probed here ($10^{42}-10^{45}\rm\,erg\,s^{-1}$). The relations between the MIR lines studied here and the X-ray emission are reported in Table\,\ref{tab:flux_lum_tbl}.
 \item The CL emission is correlated more tightly to the bolometric luminosity ($\sigma \sim 0.2 - 0.3$\,dex), calculated from careful analysis of the spectral energy distribution \citep{Gupta:2024bp}, than to the X-ray luminosity (Figure\,\ref{fig:lbol}). The relations between the MIR lines studied here and the bolometric emission are reported in Table\,\ref{tab:lineVSbol}.
    \item The ratio between line flux/luminosity and hard X-ray flux/luminosity does not depend on N$_\textrm{H}$, M$_\textrm{BH}$, L$_\textrm{X-ray}$, or $\lambda_{\rm Edd}$ (Figure\,\ref{fig:fluxratio}),  within the range of parameters studied here. A possible decrease of the ratio between the line and bolometric luminosity and the black hole mass is found above $\sim 10^{7.5}\rm M_{\odot}$ for [{\textsc Ne\,V}] 14.32\,\micron, [{\textsc Ne\,V}] 24.32\,\micron, [{\textsc O\,IV}] 25.89\,\micron, and [{\textsc S\,III}] 33.48\,\micron.
\end{itemize}

Our results show that these lines can reliably serve as proxies for X-ray luminosity, without bias across black hole mass, column density, Eddington ratio, and X-ray luminosity, at least within the ranges of these parameters probed here. Future studies of AGN with \textit{JWST} will be able to use these CLs to study obscured growth across cosmic time.

\begin{acknowledgments}

We thank the referee for their suggestions and very detailed comments, which helped us improve the manuscript.
CR acknowledges support from Fondecyt Regular grant 1230345, ANID BASAL project FB210003 and the China-Chile joint research fund.
MJT acknowledges support from a FONDECYT postdoctoral fellowship (3220516). YD acknowledges support from a FONDECYT postdoctoral fellowship (3230310).
J.M.C.'s contribution was supported by an appointment to the NASA Postdoctoral Program at
the NASA Goddard Space Flight Center, administered by Oak Ridge
Associated Universities under contract with NASA. RR acknowledges support from the Fundaci\'on Jes\'us Serra and the Instituto de Astrof{\'{i}}sica de Canarias under the Visiting Researcher Programme 2023-2025 agreed between both institutions. RR also acknowledges support from the ACIISI, Consejer{\'{i}}a de Econom{\'{i}}a, Conocimiento y Empleo del Gobierno de Canarias and the European Regional Development Fund (ERDF) under grant with reference ProID2021010079, and the support through the RAVET project by the grant PID2019-107427GB-C32 from the Spanish Ministry of Science, Innovation and Universities MCIU. This work has also been supported through the IAC project TRACES, which is partially supported through the state budget and the regional budget of the Consejer{\'{i}}a de Econom{\'{i}}a, Industria, Comercio y Conocimiento of the Canary Islands Autonomous Community. RR also thanks to Conselho Nacional de Desenvolvimento Cient\'{i}fico e Tecnol\'ogico  ( CNPq, Proj. 311223/2020-6,  304927/2017-1 and  400352/2016-8), Funda\c{c}\~ao de amparo \`{a} pesquisa do Rio Grande do Sul (FAPERGS, Proj. 16/2551-0000251-7 and 19/1750-2), Coordena\c{c}\~ao de Aperfei\c{c}oamento de Pessoal de N\'{i}vel Superior (CAPES, Proj. 0001). This work was funded by ANID through CATA-BASAL FB210003 (FEB); FONDECYT Regular 1200495 (FEB) and 1241005 (FEB); and Millennium Science Initiative AIM23-0001  and ICN12\_009 (FEB). KKG thanks the Belgian Federal Science Policy Office (BEL- SPO) for the provision of financial support in the framework of the PRODEX Programme of the European Space Agency (ESA). ARL acknowledges support from a FONDECYT postdoctoral fellowship (3210157). The work of DS was carried out at the Jet Propulsion Laboratory, California Institute of Technology, under a contract with NASA.

\end{acknowledgments}

\facilities Swift, WISE, IRAS, Akari, Spitzer

\appendix

\section{Excluded sources}
\label{app:exc}
There are some cases that required us to exclude the spectra for source(s) from analyses. The source SDSS\,J130005.35$+$163214.8 only had a high-resolution spectrum with the range of 9-18\,\micron, and thus could not be analyzed for [{\textsc S\,III}] 18.7078\,\micron, [{\textsc Ne\,V}] 24.32\,\micron, [{\textsc O\,IV}] 25.89\,\micron, and [{\textsc S\,III}] 33.48\,\micron. When evaluating [{\textsc S\,III}] 18.7078\,\micron, three other sources (NGC\,985, IRAS\,05189$-$2524, and UGC\,5101) had missing pixels in their \textit{Spitzer} spectra that greatly affected the fit. Therefore, these were also excluded from analyses. When evaluating [{\textsc S\,III}] 33.48\,\micron, we noticed quite a few problems with some spectra. For Mrk\,352, most of the flux density pixels were "zero", thus a model fit was incalculable. Additionally, the spectra for PG\,0026$+$129, 3C\,234, 3C\,273, Q\,1821$+$643, and PG\,2349$-$014 had red-shifted out enough to not include the rest-frame 25.89\,\micron\, CL. Lastly, for the sources PG\,0804$+$761 and 4C\,$+$74.26, their spectra were partially redshifted out, but did not have enough flux density for a proper fit, and therefore were excluded from analyses.

\section{Coronal line fluxes}
\label{app:fluxes}
In Table\,\ref{tab:Spitzer_tbl1} we report the fluxes of all MIR lines studied here, while the list of sources for which only upper limits are available are shown in Table\,\ref{tab:UL_list}.

\begin{longrotatetable}
\begin{deluxetable*}{ccccccccc}
\tablecaption{Table of Spitzer IRS Fluxes and model information for [{\textsc Ne\,V}] 14.32\,\micron\,, [{\textsc Ne\,V}] 24.32\,\micron\, and [{\textsc O\,IV}] 25.89\,\micron\,. Flux Values in units of $10^{-14}$ erg$\cdot$cm$^{-2}$$\cdot$s$^{-1}$}
\tablewidth{0pt}
\tablehead{
\colhead{(1) } & \colhead{(2)} & \colhead{(3)} & \colhead{(4) } & \colhead{(5)}  & \colhead{(6)} & \colhead{(7)} &  \colhead{(8) } &  \colhead{ (9) }\\
\noalign{\smallskip}
\colhead{\textit{SWIFT} ID } &\colhead{Source Name} &\colhead{z} &\colhead{[{\textsc Ne\,V}]  14.32\,\micron} & \colhead{Model} & \colhead{ [{\textsc Ne\,V}] 24.32\,\micron} &  \colhead{Model } &  \colhead{ [{\textsc O\,IV}] 25.89\,\micron}  &  \colhead{Model }\\
\colhead{  } &\colhead{ } &\colhead{ } &\colhead{d.o.f} & \colhead{$\chi^2$} & \colhead{d.o.f}  & \colhead{$\chi^2$} & \colhead{d.o.f} &  \colhead{ $\chi^2$}
}
  \startdata
  	SWIFT\,J0006.2$+$2012 	&    Mrk\,335 	&    0.0258 	&    	1.1 $\pm$ 0.2	&	 S	&	2.2 $\pm$ 0.3	&	 S	&	4.9 $\pm$ 0.3	&	 S	\\
        &	 &	 &	 17 &	 0.8 &	 18 &	 1.1 &	 21 &	 1.6\\
        SWIFT\,J0029.2$+$1319 	&    PG\,0026$+$129 	&    0.142 	&    	0.7 $\pm$ 0.1	&	 S	&	 $\leq$ 0.2	&	 S	&	0.6 $\pm$ 0.0	&	 S	\\
        &	 &	 &	 17 &	 1.7 &	 18 &	 3.3 &	 19 &	 12.5\\
        SWIFT\,J0042.9$-$2332 	&    NGC\,235A 	&    0.0222 	&    	3.5 $\pm$ 0.2	&	 S	&	5.1 $\pm$ 0.2	&	 S	&	15.2 $\pm$ 0.5	&	 D, R	\\
        &	 &	 &	 16 &	 0.4 &	 19 &	 0.8 &	 16 &	 0.7\\
        SWIFT\,J0048.8$+$3155 	&    Mrk\,348 	&    0.015 	&    	6.2 $\pm$ 0.2	&	 S	&	4.5 $\pm$ 0.8	&	 S	&	15.8 $\pm$ 0.9	&	 S	\\
        &	 &	 &	 16 &	 2.1 &	 19 &	 0.4 &	 19 &	 4.1\\
\enddata

    \tablecomments{Table of Spitzer IRS Fluxes. }
\tablenotetext{*}{For the Model column, "S", "D", and "T" indicate when a single, double, and triple Gaussian was used, respectively. Furthermore, "O" indicates when an additional Gaussian was included to account  (with the exception of NGC\,1068). Gaussian was included to reproduce a nearby emission line at longer (redder) wavelengths, "B" when an additional broad Gaussian line centered at the same wavelength as the expected emission line was included (with the exception of NGC\,1068).}
	\label{tab:Spitzer_tbl1}
\end{deluxetable*}
\end{longrotatetable}

\begin{longrotatetable}
\begin{deluxetable*}{ccccccc}
\tablecaption{Table of Spitzer IRS Fluxes and model information for [{\textsc S\,III}] 18.7078\,\micron\, and [{\textsc S\,III}] 33.48\,\micron. Flux Values in units of $10^{-14}$ erg$\cdot$cm$^{-2}$$\cdot$s$^{-1}$}
\tablewidth{0pt}
\tablehead{
\colhead{(1) } & \colhead{(2)} & \colhead{(3)} & \colhead{(4) } & \colhead{(5)}  & \colhead{(6)} & \colhead{(7)}\\
\noalign{\smallskip}
\colhead{\textit{SWIFT} ID } &\colhead{Source Name} &\colhead{z} &\colhead{[{\textsc S\,III}]  18.7078\,\micron} & \colhead{Model} & \colhead{ [{\textsc S\,III}] 33.48\,\micron} &  \colhead{Model }\\
\colhead{  } &\colhead{ } &\colhead{ } &\colhead{d.o.f} & \colhead{$\chi^2$} & \colhead{d.o.f}  & \colhead{$\chi^2$}
}
  \startdata
  SWIFT\,J0006.2$+$2012 	&    Mrk\,335 	&    0.0258 	&    	1.5 $\pm$ 0.4	&	 S	&	1.9 $\pm$ 0.5 	&	 S	\\
&	 &	 &	 16 &	 4.7 &	 13 &	 1.9\\
SWIFT\,J0029.2$+$1319 	&    PG\,0026$+$129 	&    0.142 	&    	 $\leq$ 0.5	&	 S	&	 N/A 	&	 N/A	\\
&	 &	 &	 16 &	 1.5 &	 N/A &	 N/A	\\
SWIFT\,J0042.9$-$2332 	&    NGC\,235A 	&    0.0222 	&    	4.4 $\pm$ 0.6	&	 S	&	6.0 $\pm$ 1.0 	&	 S	\\
&	 &	 &	 17 &	 1.5 &	 13 &	 8.0\\
SWIFT\,J0048.8$+$3155 	&    Mrk\,348 	&    0.015 	&    	8.0 $\pm$ 0.3	&	 S	&	7.8 $\pm$ 0.4 	&	 S	\\
&	 &	 &	 16 &	 2.1 &	 15 &	 7.0\\
\enddata

    \tablecomments{Table of Spitzer IRS Fluxes. }
\tablenotetext{*}{The representations for this table resembles that of Table \ref{tab:Spitzer_tbl1}.}
	\label{tab:Spitzer_tbl2}
\end{deluxetable*}
\end{longrotatetable}

\begin{table*}
 	\centering
	\begin{tabular}{ccccc}
		\hline
		\hline
		{[{\textsc Ne\,V}]} 14.32\,\micron & {[{\textsc S\,III}]} 18.71\,\micron & {[{\textsc Ne\,V}]} 24.32\,\micron & {[{\textsc O\,IV}]} 25.89\,\micron & {[{\textsc S\,III}]} 33.48\,\micron\\
        \hline
        SWIFT\,J0059.4$+$3150	&	SWIFT\,J0029.2$+$1319	&	SWIFT\,J0029.2$+$1319	&	SWIFT\,J0255.2$-$0011	&	SWIFT\,J0152.8$-$0329 	\\
        SWIFT\,J0124.5$+$3350	&	SWIFT\,J0059.4$+$3150	&	SWIFT\,J0059.4$+$3150	&	SWIFT\,J1206.2$+$5243	&	SWIFT\,J0255.2$-$0011 	\\
        SWIFT\,J0230.2$-$0900	&	SWIFT\,J0214.6$-$0049	&	SWIFT\,J0138.6$-$4001	&	SWIFT\,J1305.4$-$4928	&	SWIFT\,J0304.1$-$0108 	\\
        SWIFT\,J0238.2$-$5213	&	SWIFT\,J0736.9$+$5846	&	SWIFT\,J0152.8$-$0329	&	SWIFT\,J2217.0$+$1413	&	SWIFT\,J0451.4$-$0346 	\\
        SWIFT\,J0241.3$-$0816	&	SWIFT\,J1001.8$+$2848	&	SWIFT\,J0241.3$-$0816	&		&	SWIFT\,J1209.4$+$4340 	\\
        SWIFT\,J0255.2$-$0011	&	SWIFT\,J1921.1$-$5842	&	SWIFT\,J0255.2$-$0011	&		&	SWIFT\,J1246.6$+$5435 	\\
        SWIFT\,J0451.4$-$0346	&	SWIFT\,J2042.3$+$7507	&	SWIFT\,J0508.1$+$1727	&		&	SWIFT\,J1341.9$+$3537 	\\
\hline
	\end{tabular}
	\caption{List of upper limits calculated for all MIR lines and corresponding SWIFT IDs. See \S\ref{sec:ULCalc} for further details.  }
    \label{tab:UL_list}
\end{table*}

\section{Best-Fit Centroid Wavelength}\label{app:bestlamb}
\begin{figure*}
    \centering
    \includegraphics[width=0.33\textwidth]{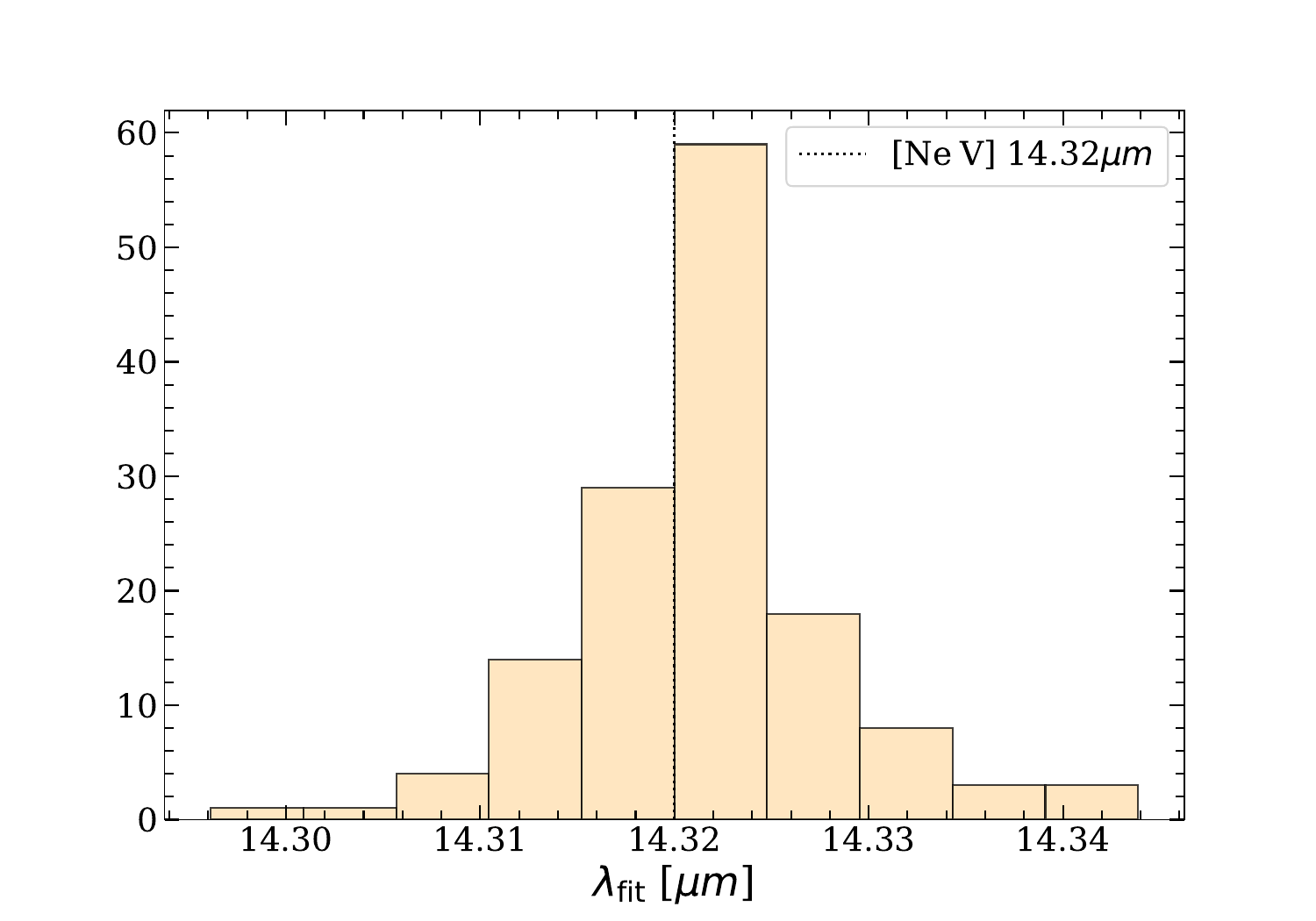}
    \includegraphics[width=0.33\textwidth]{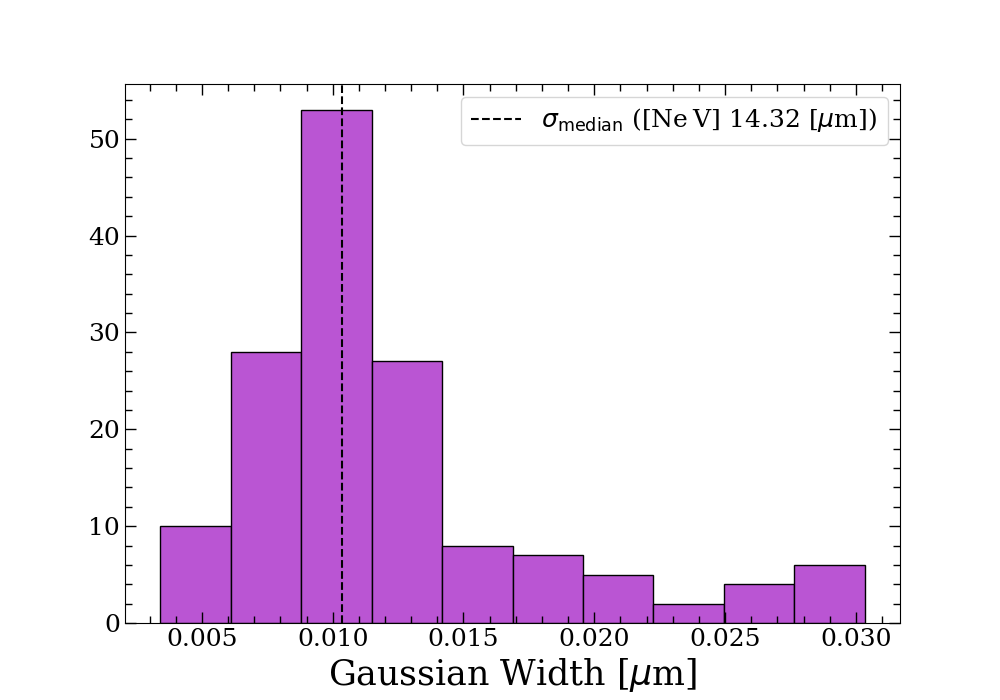}
    \includegraphics[width=0.33\textwidth]{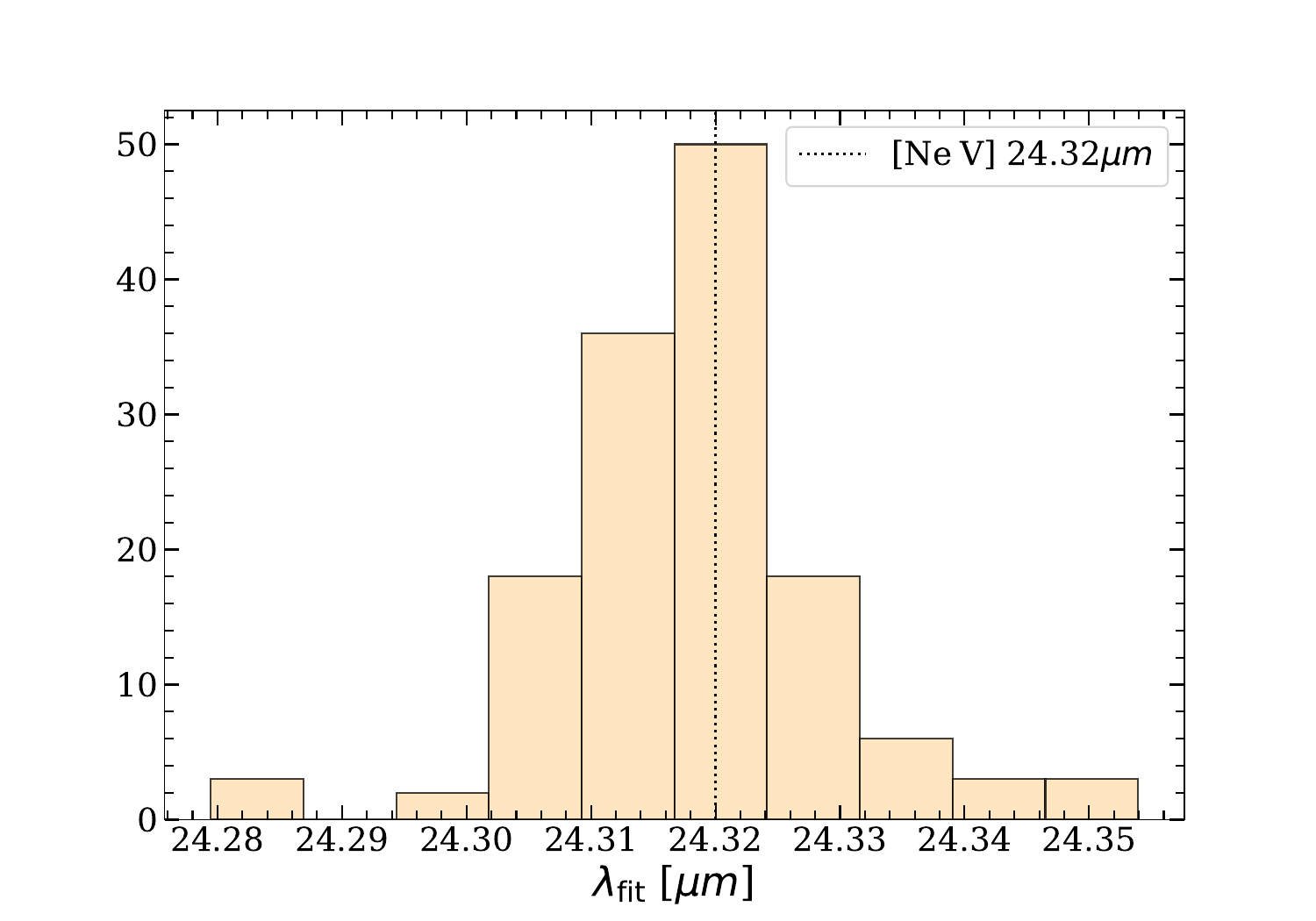} 
    \includegraphics[width=0.33\textwidth]{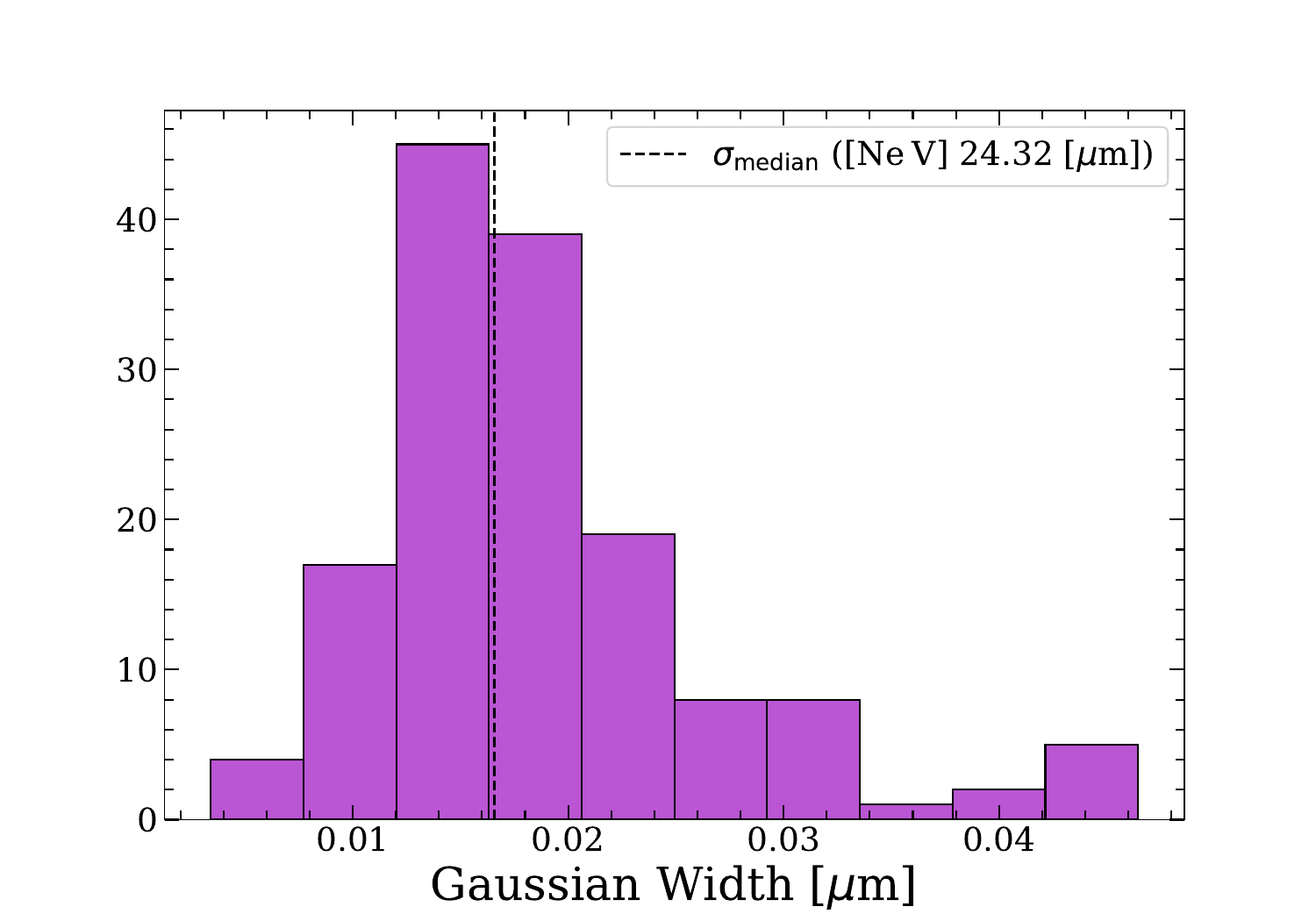}
    \includegraphics[width=0.33\textwidth]{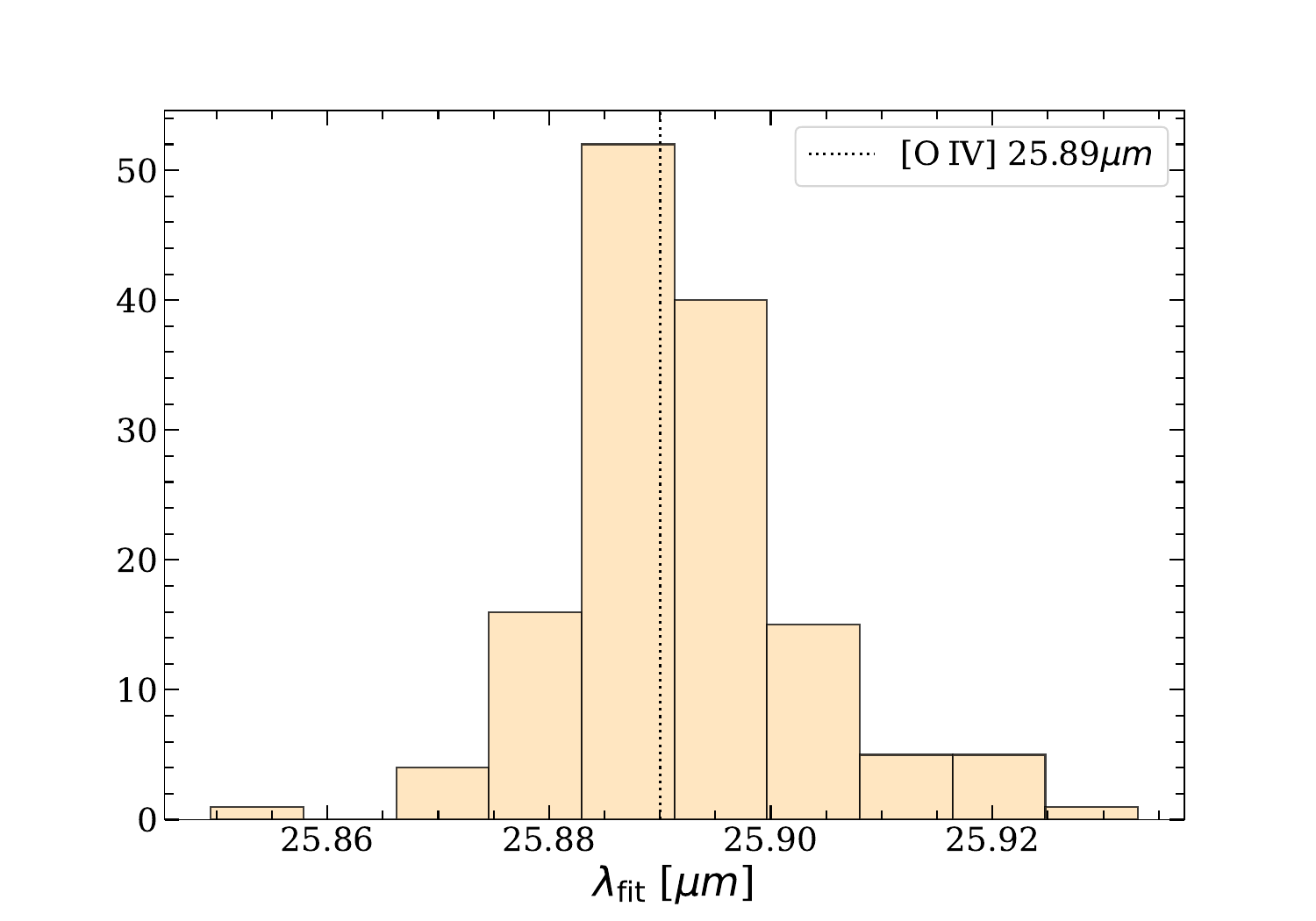} 
    \includegraphics[width=0.33\textwidth]{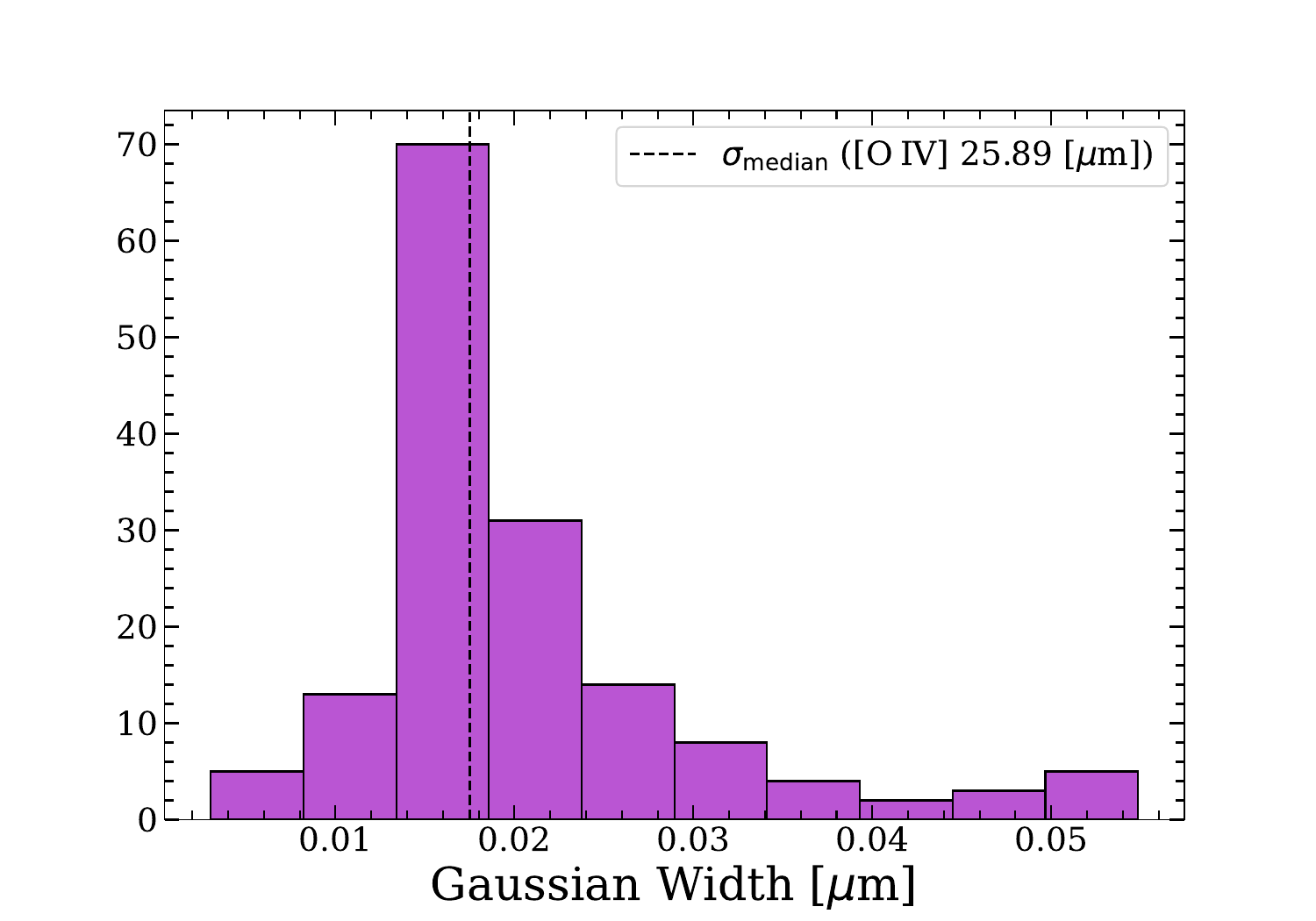}
    \includegraphics[width=0.33\textwidth]{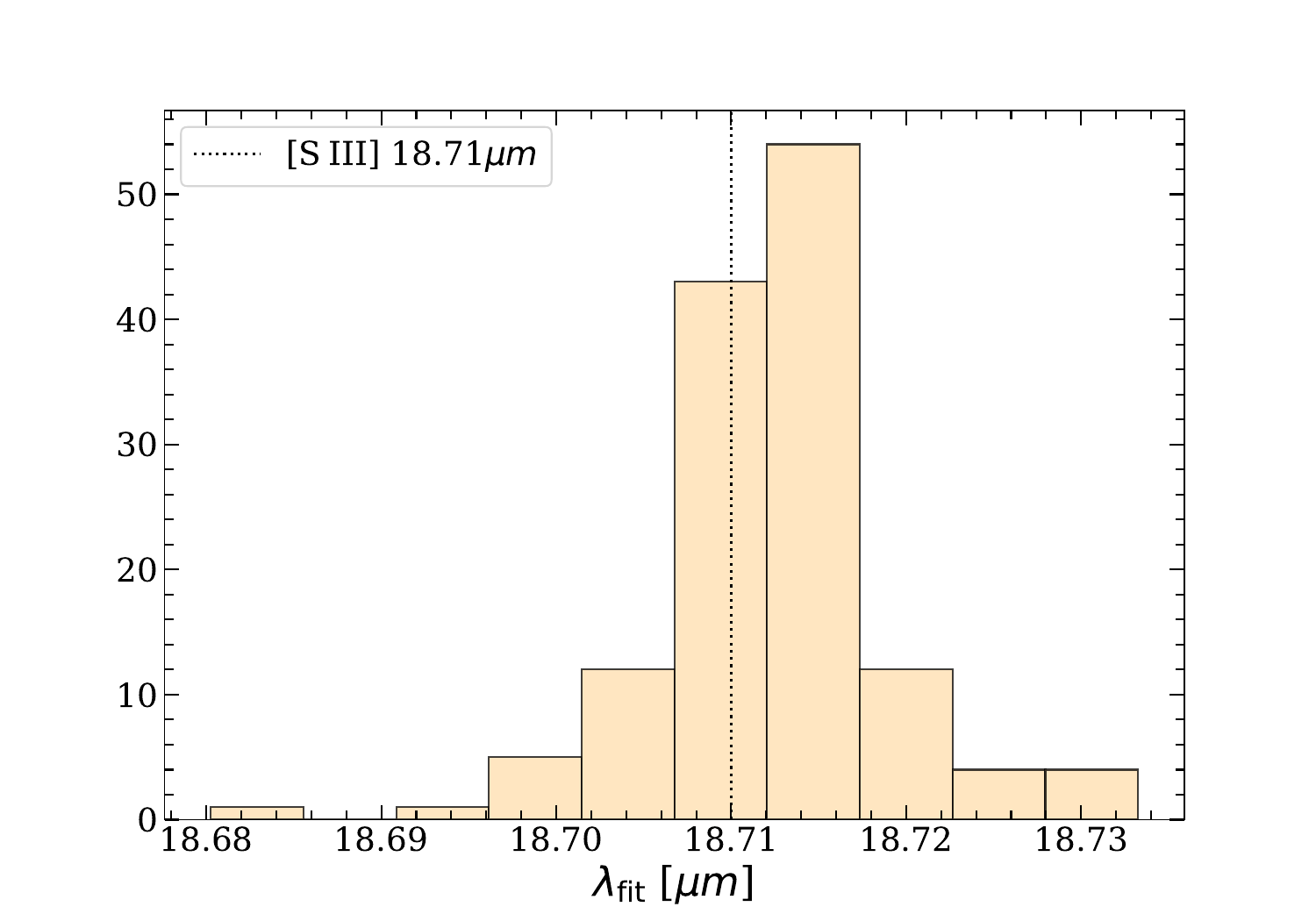} 
    \includegraphics[width=0.33\textwidth]{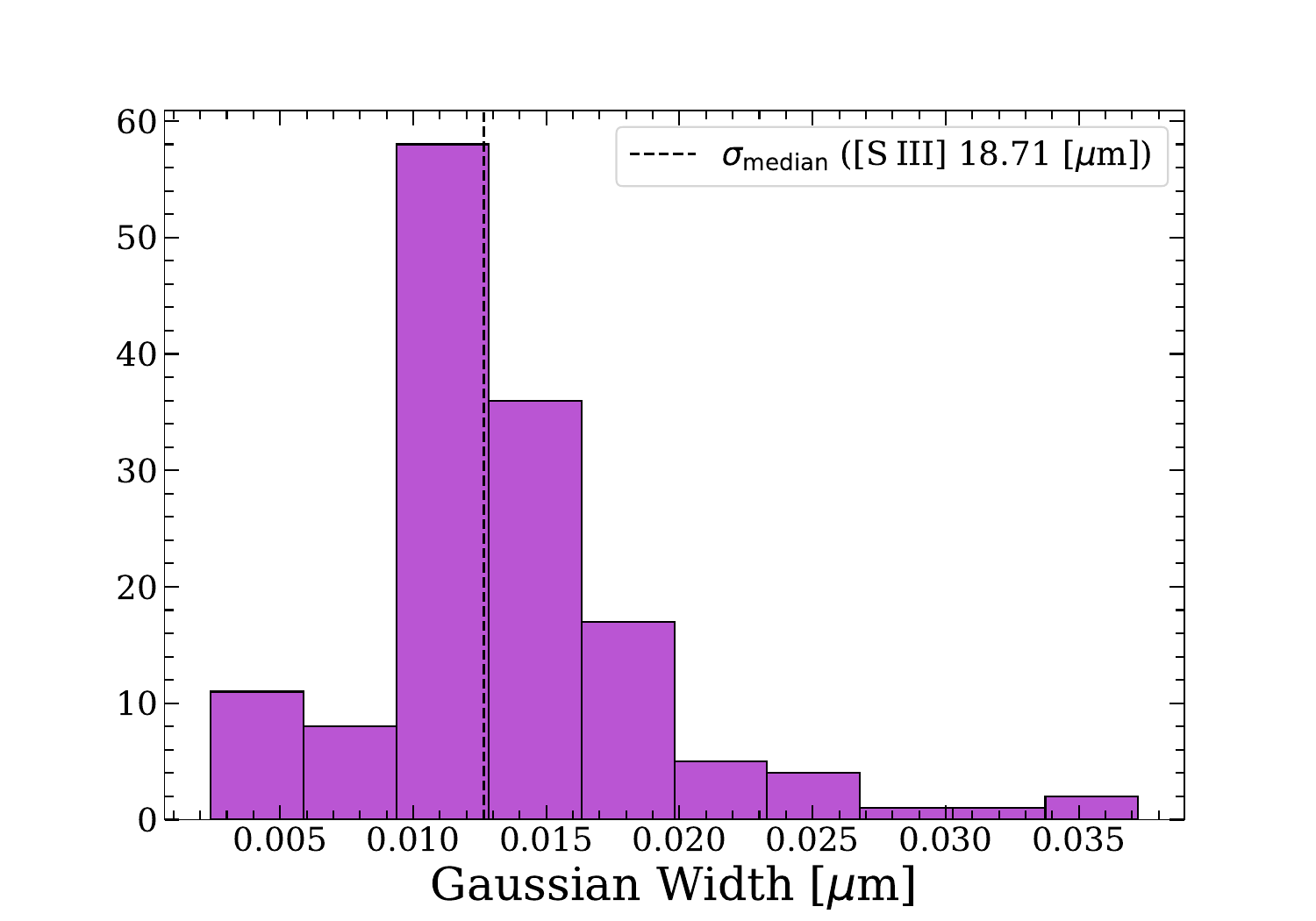}
    \includegraphics[width=0.33\textwidth]{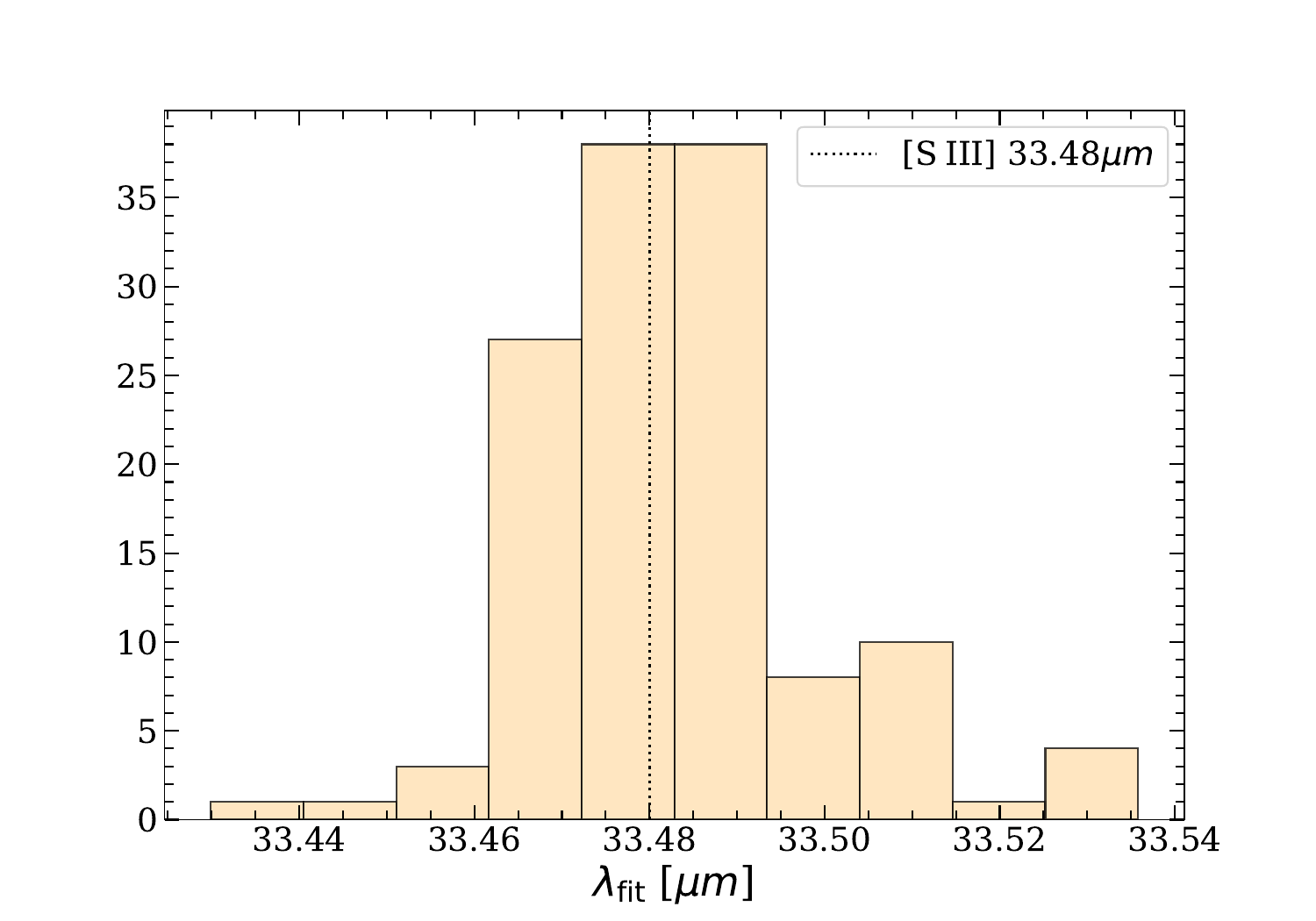} 
    \includegraphics[width=0.33\textwidth]{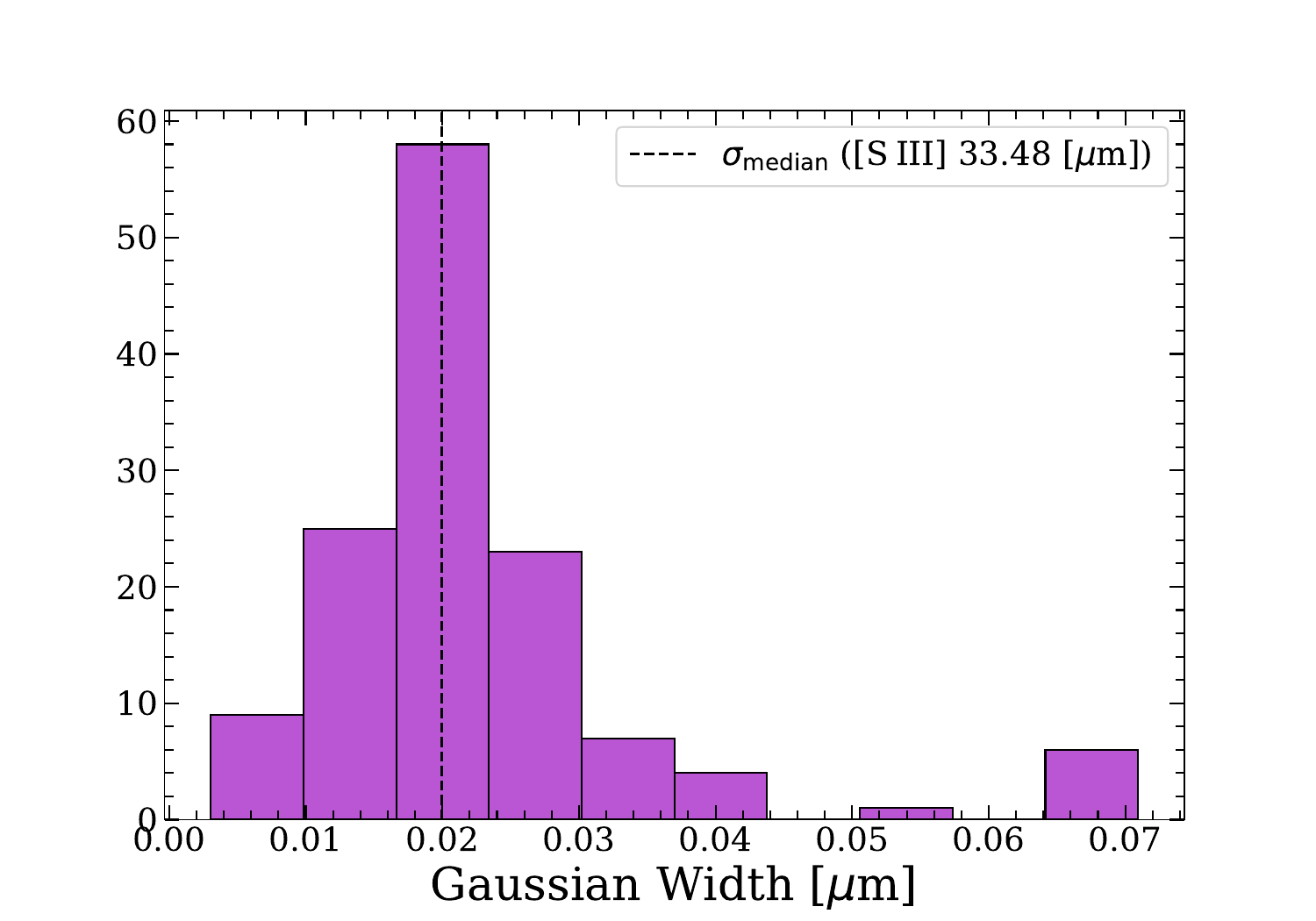}
    \caption{\textit{Left panels:} Histograms of the best fit, $\lambda_{\rm{fit}}$, centroid wavelength  from the line fitting method (see \S\ref{sec:Fitting}) for the forbidden emission lines [{\textsc Ne\,V}] 14.32/24.32\,\micron, [{\textsc S\,III}] 18.71/33.48\,\micron, and [{\textsc O\,IV}] 25.89\,\micron. The distribution of wavelengths peak at the expected wavelength for all five lines (dotted vertical lines). \textit{Right panels:} Histograms of the characteristic width of the Gaussian fits for the five MIR lines. If a source was considered to contain a blue-shifted line and/or a broad emission line, where the second/third gaussian flux was included in the total reported flux (see \S\ref{sec:blue}, \S\ref{sec:broad}, and \S\ref{sec:tripgauss}), then the corresponding $\sigma$ was included as an additional data point in the histogram. } 
    \label{fig:bestlam}
\end{figure*}

In Figure~\ref{fig:bestlam}, we show the histograms of the best-fit centroid wavelengths for the [{\textsc Ne\,V}] 14.32/24.32\,\micron\, [{\textsc O\,IV}] 25.89\,\micron\ and [{\textsc S\,III}] 18.71/33.48\,\micron\ lines. We also illustrate, as dotted vertical lines, the expected centroid wavelength of the lines, showing that the vast majority of the lines are found at wavelengths consistent with the expected ones.

\section{Comparison with Weaver et al. 2010}
\label{app:Weaver}
\begin{figure*}
    \centering
    \includegraphics[scale = 0.33]{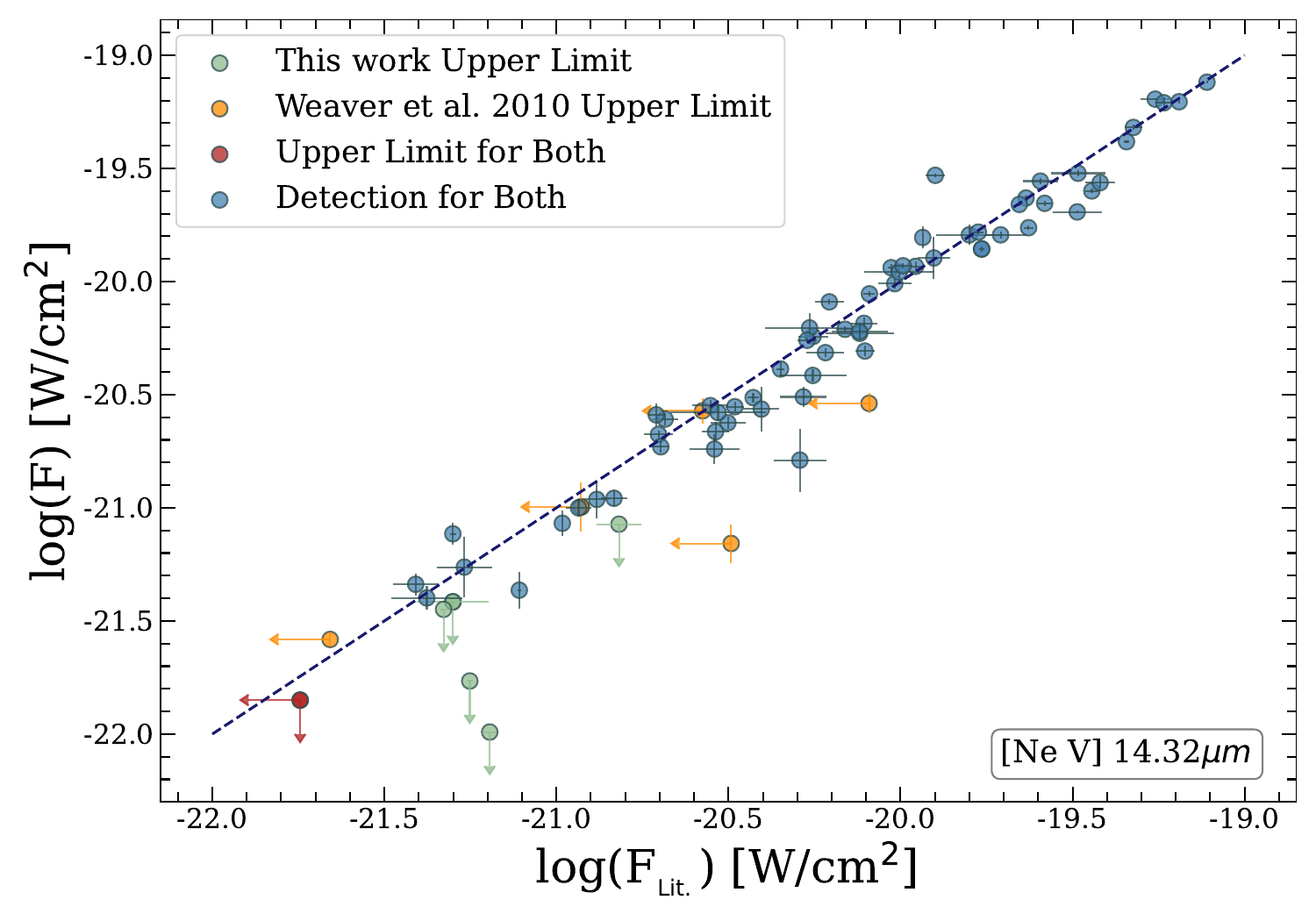}
    \includegraphics[scale = 0.33]{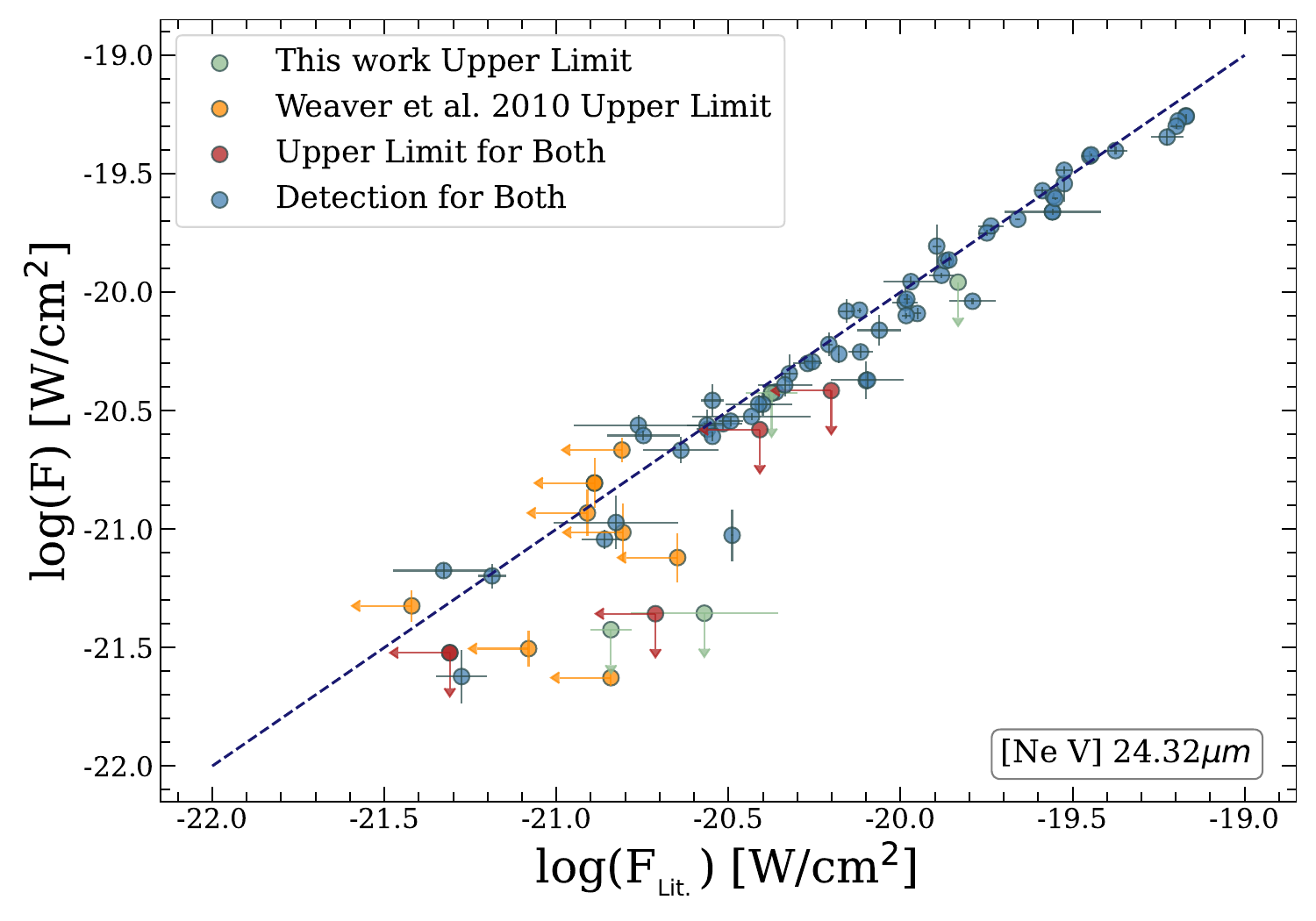} 
    \includegraphics[scale = 0.33]{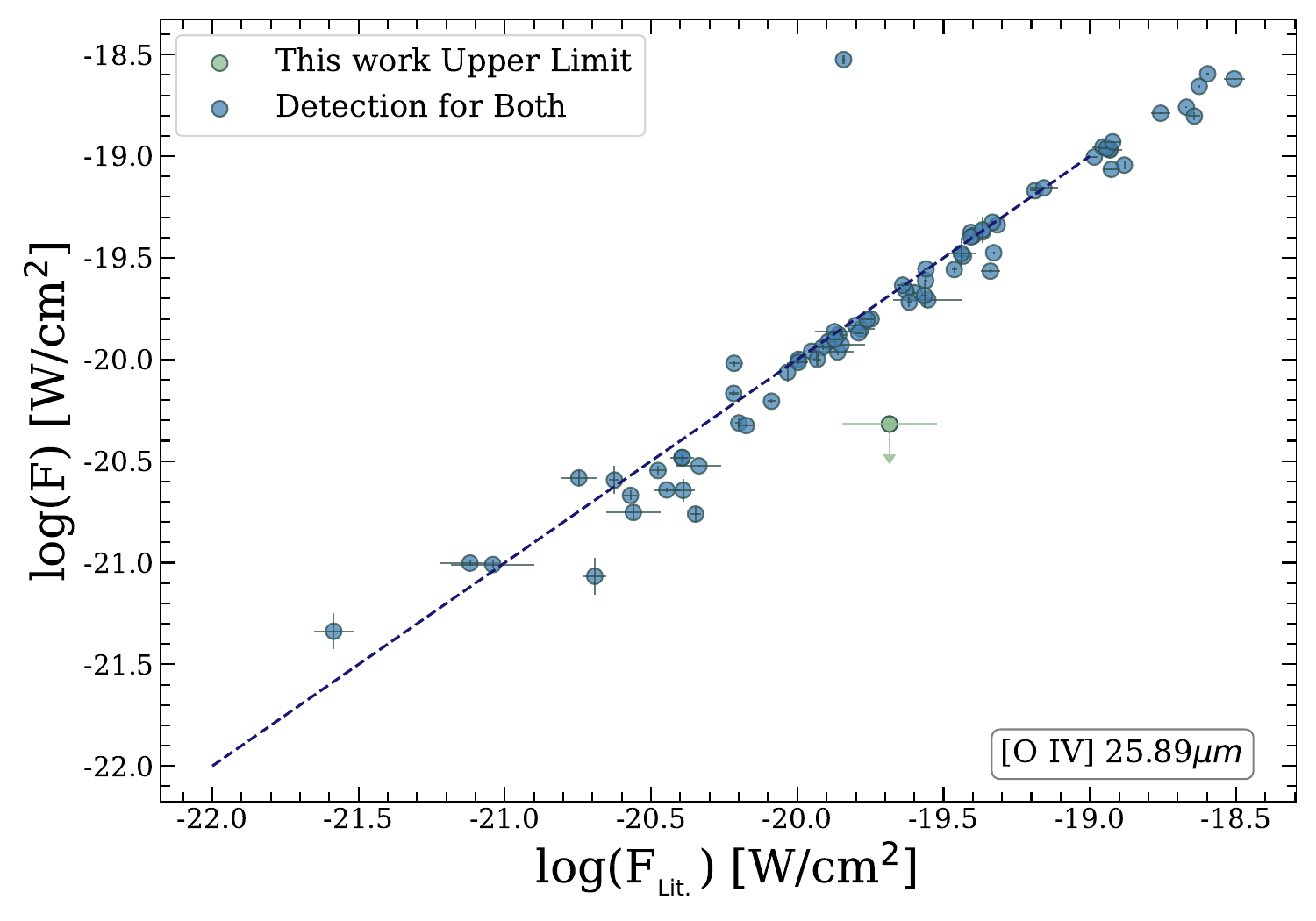} 
    \caption{For the 75 sources from  \citet{2010ApJ...716.1151W} which are also in our sample, we compare the log of their emission-line fluxes with the log of our measurements for the [{\textsc Ne\,V}] 14.32, [{\textsc Ne\,V}] 24.32,  and [{\textsc O\,IV}] 25.89\,\micron\, lines in the top left, top right and bottom panels, respectively. 
    Objects shown in blue have 3-$\sigma$ detections in both works; those in orange and red have upper limits.
    The dashed line shows the 1:1 relationship; our measurements are consistent with those of \citet{2010ApJ...716.1151W}.
    Objects for which only upper limits are derived have low signal-to-noise spectra, but even for these fainter objects our measurements are still in agreement.
    We therefore believe our results are consistent with \citet{2010ApJ...716.1151W}, but we make use of a much larger sample of AGN with \textit{Spitzer} spectra and so we perform our own emission-line measurements in a consistent manner across our whole sample. Note: the values in \citet{2010ApJ...716.1151W} were reported in units of W/cm$^{2}$. As such, we converted our units to match theirs, but only for this comparison. }
    \label{fig:weaver}
\end{figure*}

In Figure~\ref{fig:weaver}, we compare the logs of [{\textsc Ne\,V}] 14.32/24.32\,\micron\ and [{\textsc O\,IV}] 25.89\,\micron\ fluxes we obtained with the log of those reported by \citet{2010ApJ...716.1151W}.
\citet{2010ApJ...716.1151W} studied {\it Spitzer} spectra for a sample of 79 hard X-ray selected AGN from earlier versions of the {\it Swift}/BAT catalogue \citep{Tueller:2008eh,Tueller:2010po}. About half of the sources (75) of our sample overlap with \citet{2010ApJ...716.1151W}. For those sources, our log(fluxes) agree very well with those of \citealp{2010ApJ...716.1151W}, showing very little scatter (0.02, 0.03, and 0.04 dex for [{\textsc Ne\,V}] 14.32/24.32\,\micron\ and [{\textsc O\,IV}] 25.89\,\micron\,, respectively.).

\bibliography{references}
\bibliographystyle{aasjournal}

 \end{document}